# Group 13 Metals as L-Type Ligands for Transition Metals

Hellen Videa, M. Ángeles Fuentes, Antonio J. Martínez-Martínez*

Department of Chemistry and Center for Research in Sustainable Chemistry (CIQSO), University of Huelva, Huelva 21007, Spain.

*Corresponding e-mail: antonio.martinez@dqcm.uhu.es

**Abstract:** Low-valent Group 13 fragments can serve as neutral two-electron L-type metalloligands to transition-metal (TM) centers, enabling heterometallic M–TM platforms with bonding and reactivity patterns distinct from classical CO, phosphine and carbene ligation. This chapter develops a unifying, descriptor-based view of aluminylene Al(I), gallylene Ga(I), and indylene In(I) donors, and contrasts them with the limited L-type behavior of Tl(I). We map synthetic gateways to isolable M(I) donors, analyze their σ-donation/π-acceptance profiles, and extract periodic design rules in which the σ-donor strength decreases Al > Ga > In, whereas Tl(I) has not yet been convincingly shown to engage in neutral L-type Tl→TM coordination. Borderline cases that blur L-, X-, and Z-type classifications are also examined to clarify descriptors and guide consistent usage across the series. This contribution links ligand sterics/electronics, ambiphilicity at M(I), and the chosen TM fragment to guide the rational design of M–TM platforms that harness Group-13 M(I) donors for small-molecule activation and cooperative catalysis.

## 1 Introduction: Group-13 metal L-type donors

In the Covalent Bond Classification (CBC),[1] an L-type ligand donates a full lone pair (two electrons) to a metal center, an X-type ligand contributes only one electron (typically as an anion or radical), and a Z-type ligand acts as a two-electron acceptor (Figure 1).[1d] Low-valent Group 13 fragments in the +1 oxidation state (often termed aluminylenes, gallylenes, etc.) are generally L-type ligands, behaving as neutral two-electron donors akin to CO or $PR_3$. In contrast, an anionic Group 13 species (aluminyl, gallyl, etc.) acts as an X-type ligand, donating one electron to the metal–metal.[2] For example, recent catalysts have employed X-type aluminyl ligands (Al(I) anions) to achieve oxidative addition processes like alkane C–H activation and aryl–F bond cleavage.[3] Conversely, neutral L-type aluminylene ligands (Al(I) donors) coordinate to metals without changing the metal's charge, modulating reactivity in a more subtle, ambiphilic way. The choice between L- and X-type bonding has a tangible impact on reactivity: X-type aluminyls confer strong basic character and have enabled demanding transformations like alkane dehydrogenation and aryl fluoride magnesiation,[4] whereas L-type aluminylenes behave more like classical Lewis base ligands, facilitating cooperative transformations (*vide infra*).

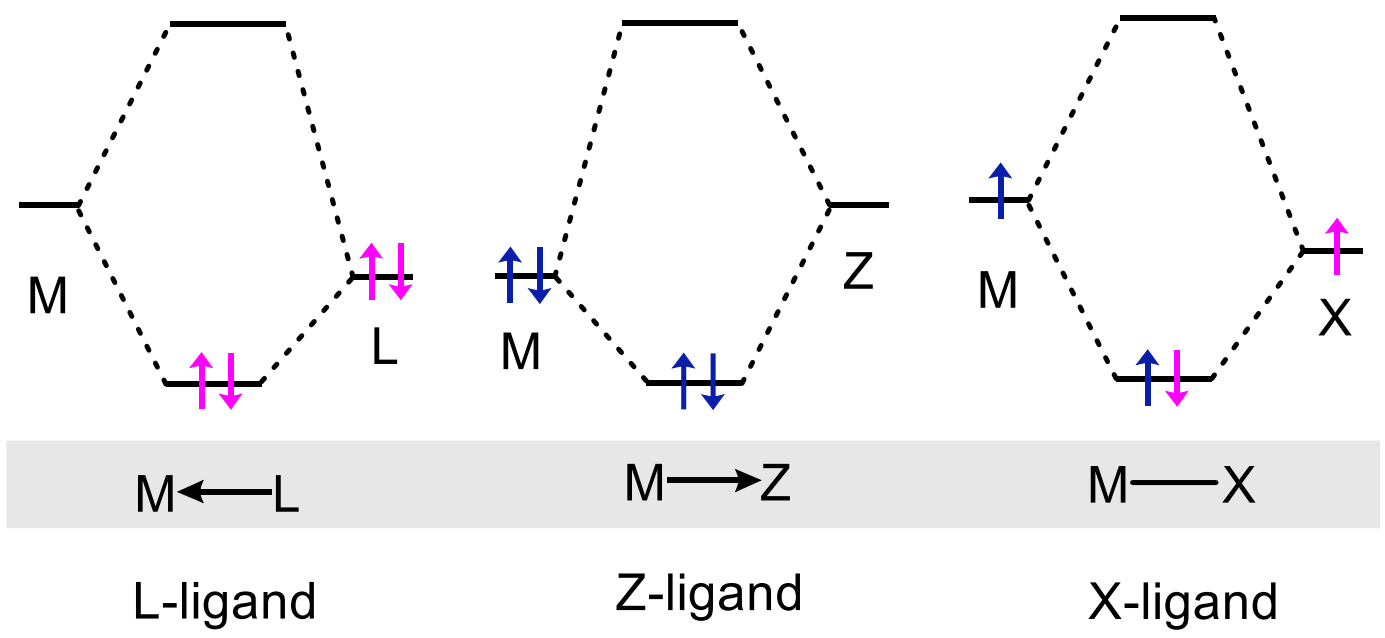

**Figure 1.** The Covalent Bond Classification (CBC) of L-, X-, and Z-type ligands.

Confirming that a given neutral monovalent Group 13 species is behaving as an L-type donor relies on showing that the ligand brings a lone pair and remains neutral upon coordination. Electron-counting and spectroscopic data should indicate that the metal's oxidation state is unchanged by binding of the Group 13 fragment. For instance, attaching a Ga(I) ligand such as Cp*Ga (Cp* = pentamethylcyclopentadienyl) to a transition metal can replace a CO or phosphine ligand without altering the metal's formal oxidation state, analogous to substituting one neutral two-electron donor for another. Computational analyses further demonstrate that these M(I) metalloligands typically donate approximately two electrons in the σ-bond[5] and accept a degree of π-back-donation (similar to CO), yet they do not withdraw enough electron density to become formally anionic.[6] Thus, the M–TM bond (where M denotes the Group 13 center and TM the transition metal) in such cases is best described as a coordinate covalent linkage with both electrons originating from the Group 13 lone pair. If the interaction were X-type instead, one would expect the metal's oxidation state to shift and often observe a shorter, more polar metal–metal bond, since anionic ligands generally produce shorter M–X bonds than neutral donors.[7] In practice, many M–TM bonds with neutral Al(I) or Ga(I) ligands are indeed relatively long and polarized, but they are still better viewed as dative bonds, often denoted M→TM, rather than true ionic M–TM$^-$ linkages. Nevertheless, some complexes blur the distinction between L- and X-type character or involve multi-center bonding that challenges simple classification. These borderline cases merit special attention and are discussed later in this chapter.

Considering the isolobal analogy between low-valent Group 13 species (M:) and carbenes ($R_2C$:), researchers have also been exploring carbene analogs that incorporate heavier main-group elements. One strategy involves replacing the carbon atom in a carbene with elements from Groups 13, 14, or 15.[8] This approach has yielded novel ligand systems such as borylenes (R–B:, boron(I) analogs of carbenes), germylenes ($R_2Ge$:, germanium(II) analogs), and nitrenes (R–N:, neutral singlet nitrogen analogs), which exhibit electronic properties similar to traditional carbenes.[9] These insights underscore the significance of studying how a low-valent main-group center and a transition metal can work in tandem.

Focusing specifically on Group 13 elements, metals like aluminum, gallium, indium, and thallium exhibit dual reactivity modes depending on their oxidation state. In the +3-oxidation state, M(III), they predominantly act as Lewis acids or Z-type acceptor ligands, readily engaging in electrophilic interactions when partnered with transition-metal centers. However, in the +1-oxidation state, M(I), these species feature a lone pair and an accessible p-orbital. This endows them with strong σ-donor and effective π-acceptor characteristics, allowing them to facilitate nucleophilic activation of otherwise inert carbon–heteroatom bonds. This dual reactivity is illustrated schematically in Figure 2.

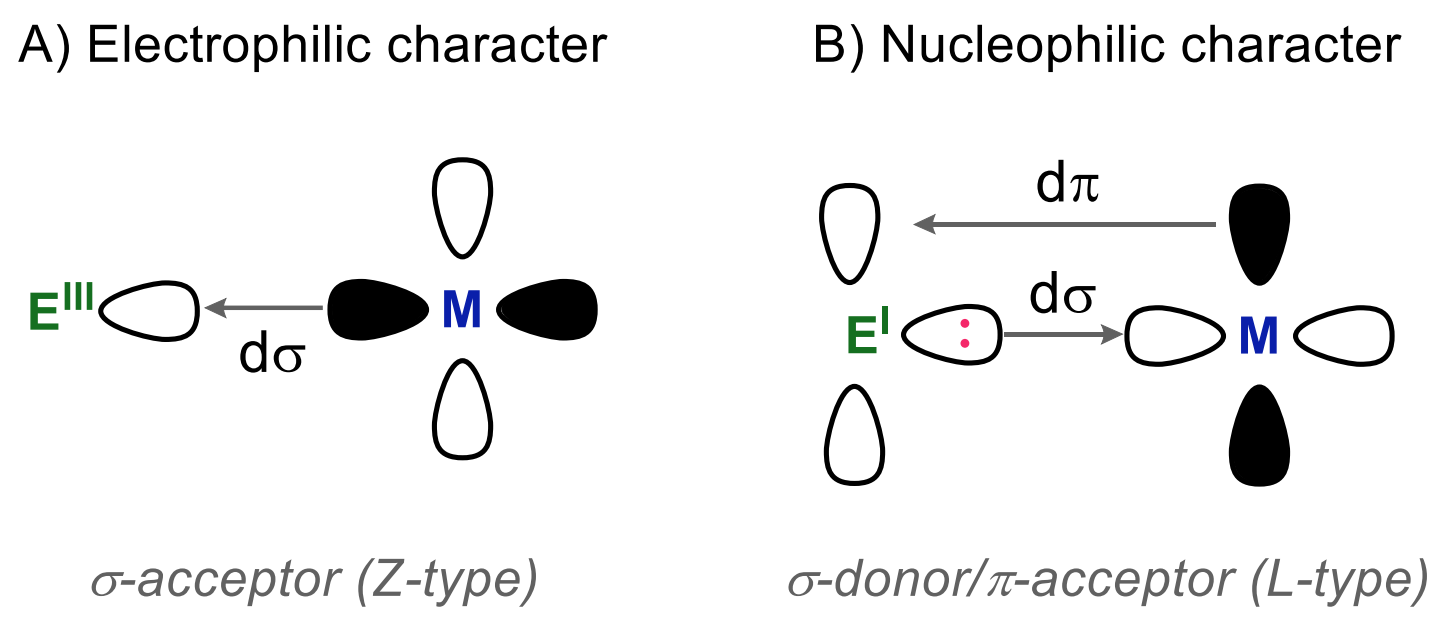

**Figure 2.** Dual electronic nature of Group 13 metalloligands.

The late-20th-century isolation of stable low-valent Group 13 metal complexes opened an innovative avenue for coordination chemistry. These M(I) organometallics, stabilized by anionic bulky ligands, derive their unique ligand properties from the combination of a filled lone-pair orbital and an orthogonal empty p-orbital on the Group 13 center. This electronic configuration imparts a blend of σ-donor and π-acceptor abilities reminiscent of classical carbenes ($R_2C$) or phosphines ($R_3P$). Consequently, when bound to transition metals, such Group 13 centers act as versatile metalloligands, often functioning as neutral two-electron donors (L-type) and forging metal–metal bonds that depart from the norms of conventional ligand systems.[6a, 10]

This chapter will specifically highlight examples of L-type coordination by Group 13 metalloligands (Al, Ga, In, Tl) to transition metals. Both instances of terminal coordination (a single metal–Group 13 bond) and bridging coordination (where a Group 13 donor spans two or more metal centers) are covered. Borderline cases and debated bonding descriptions are noted where relevant. However, complexes in which the Group 13 fragment clearly function as an X-type nucleophile or as a Z-type acceptor are outside the scope of this discussion. The sections that follow are organized by element to systematically explore each family of metalloligands. In turn, Al(I), Ga(I), In(I), and Tl(I) complexes are discussed, detailing the synthesis of these low-valent Group 13 ligands, their typical structural motifs (terminal vs. bridging coordination to metals), electronic structure and bonding analyses, and notable reactivity patterns including any catalytic applications. Key literature milestones, from the first isolable examples to recent cutting-edge developments, are highlighted to illustrate the evolution of this field.

## 2 Aluminum as an L-type ligand in transition-metal complexes

### 2.1 Monomeric neutral Al(I) species: Design and metrics

Historically, hints of monovalent aluminum chemistry can be traced back to the 1940s with observations of partially reduced aluminum halides (AlX, X = Cl, Br, I).[11]

Modern chemistry of neutral Al(I) centers, however, has been driven largely by organometallic frameworks such as pentamethyl cyclopentadienyl (Cp*) and β-diketiminate (BDI) ligands, which can stabilize the elusive Al(I) species. The quest for isolable organoaluminium(I) compounds began in earnest with a breakthrough by Schnöckel in 1991,[12] who reported the first room-temperature stable Al(I) complex, the tetrameric $[Cp^*Al]_4$ species **1** by reacting AlCl with $(Cp^*)_2Mg$ ($Cp^* = C_5Me_5$, Figure 3). In the solid state, **1** adopts a tetrahedral $Al_4$ core supported by four $\eta^5$-Cp* ligands; notably, in solution it readily dissociates into monomeric Cp*Al units, enabling the monomer's rich reactivity as a donor ligand.[10a, 10c, 13] An alternative route to the same Al(I) tetramer was soon developed via reductive dehalogenation: treating the Al(III) precursor $Cp^*AlCl_2$ with potassium metal cleanly furnishes $[Cp^*Al]_4$.[14] These methods provided access to Cp* stabilized Al(I) and opened the door to further exploration of monovalent aluminum chemistry.

Subsequent modifications to the cyclopentadienyl ligand demonstrated ways to shift the monomer–oligomer equilibrium more towards "free" Al(I) centers. For example, employing bulkier cyclopentadienyl ligands in place of Cp* yielded $(C_5(CH_2Ph)_5)Al$ **2**,[13d] $(C_5H_2(1,2,4\text{-}(SiMe_3)_3))Al$ **3**[13c] and $(C_5H_2(1,2,4\text{-}tBu_3))Al$ **4** (Figure 3),[13b] showing a greater tendency to exist as discrete monomers in solution and preventing Al–Al aggregation. These results hinted at "masked" monovalent Al(I) centers, analogous to carbenes in the sense that they are lone-pair donors, stabilized by substantial steric or electronic protection from the ligands. More recently, an extremely bulky pentaisopropyl cyclopentadienyl ligand was used to isolate monomeric $(Cp^{iPr})Al$ **5** ($Cp^{iPr} = C_5iPr_5$, iPr = isopropyl group). Complex **5** was obtained by breaking apart the tetramer **1** via reaction with $(Cp^{iPr})Li(OEt_2)$, thereby trapping the Al(I) as a monomeric Cp derivative.[15] Each of these cyclopentadienyl variants provided further evidence that appropriate ligand design can render Al(I) stable in isolation, much like a classical carbene.

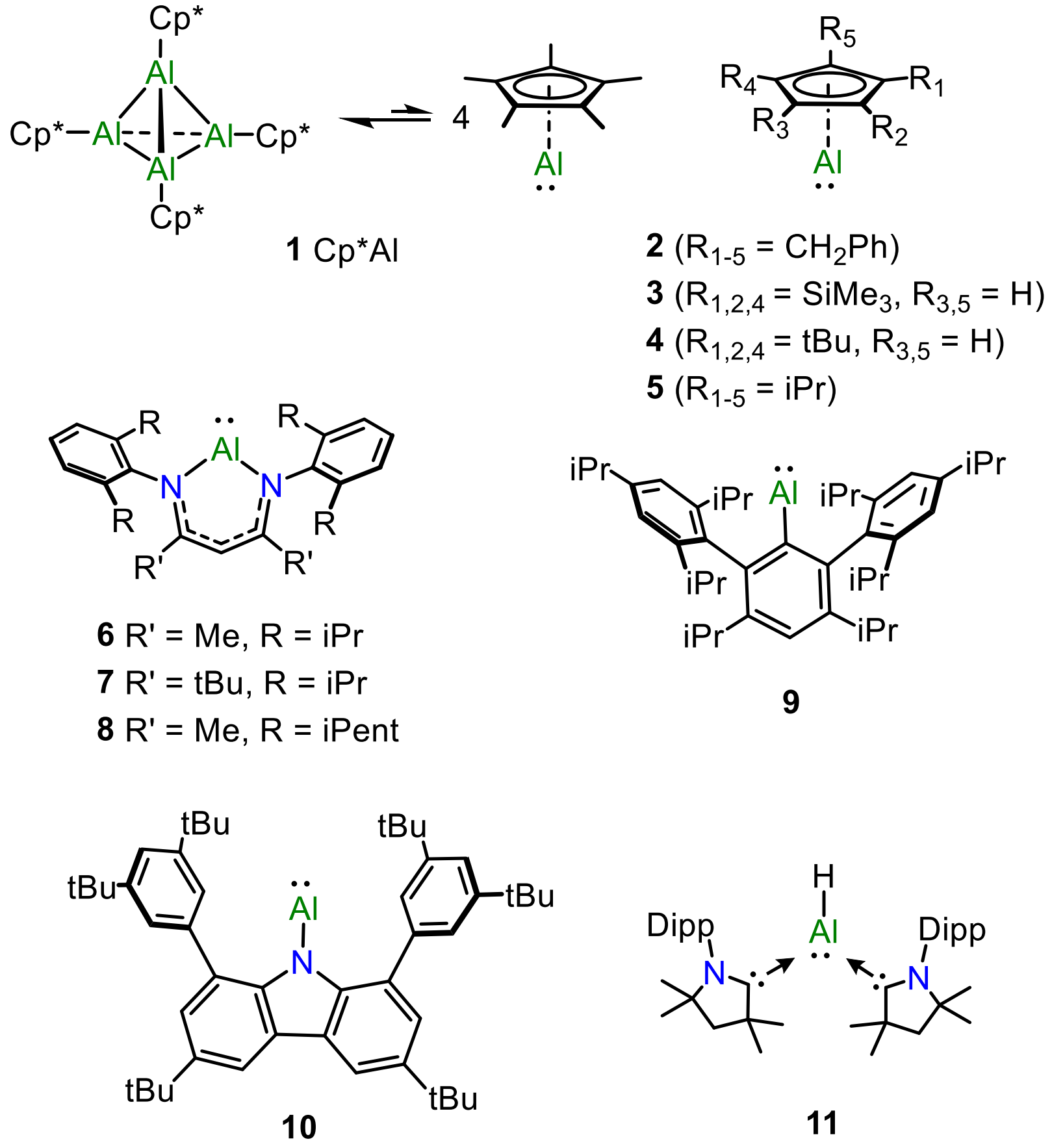

**Figure 3.** Selected examples of neutral Al(I) compounds.

In parallel with the Cp-based approaches, Roesky and colleagues introduced a different scaffold that led to truly monomeric Al(I) species: the β-diketiminate ligand. They described the monovalent aluminum complex (BDI)Al(I) **6** (Figure 3), obtained by reducing a diiodo-derivative (BDI)$AlI_2$ with potassium metal.[16] Complex **6**, often termed an “aluminylene” is analogous to an N-heterocyclic carbene: the BDI ligand confers both strong σ-donor support and π-delocalization, stabilizing the low-coordinate Al center. The choice of a bulky β-diketiminate was strategic, as such ligands are well-known to coordinate many metals across the periodic table in various oxidation states.[17] Roesky's aluminylene **6** was among the first bona fide monomeric Al(I) compounds crystallized in the solid state, proving that a robust chelating ligand could anchor a low-valent aluminum center in isolation. Building on this success, several variants have been prepared by modifying the β-diketiminate substituents. For instance, the tert-butyl-substituted analog **7** was synthesized by reduction of the corresponding diiodide,[18] and a version with bulkier 2,6-diisopentylphenyl aromatic flanks **8** was similarly realized via $KC_8$ reduction of an Al(III) diiodide precursor.[19] These BDI-supported aluminylenes solidified the analogy between Al(I) and carbene chemistry, establishing a family of Al(I) complexes that are isolable, monomeric, and nucleophilic at aluminum.

Despite these advances, one recurring challenge is that many neutral Al(I) compounds tend to exist in equilibrium with dimeric or oligomeric forms or can undergo facile disproportionation to Al(III) and Al metal. To suppress such pathways, extremely bulky ancillary ligands or rigid frameworks are often required (e.g., terphenyl ligands or amide backbones have been employed in some cases). Roesky's BDI complex **6** was a breakthrough in achieving a "true" monomeric Al(I), and recent research has pushed the limits of steric protection even further. For example, Power reported an unsupported monovalent Al center stabilized solely by very bulky aryl groups in complex **9**, achieved by reducing an aryl–Al(III) diiodide with a Na/NaCl alloy.[20] Similarly, the one-coordinate aluminylene **10** featuring a carbazolyl-derived ligand can be obtained via $KC_8$, K/KI $(Cp^*)_2Co$ reduction of a diiodo-alane.[21] These species **9** and **10** (Figure 3) essentially represent monomeric Al(I) trapped by massive ligands, forgoing the need for conventional multidentate chelation. The continued emergence of such compounds underscores the theme that increasingly large or electron-releasing ligands can “free” the Al(I) center while keeping it kinetically protected.

Another remarkable development in Al(I) chemistry is the isolation of aluminum(I) hydride complexes stabilized by carbenes. Braunschweig and coworkers reported the complex $(CAAC^{Me4})_2AlH$ **11**, which features a formal Al(I) center bonded to a hydride and supported by two cyclic (alkyl)(amino)carbene (CAAC) ligands.[22] Complex **11** was synthesized by reducing a dihalogenoalane precursor ($RAlX_2$) with $KC_8$ in the presence of CAAC ligands, which donate strongly into Al and also accommodate its low coordination number. The combination of steric bulk and potent σ-donation from two CAACs allowed the Al center to accept a hydride and remain monovalent. This species can be viewed as the Al(I) analog of a hydrido-carbene complex, and it highlights the expanding repertoire of low-valent Al compounds with unusual substituents ($H^-$ in this case). Although still relatively few, these recent low-coordinate Al(I) complexes (including **9**, **10**, **11** and others) show great promise: they have already begun to exhibit novel reactivity patterns and are likely to play key roles in advancing new chemistry of aluminum in its +1-oxidation state.

## 2.2 Cp*Al as a carbonyl-analog L-type donor: Bonding and reactivity

With reliable sources of monovalent aluminum in hand, such as the Cp*-based tetramer **1,** chemists turned to exploring Al(I) as a two-electron donor ligand to transition metals, akin to an L-type “aluminylene” ligand analogous to a phosphine or NHC. The earliest examples of aluminum(I) coordinated to transition metals appeared in the mid-1990s. Schnöckel reported the first Al–TM complex in 1995, wherein the Cp*Al fragment acts in a bridging capacity between metal centers.[23] By reacting the Ni(0) complex $(Cp)_2Ni$ with tetrameric $[Cp^*Al]_4$ **1**, they obtained a bimetallic species formulated as $(CpNi)_2(\mu\text{-}Cp^*Al)_2$ **12** was carried out using **1** as a reducing agent in the presence of $(Cp)_2Ni$ (Figure 4) featuring two nickel atoms bridged by two Al(I) ligands. This demonstrated that the Al(I) unit could coordinate to metals, although in this case each Al was shared by two Ni atoms (forming Ni–Al–Ni bridges). Shortly thereafter, Fischer and Frenking achieved the first complex with a terminal Al(I) ligand bound to a single transition metal center.[5a] Using a direct salt-metathesis route, they reacted the dichloride $[Cp^*AlCl_2]_2$ with an excess of Collman's reagent $K_2[Fe(CO)_4]$, which led to the isolation of a carbonyl complex characterized as $(Cp^*Al)Fe(CO)_4$ **13**. In complex **13**, the $Fe(CO)_4$ fragment is bound by a lone-pair donor Cp*Al unit, unequivocally establishing that Cp*Al can serve as a one-electron donor into a metal center (Al–Fe single bond). Around the same time, Schnöckel reported that treating $Co_2(CO)_8$ with $[CpAl]_4$ yields a tetra-nuclear cluster $(\mu\text{-}Cp^*Al)_2Co_2(CO)_6$ **14**.[24] The structure of **14** features an $Al_2Co_2$ core where two Cp*Al ligands bridge a Co–Co bond (each Al bonded to both Co atoms), forming a distorted tetrahedron. Notably, the Al atoms in **14** do not retain a purely "carbene-like" coordination to any single metal, instead participating in multicenter bonding within the $Al_2Co_2$ framework (DFT calculations confirmed the Al centers in **14** are better described as part of a Co–Al–Al–Co cluster, rather than isolated Al→Co donors). Nonetheless, complex **14** expanded the scope of aluminylene chemistry to bimetallic clusters. Yet another early example was provided by Schnöckel's group in 1999, who prepared the aluminum–chromium complex $(Cp^*Al)Cr(CO)_5$ **15** by reacting Cp*Al with $Cr(CO)_5(COT)$ (COT = 1,3,5,7-Cyclooctatetraene).[5c] Complex **15** contains a bona fide Al–Cr bond, further showcasing that even hard, electron-rich metals like Cr(0) can accept coordination by the soft Al(I) donor. These pioneering studies (summarized in Figure 4) established that Cp*Al could bind to transition metals in various modes, terminally in **13** and **15** or in bridging positions in **12** and **14**, laying the groundwork for a new domain of organometallic chemistry.

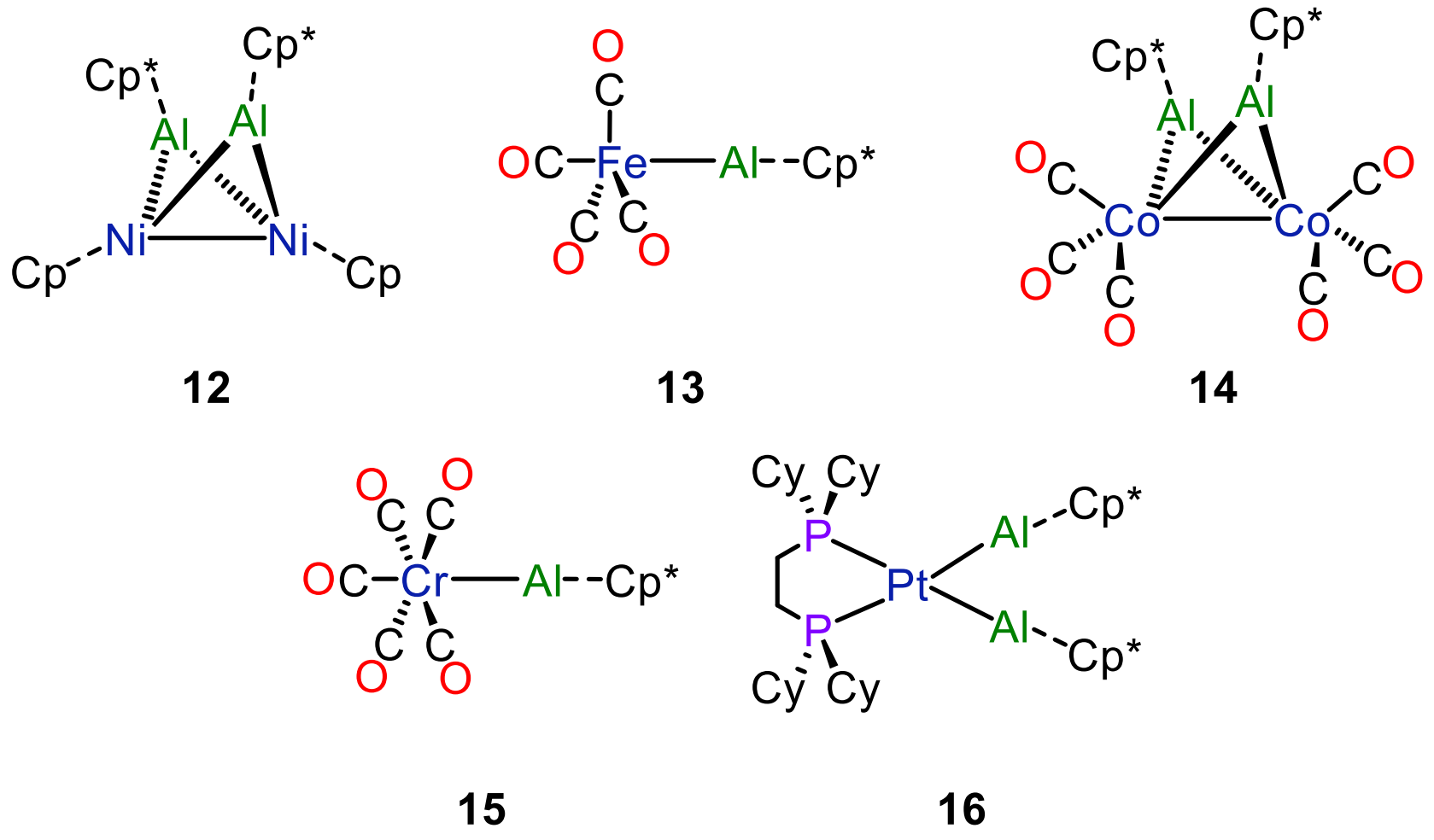

**Figure 4.** Early examples of Al→TM L-type compounds.

After the initial carbonyl and sandwich complexes, the field evolved toward the exploration of aluminylene coordination in other ligand environments to understand factors affecting Al–TM bond stability. Treating a σ-bonded platinum(II) complex (dcpe)Pt(H)($CH_2$tBu) (dcpe = 1,2-bis(dicyclohexylphosphino)ethane), with $[CpAl]_4$ led to the formation of a tetrahedral Pt(0) species $(Cp^*Al)_2$Pt(dcpe) **16** (Figure 4), in which two terminal Cp*Al ligands coordinate to a platinum center.[25] Compound **16** was crystallographically characterized as a $PtAl_2$ unit with Pt–Al single bonds and a chelating dcpe ligand completing the coordination sphere. The successful isolation of **16** highlighted how ancillary ligands on the metal can influence the stability of Al–TM bonds: the strong donor phosphine (dcpe) helps electron-rich Pt stabilize the Al ligands, and the initial presence of a hydride and alkyl on Pt (which eliminate as iso-butane) provides a convenient entry to the Pt(0) state bonded to Al(I). This example underscored that both the electronic properties of the metal complex and the nature of substituents on aluminum (here Cp*) are crucial in obtaining robust Al–metal linkages. In essence, using a supporting ligand framework (like phosphines or carbonyls) on the transition metal can mitigate the tendency of the aluminylene either to abstract or to oligomerize, thereby yielding stable, isolable Al–TM complexes.

Fischer and co-workers further demonstrated that it is possible to coordinate multiple Al(I) ligands to a single transition metal center, achieving highly unusual homoleptic complexes. Reacting zero-valent Ni with Cp*Al produces a four-coordinate Ni–Al complex. Specifically, the reaction of $Ni(COD)_2$ (COD = 1,5-cyclooctadiene) with $[Cp^*Al]_4$ resulted in the compound $(Cp^*Al)_4Ni$ **17** (Figure 5A), in which a Ni(0) center is bound by four Cp*Al ligands arranged tetrahedrally.[26] Similarly, using a palladium(II) precursor, $(tmeda)PdCl_2$ (tmeda = N,N,N',N'-tetramethylethylenediamine) the species $(Cp^*Al)_4Pd$ **18** is generated (Figure 5A).[26] In the latter case, the formation of **18** implies that the Pd(II) is reduced to Pd(0) (likely with concomitant oxidation of Al(I) to Al(III) chloride species as byproducts) and that four Cp*Al ligands then coordinate the resulting Pd center. Species **17** and **18** are remarkable as they contain homoleptic aluminylene coordination (no conventional ligands like CO or phosphine remain) and feature metals in a highly electron-rich environment. Such compounds are inherently interesting due to their high Al:TM ratio and as potential precursors to multimetallic clusters or catalytically active species. Indeed, complex **17** proved to be a versatile starting material in subsequent reactions. For example, $(CpAl)_4Ni$ **17** reacts with the gold(I) complex $ClAuPPh_3$ via ligand substitution and reduction to produce heterometallic Au–Ni–Al cluster compounds, giving heterometal-doped gold clusters, $[Ni(AuPPh_3)_6(AuCl)_3(Cp^*Al)]$ **19** and $[Ni(AuPPh_3)_4(AuCl)_3(Cp^*Al)_2]$ **20** (Figure 5B).[27] These can be described as $Ni@Au_9Al$ and $Ni@Au_7Al_2$ core-shell clusters, respectively, where a central Ni atom is encapsulated in a shell of Au atoms, with one or two Cp*Al units also bound in the outer shell. The formation of **19** and **20** demonstrates that Al(I) ligands can facilitate the assembly of complex multimetallic architectures. The aluminylene not only binds to Ni but can also bridge or cap Au–Ni or Au–Au interactions during cluster growth. This illustrates how Cp*Al can act as a flexible connector or "glue" in heterometallic clusters, enabling combinations of metals that might be difficult to achieve through direct synthesis.

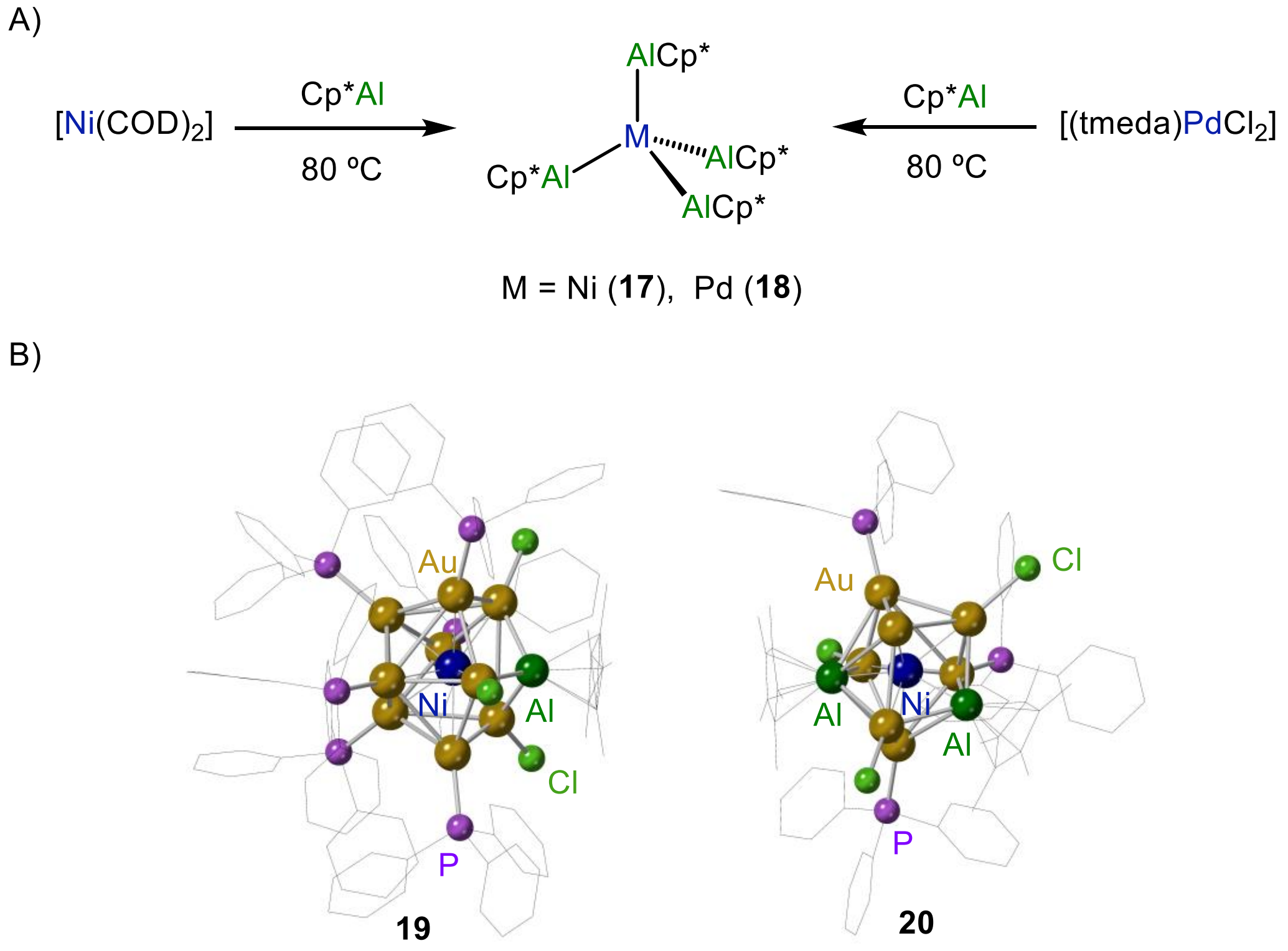

**Figure 5.** A) Synthesis of homoleptic Ni–Al and Pd–Al complexes. B) Molecular structures of the resulting clusters from the $(CpAl)_4Ni$ **17** reaction with the gold(I) complex.

The coordination chemistry of Cp*Al with nickel is particularly rich. In addition to the tetrahedral complex $(Cp^*Al)_4Ni$ **17** a dinuclear Ni–Al complex $(Cp^*Al)_5Ni_2$ **21** can form under certain conditions. Saillard and Fischer reported that treating a nickel(0) source $Ni_2(dvds)_3$ (dvds = 1,3-divinyltetramethyldisiloxane, a stabilizing ligand) with an excess of $(Cp^*Al)_4$ in toluene leads primarily to $(Cp^*Al)_4Ni$ **17** but also to the minor bimetallic byproduct $(Cp^*Al)_5Ni_2$ **21**.[28] Compound **21** contains a Ni–Ni bond with five Cp*Al ligands bridging and terminally coordinated in a mixed fashion (likely three Al bridging the Ni–Ni and two terminal, in analogy to known $Pd_2Al_5$ clusters). Recrystallization allowed **21** to be isolated pure, indicating that even with the same reactants the nuclearity of the product can vary, possibly controlled by subtle kinetic factors or stoichiometry. The isolation of **21** alongside **17** underscores that Cp*Al ligands can support metal–metal bonded complexes (in this case a Ni–Ni core) by saturating the coordination environment and preventing further cluster growth.

By leveraging the lability of certain Ni–Al bonds, researchers have developed strategies to introduce additional ligands or to activate small molecules. For instance, addition of a neutral two-electron donor like $PEt_3$ triggers reductive elimination of triethylsilane ($Et_3Si$–H) from the hydrido(silyl) complex $(Cp^*Al)_3Ni(H)(SiEt_3)$ and coordination of the phosphine (Figure 6A), yielding the tris(aluminylene) nickel hydride species $(Cp^*Al)_3Ni(PEt_3)$ **22**.[29] In **22**, the Ni center retains three Cp*Al ligands and now binds a phosphine, demonstrating that not only can Cp*Al donate to Ni, but the Ni–Al framework can accommodate classical ligands via controlled ligand exchange. This experiment also highlights that Ni–H and Ni–Si bonds in such aluminylene complexes are reactive and can be exploited to tailor the coordination sphere (for example, removing H/$SiEt_3$ to allow a new donor to bind Ni).

Perhaps even more striking is the ability of Ni–Al complexes to mediate C–H and Si–H bond activation, showcasing unique cooperative reactivity. In 2004, Fischer reported that the intermediate $(Cp^*Al)_3Ni$ **23** (Figure 6B), likely a 16-electron Ni(0) species stabilized by three Al(I) ligands that can activate a silane and even an arene C–H bond.[30] When $(Cp^*Al)_3Ni$ **23** was treated with a slight excess of Cp*Al (effectively adding one more Cp*Al unit) in the presence of triethylsilane ($Et_3SiH$), two novel complexes were isolated. One is the monohydrido complex $(Cp^*Al)_3Ni(H)(SiEt_3)$ **24**, which contains a Ni–H bond and a Ni–$SiEt_3$ bond (the $SiEt_3$ likely arising from oxidative addition of $Et_3Si$–H to Ni, with Al ligands stabilizing the Ni center).[30] The other is a species formulated as $(Cp^*Al)_3Ni(\mu_2\text{-}H)((C_6H_5)Cp^*Al$ **25**. Species **25** features a bridging hydride ($\mu^2$-H) shared between Ni and one Al center, and significantly, a $(C_6H_5)Cp^*Al$ unit, which implies that a phenyl group has become attached to one of the Cp*Al ligands. The presence of the $C_6H_5$ substituent suggests that a C–H bond (most plausibly from an aromatic solvent or from a Cp* methyl group) was cleaved, and the phenyl fragment bonded to Al, forming an anionic ($C_6H_5$-Cp*)Al species, while the liberated hydrogen ended up as the Ni–Al bridging hydride. These reactions demonstrate a cooperative effect. Ni(0) alone might activate $Et_3Si$–H or benzene only under harsh conditions, but in the presence of aluminylene ligands, oxidative addition occurs at Ni with the Al ligands helping to stabilize the fragments (e.g., by accepting a phenyl group or a hydride). The result is that the Ni–Al complex can activate strong σ-bonds (Si–H, aryl C–H) at mild conditions, a reactivity pattern that parallels what has been observed in some Ga–Ni systems and even Mg–Ni (Grignard) chemistry, but here the Al(I) ligand plays a dual role of both donor and participant in bond cleavage. This unusual reactivity of $(Cp^*Al)_3Ni$ highlights the potency of aluminylene ligands in facilitating oxidative addition processes at transition-metal centers, a theme that mirrors other systems described below.

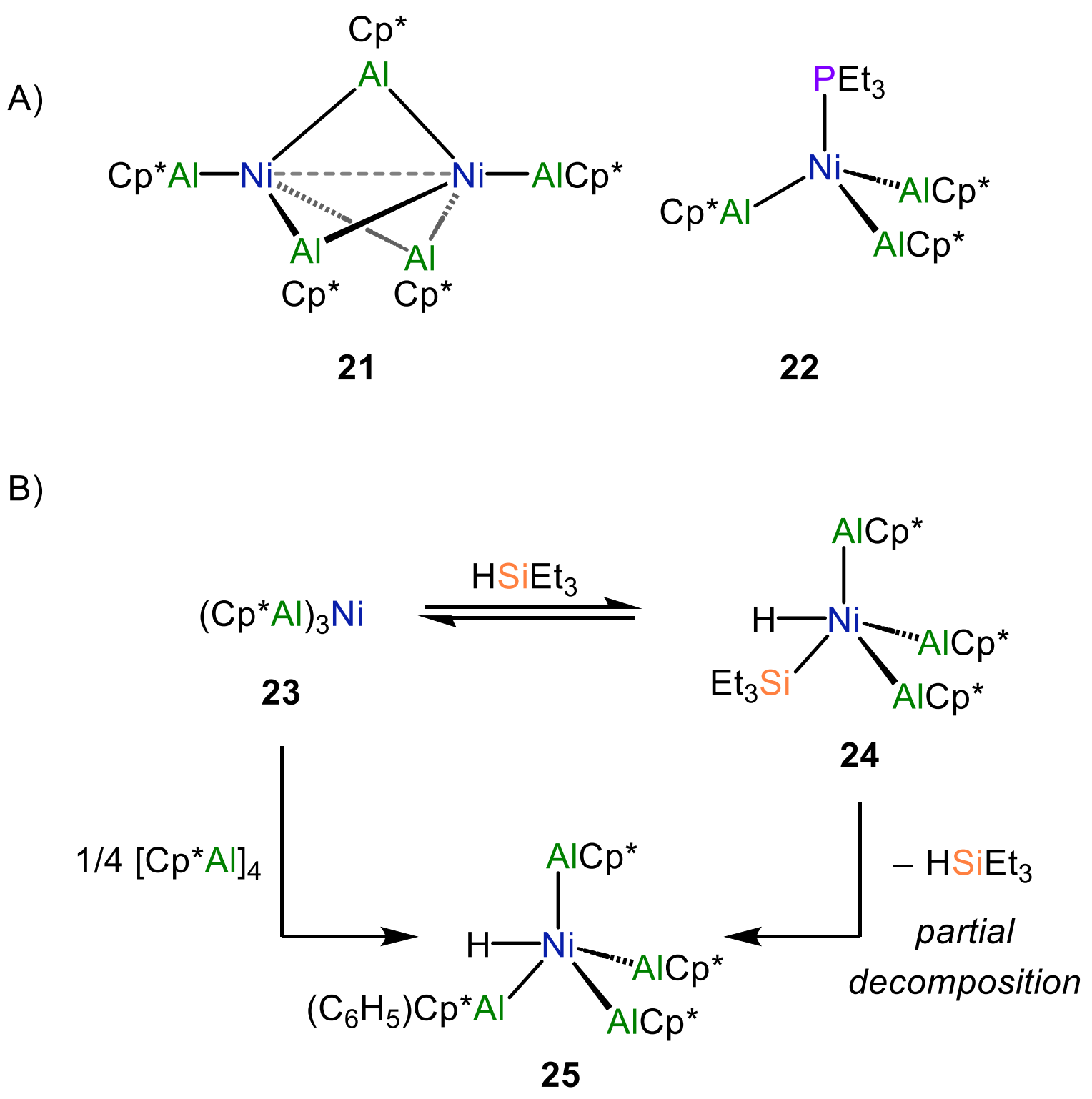

**Figure 6.** A) Coordination chemistry and structures of Cp*Al with Ni. B) Applications on C–H and Si–H bond activation using the intermediate Ni(Cp*Al)$_3$.

Beyond simple ligand substitution and E–H activation, Ni–aluminylene complexes have also been shown to promote carbon–carbon coupling of unsaturated molecules. For instance, of Ni(COD)$_2$ with Cp*Al in the presence of various alkynes generates a Ni–Al system can induce the dimerization of alkynes to form butadiene diyl ligands. Specifically, adding 3-hexyne (an aliphatic alkyne) to a Cp*Al/Ni(0) mixture yielded (Cp*Al)Ni(tebd)(COD) **26** (tebd = tetraethylbutadiene) as the main product (Figure 7).[31] **26** contains a C$_4$ fragment (tetraethylbutadiene) that results from the coupling of two 3-hexyne molecules, coordinated to Ni, alongside one Cp*Al ligand and a remaining cod ligand. Similarly, using dpa (diphenylacetylene) as an aromatic alkyne instead led to (Cp*Al)Ni(tpbd)(dpa) **27** where tpbd (tetraphenylbutadiene) is the coupled diacetylene and dpa indicates a still-coordinated diphenylacetylene ligand in the complex.[31] These outcomes suggest that the Al–Ni framework can facilitate reductive coupling of alkynes: the Ni center likely mediates the C–C bond formation between two alkyne units, while the aluminylene ligands stabilize the intermediate organometallic species and perhaps modulate electron flow. The ability to form 1,3-butadiene species from alkynes in the coordination sphere is reminiscent of known nickel catalysis for the dimerization or oligomerization of alkynes, but here the presence of Al(I) may influence the selectivity or stability of the resulting complexes.

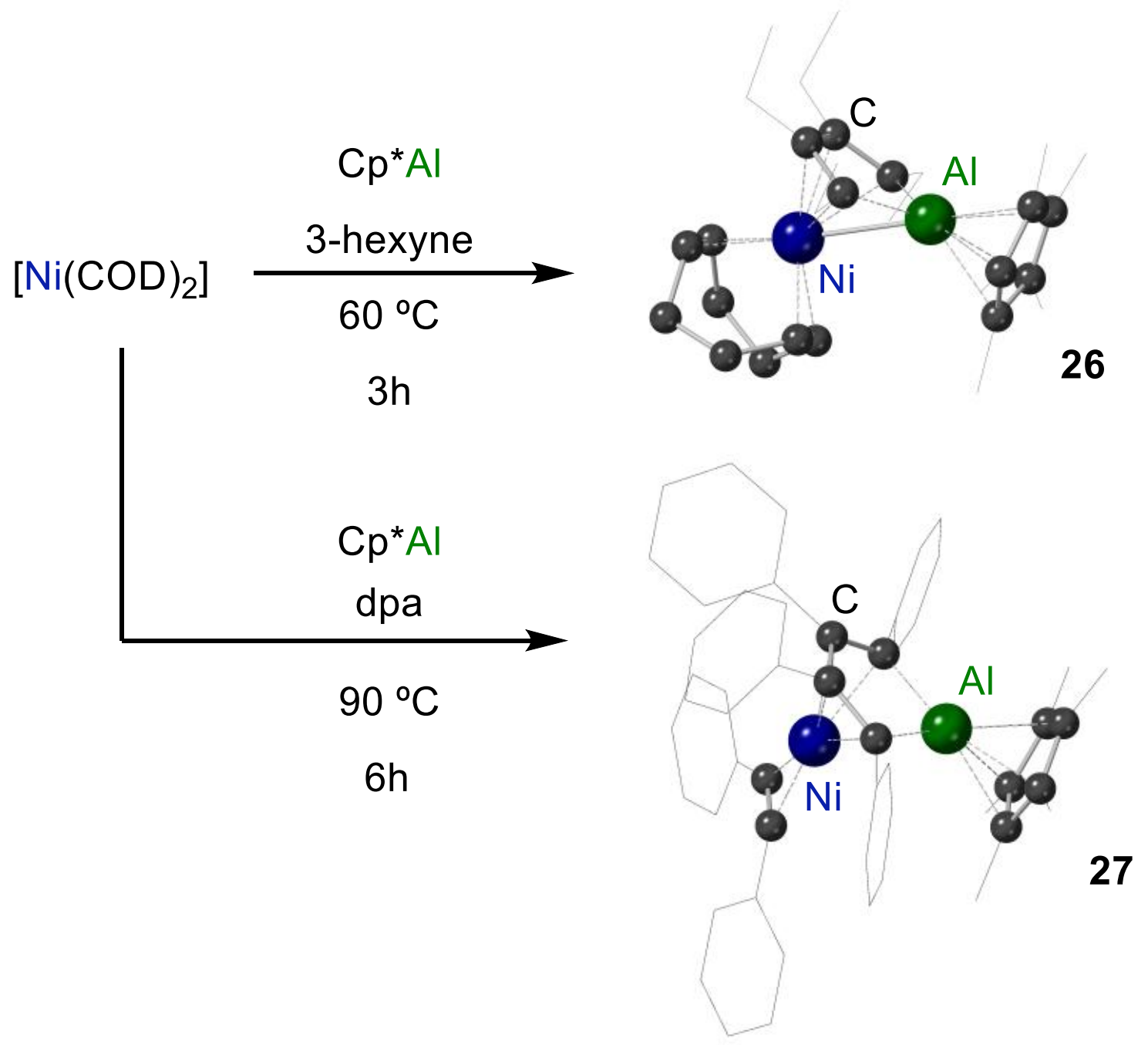

**Figure 7.** Reductive coupling of alkynes: the Ni center likely mediates the C–C bond formation between two alkyne units

Turning to other transition metals, group 6 metals have provided further insight into Al(I) L-type coordination. The reaction of W(0) species with Cp*Al (Figure 8) led to the displacement of two $PMe_3$ and formation of a bis(aluminylene) adduct $(Cp^*Al)_2W(PMe_3)_2(C_2H_4)_2$ **28**.[32] In this complex, two Cp*Al ligands occupy *trans* positions in the octahedral coordination sphere of W, and the W center still binds two ethylenes and two phosphines. In contrast, when a larger excess of Cp*Al (approximately six equivalents) was used, the product was **29**. Species **29** contains six Al(I) ligands around tungsten forming a pseudo-octahedral $WAl_6$ core and retains two ethylene ligands. More remarkably, **29** features both terminal and bridging hydrides in its structure, as well as evidence of C–H activation of an ethylene ligand. The presence of hydride ligands indicates that some ethylene ($C_2H_4$) underwent oxidative addition. A hydrogen atom from an ethylene C–H bond has been transferred, likely ending up bridging between W and an Al center, while the ethylene is converted to an alkyl group bound to Al, formally a metallacycle or an "aluminacycle" with a $–CH_2–Al–$ moiety. The ability to isolate **29** demonstrates that heavy loading of Al(I) L-type donors can dramatically alter the reactivity at the metal center, pushing W(0) into oxidative addition chemistry that might not perform with only classical ligands. It appears that the electron-rich environment provided by multiple aluminylene ligands, possibly combined with their ability to accept H atoms or form bridges, enables the W center to break C–H bonds (of ethylene) and form W–H and Al–C bonds. **29** thus stands as a vivid example of cooperativity: multiple Al(I) ligands acting in concert to induce and stabilize unusual organometallic transformations at a transition metal center (in this case, ethylene C–H activation at W).

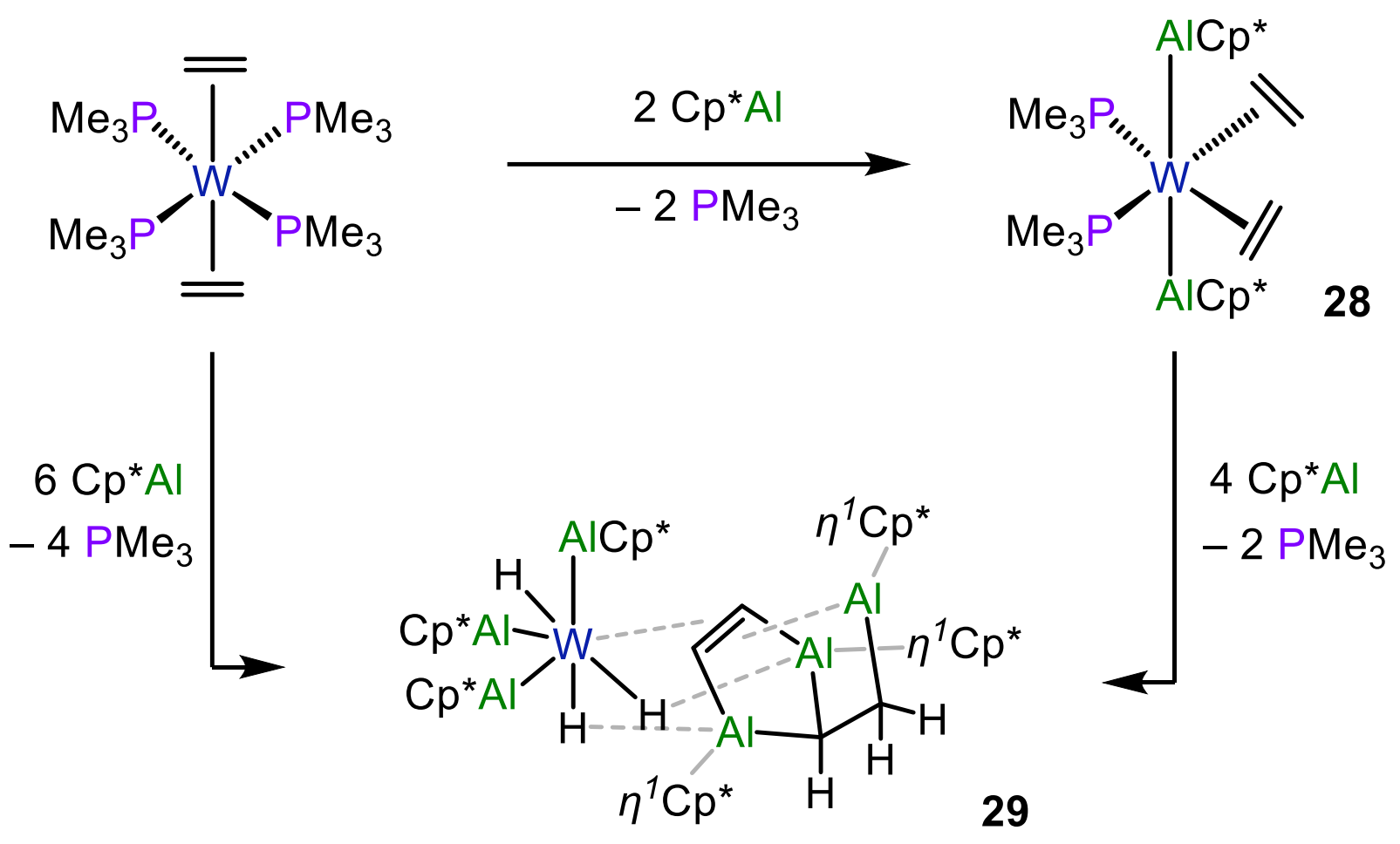

**Figure 8.** Reactivity of Al–W species via coordination and C–H bond activation.

Reactions of Cp*Al with Fe and Ru, both in low formal oxidation states, have also yielded noteworthy complexes featuring hydride and alkyl bridging motifs via L-type ligation of Cp*Al with Fe(0) and Ru(0) π-complexes. When the arene/diene complex Fe($\eta^6$-toluene)($\eta^4$-1,3-butadiene) was treated with six equivalents of $(Cp^*Al)_4$ (providing approximately six monomeric Cp*Al units), a mixture of Al–Fe hydride clusters was obtained.[5b] The major product was identified as $(Cp^*Al)_3Fe(H)_2(CH_2C_5Me_4Al)_2$ **30** (Figure 9). In this species, the Fe center is bonded to three terminal Cp*Al ligands and additionally two Al atoms that are each bound via a $CH_2C_5Me_4$ group, along with two bridging hydrides (Fe–H–Al). The $CH_2C_5Me_4Al$ ligands can be thought of as resulting from deprotonation of two Cp* ligands: essentially, two Cp*Al units have had one methyl group on their Cp* ring activated (deprotonated) by the Fe (with assistance from Al), creating a methylene-bridged Al–Fe linkage (Al–$CH_2$–Fe) and a hydride that bridges Fe to that Al. A minor product from the same reaction was $(Cp^*Al)_3Fe(H)_3(CH_2C_5Me_4Al)_3$ **31**, which contained even more hydrides (three) and three bridging $CH_2C_5Me_4Al$ units with only two terminal Cp*Al. Species **30** and **31** collectively reveal that multiple Cp*Al ligands on Fe can synergistically activate the C–H bonds of Cp* methyl groups, transferring hydrogen to the metal and bridging to Al while binding the remainder of the organic fragment as an alkyl group on Al. In essence, the normally inert methyl groups of Cp* became participants in metalation, thanks to the mediation by Fe and the Lewis acidic/basic cooperation between Fe and Al centers. An analogous reaction was observed with ruthenium: the Ru(0) complex Ru($\eta^4$-COD)($\eta^6$-COT) (COD = 1,5-cyclooctadiene, COT = cyclooctatetraene) reacts with about five equivalents of Cp*Al to yield with five equivalents of Cp*Al and $(Cp^*Al)_3Ru(H)_2(CH_2C_5Me_4Al)_2$ **32**.[5b] Species **32** is the Ru analog of **30**, featuring two bridging methylene-aluminylene units and two hydrides. This suggests that Fe and Ru, both 18-electron complexes initially, undergo similar processes of Cp* methyl C–H activation and hydride formation when faced with an excess of Al(I) ligands. Interestingly, a different outcome was seen with Rh(I), which has a 14-electron count in the precursor used. In analogy to these species, reacting the Rh(I) diene cation $[Rh(\eta^4\text{-COD})_2][BAr^F{}_4]$ (where $BAr^F{}_4$ is a weakly coordinating anion featuring $Ar^F{}_4$ groups 3,5-$(CF_3)_2C_6H_3$), with three equivalents of Cp*Al led to a complex cation formulated as $[(Cp^*Al)_3(Rh(\eta^4\text{-COD})]^+$ **33** (Figure 9).[33] In **33**, the Rh(I) center is bound to three AlCp* ligands and a COD ligand, with a charge-balancing $BAr^F{}_4$ anion. Unlike the Fe and Ru cases **30**, **31** and **32**, no hydrides or activated Cp* fragments were reported in the Rh complex **33**. This could be due to the different electron count and oxidative addition propensity of Rh(I) versus Fe(0)/Ru(0). Nonetheless, **33** confirms that even cationic late-transition metals can coordinate multiple aluminylenes Cp*Al ligands in an L-type fashion. The series of Al–Fe, Al–Ru, and Al–Rh complexes **29**–**32** emphasizes that aluminylenes Cp*Al ligands can enable unconventional reactions such as intramolecular C–H activation of ancillary ligands, and that the extent of such reactivity is highly dependent on the metal's identity and electron count.

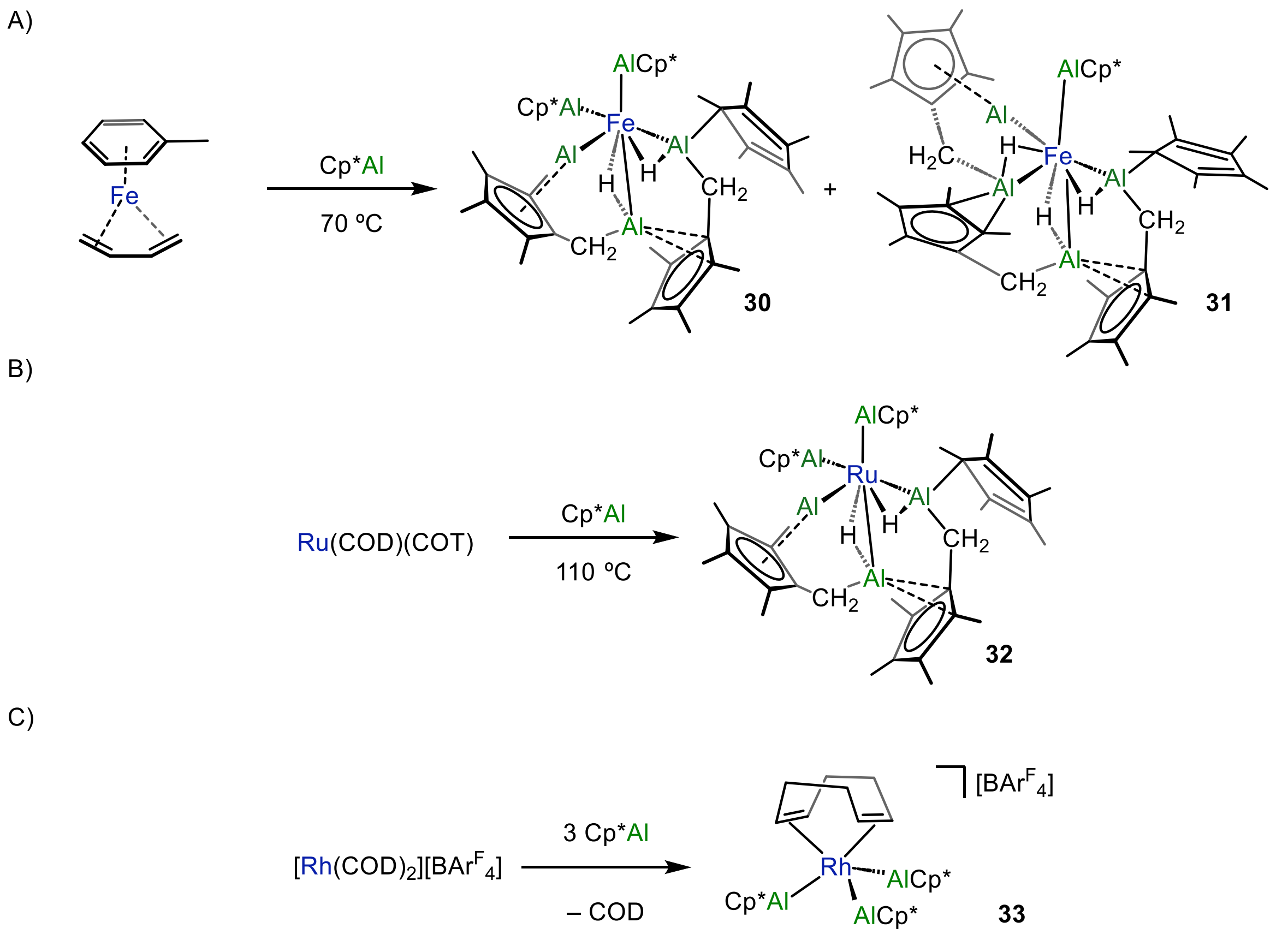

**Figure 9.** Synthesis of Al–M complexes (M= Fe, Ru, Rh) **30–33**.

Aluminum(I) L-type metalloligands have also shown a strong aptitude for stabilizing higher-nuclearity clusters containing mixed metals. Fischer's group, for instance, explored reactions of Cp*Al with group 10 metal clusters and observed formation of intricate Al–M cluster compounds. One example is the reaction of $Pd_2(dvds)_3$ with Cp*Al.[34] This produced an Al–Pd cluster formulated as $(Cp^*Al)_2(\mu_2\text{-}Cp^*Al)_2(\mu_3\text{-}Cp^*Al)_2Pd_3$ **34** (Figure 10). In **34**, three Pd atoms form a triangular arrangement, and the six Cp*Al ligands occupy different roles. Two are terminally bound to individual Pd atoms, two are bridging between pairs of Pd($\mu_2$-bridging), and two are capping the $Pd_3$ face in a three-center fashion ($\mu_3$ bridging each Al to all three Pd). The result is a $Pd_3Al_6$ **34** cluster where aluminylenes Cp*Al L-type ligands encapsulate the $Pd_3$ core, effectively saturating all coordination positions and bridging Pd–Pd interactions. The formation of **34** alongside a minor byproduct, identified as a ($\mu_2$-Cp*Al)Pd(dvds) species indicates that Cp*Al can displace organic ligands on Pd and simultaneously promote metal aggregation by bridging multiple metal centers. In a related vein, the concept of ligand exchange between different group 13 metalloligands was demonstrated by the Ga–Pd cluster **35**, that can undergo full substitution of its Cp*Ga L-type ligands by Cp*Al (Figure 10).[34] The reaction yielded $(Cp^*Al)_5Pd_2$ **37**, isostructural to the original but with all Cp*Ga metalloligands replaced by Cp*Al. Conversely, when a similar Ga–Pt cluster $(GaCp^*)_5Pt_2$ **36** was treated with Cp*Al, the product was $(Cp^*Ga)(\mu_2\text{-}Cp^*Al)_3Pt_2$ **38**, in which only a partial exchange occurred, three of the Ga*Cp L-type units were replaced by Cp*Al, while one GaCp* remained bound to Pt. The heterometallic species **38** thus contains a mix of Ga and Al ligands around a Pt–Pt core. These experiments highlight that Cp*E (E = Al, Ga) ligands can be swapped on metal clusters, a process that might be driven by differences in binding strength or steric effects between aluminylene and gallylene Cp* L-type metalloligands. Such ligand exchange reactions are valuable for tuning cluster composition and properties, effectively allowing one to build heterometallic clusters in a stepwise fashion. Moreover, the ability of Cp*Al to form stable clusters with Pd, Pt (and by extension Ni, *vide supra*) underlines its capacity to stabilize multiple metal-metal bonds simultaneously, acting as both a terminal donor and a bridging ligand.

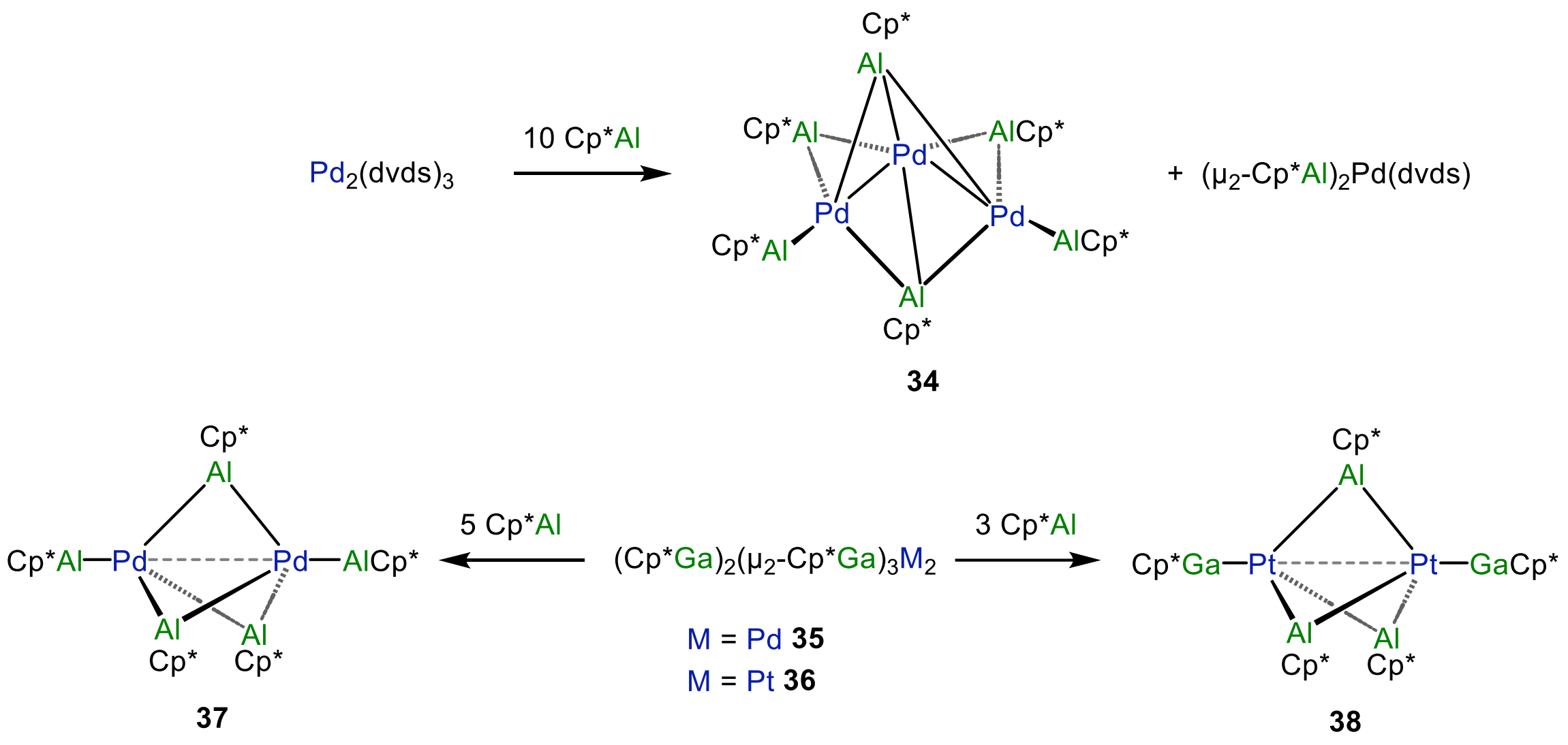

**Figure 10.** Structures of clusters $M_a(Cp^*E)_b$ (M = Pd, Pt; E = Al, Ga) **34**–**38**.

The group 11 metals (Cu, Ag, Au) have also been combined with aluminylene ligands, leading to fascinating clusters that blur the lines between molecular and metallic bonding. Cu has yielded a rich family of Al/Cu clusters. For example, reacting the hexanuclear Cu hydride $[Cu(H)PPh_3]_6$ with Cp*Al produces a high-nuclearity Al–Cu hydride cluster $(Cp^*AlCu)_6H_4$ **39** consisting of a $Cu_6$ octahedron with each of the 12 edges of the octahedron bridged by a Cp*Al ligand, thus forming an $Al_6$ octahedron encapsulating the Cu core. In addition, four hydride ligands were found, each bridging the faces of the $Cu_6$ octahedron (each hydride is $\mu_3$, capping a triangular face of $Cu_3$). This symmetric $Cu_6Al_6H_4$ structure can be seen as an “aluminated” analog of octahedral hexacuprate hydrides, where Cp*Al replaces typical phosphine or other ligands. Significantly, **39** is not a dead-end cluster, undergoing further reactivity. Upon exposure of **39** to benzonitrile (PhC≡N), an insertion reaction occurs, yielding the cluster $(Cp^*AlCu)_6(H)_3(N{=}CHPh)$ **40**.[35] In **40**, one of the hydrides has added across the C≡N bond of PhCN, resulting in an imine (–N=CHPh) that is now bound within the cluster (with the nitrogen likely coordinated to a Cu and the newly formed C–H fragment bridging to Al or Cu). This transformation is remarkable as it shows the Al–Cu hydride cluster activating a small inorganic molecule (a nitrile) to form a new C–H and C–N bond, a reaction somewhat analogous to surface hydrogenation of nitriles to imines, but here achieved in a discrete molecular cluster. It exemplifies the concept of frustrated Lewis pair (FLP) reactivity embedded in a cluster: the Lewis acidic Cu–Al framework and the hydrides together facilitate addition to a polar triple bond.

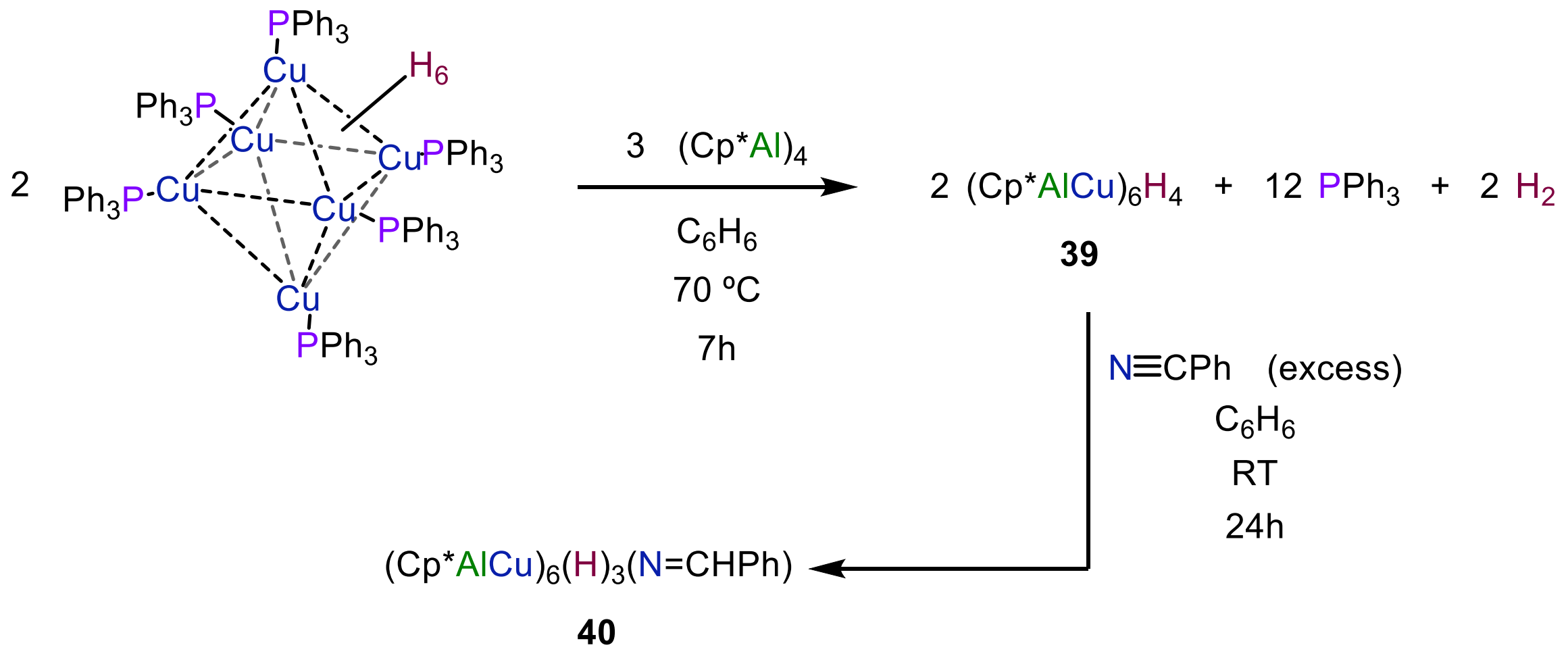

**Figure 11.** Reaction of the hexanuclear Cu hydride $[Cu(H)PPh_3]_6$ with Cp*Al.

Pushing the nuclearity even higher, combining Cu(I) mesityl $(CuMes)_5$ with Cp*Al L-donation can produce an unprecedented 55-atom cluster containing 43 Cu and 12 Al centers. The resulting cluster $(Cp^*Al)_{12}Cu_{43}$ **41** was reported in 2018 as the largest molecular heterometallic containing aluminum(I) ligands.[36] The structure of **41** is described as a "Mackay-type" icosahedral core-shell.[37] It can be viewed as concentric shells of metal atoms, reminiscent of geometric patterns found in metallic nanoparticles, but here fully ligated by organometallic groups (Cp* on Al). Such nested icosahedral architecture (with an inner Cu core and an outer mixed Cu/Al shell) highlights the tendency of Cu and Al to form super atomic clusters under the right conditions. The formation of **40** was found to proceed through smaller intermediate clusters, depending on reaction conditions and stoichiometry. From similar reactions, the species $Cu_4Al_4(Cp^*)_5(Mes)$ **42** and $Cu_2Al(Cp^*)_3$ **43** were isolated, which could be seen as fragments on the way to larger clusters. A series of mid-size clusters,[37] $Cu_7Al_6(Cp^*)_6$ **44** $HCu_7Al_6(Cp^*)_6$ **45** and $Cu_8Al_6(Cp^*)_6$ **46**, were also isolated. These clusters, **44–46** contain $Cu_7Al_6$ or $Cu_8Al_6$ cores and can eventually aggregate into the massive $Cu_{43}Al_{12}$ cluster **41** under certain conditions. The isolation of **44** and **46** indicated that not all clusters contained hydride (e.g., **44** and **46** have none, whereas **45** has one hydride), suggesting that hydride presence can vary with synthetic conditions. Overall, the Al–Cu cluster system demonstrates a living cluster growth process, where small $Al_xCu_y$ units assemble into larger ones. The Cp*Al ligand is crucial in this context: it stabilizes each step of growth and prevents uncontrolled coalescence into bulk metal. This body of work not only expands the known chemistry of Cu–H clusters but also positions Cp*Al as a powerful tool for synthesizing novel coinage-metal architectures that bridge discrete molecules and nanoclusters.

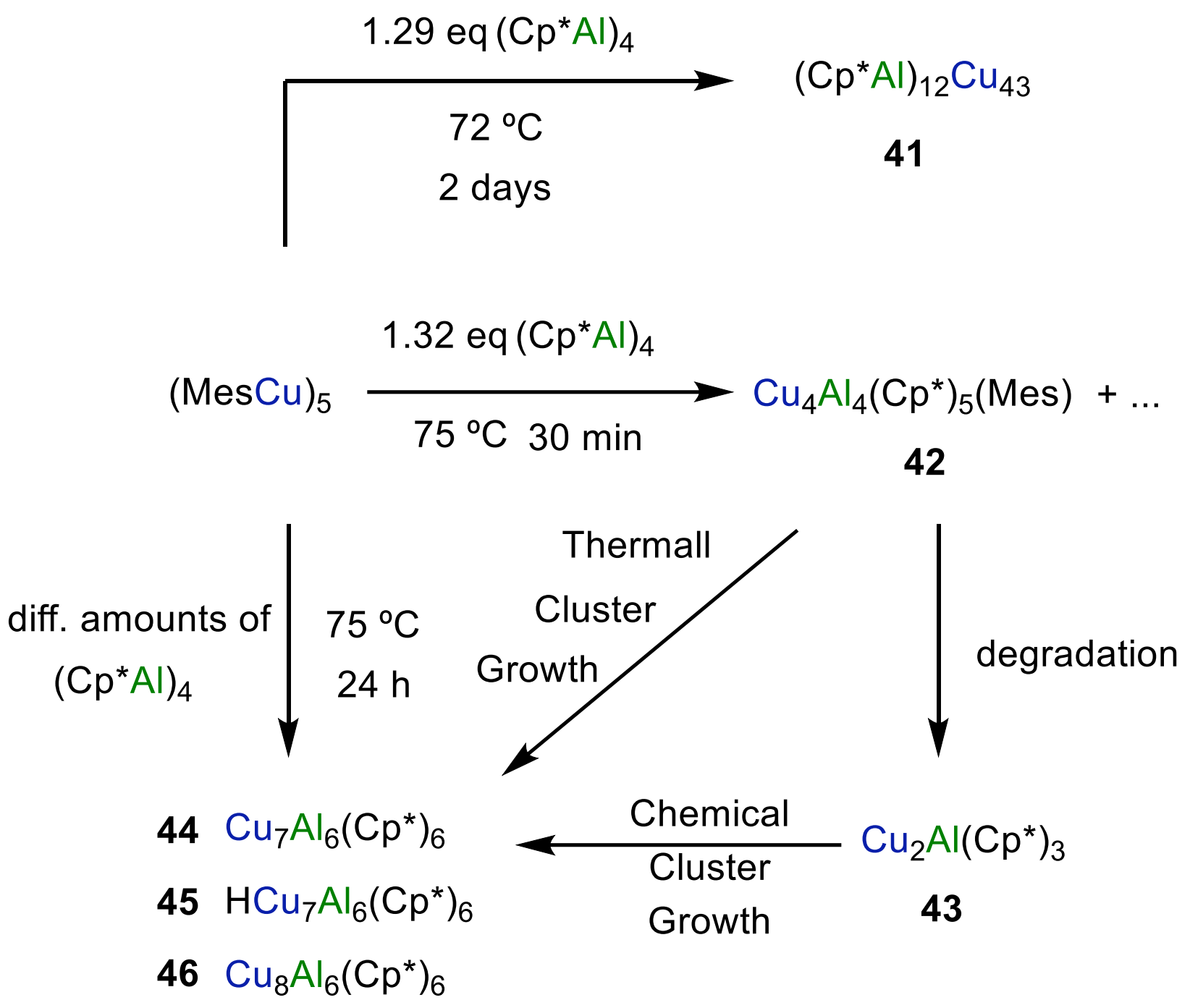

**Figure 12.** Unprecedented clusters from the reaction of $(CuMes)_5$ with Cp*Al.

Al–Au clusters form another intriguing species in Cp*Al L-coordination chemistry. The reaction of an NHC-stabilized Au(I) hydride (iDippAuH, where iDipp = 1,3-bis(2,6-diisopropylphenyl)-imidazol-2-ylidene) with Cp*Al, leads to the isolation of a dinuclear Au(0) complex formulated as $(Cp^{*}Al)_5Au_2$ **47** (Figure 13).[28] In **47**, two Au atoms are directly bonded (Au–Au) and are bridged by three Cp*Al ligands, with each Au also bearing one terminal Cp*Al (total of five Al units for two Au). The geometry can be seen as an $Au_2$ core with an Cp*Al belt, a rare example of gold bound by multiple main-group ligands in both bridging and terminal modes.[28] The presence of the iDipp carbene in the precursor likely helped to form a reactive Au(I)/Al(I) intermediate that reductively eliminated the hydride and allowed Au–Au bond formation with aluminylene coordination. Following this initial discovery, further investigations revealed that cluster **47** is a key intermediate on the way to larger gold/aluminum clusters. By adjusting reaction conditions (such as using a slight excess of Cp*Al or prolonged heating), the team obtained two higher-nuclearity clusters $(Cp*Al)_6Au_6$ **48** and $(Cp*Al)_6Au_7(H)$ **49**.[38] Cluster **48** consists of six Au atoms arranged octahedrally, each face of the octahedron capped by an Cp*Al, so that each Au is bonded to three Al, and each Al bridges a triangular face of $Au_3$. This $Au_6Al_6$ structure is notable for having no hydrides and representing a fully carbonaceous coordination sphere (Cp* rings) around the metal core. Cluster **49** contains seven Au atoms with one interstitial hydride, forming essentially a $Au_7$ core with one $\mu_4$-H and six Cp*Al around the exterior. The hydride in **49** likely comes from residual Au–H or Al–H species during cluster assembly. The sequential formation **47** → **48** → **49** parallels the Cu/Al cluster chemistry in some ways. Smaller units **47** with $Au_2Al_5$ can serve as seeds that merge into larger clusters $Au_6Al_6$ and $Au_7Al_6H$. These Al–Au L-type clusters are conceptually important because they have potential avenues for discovering new catalytic or electronic materials based on such units.

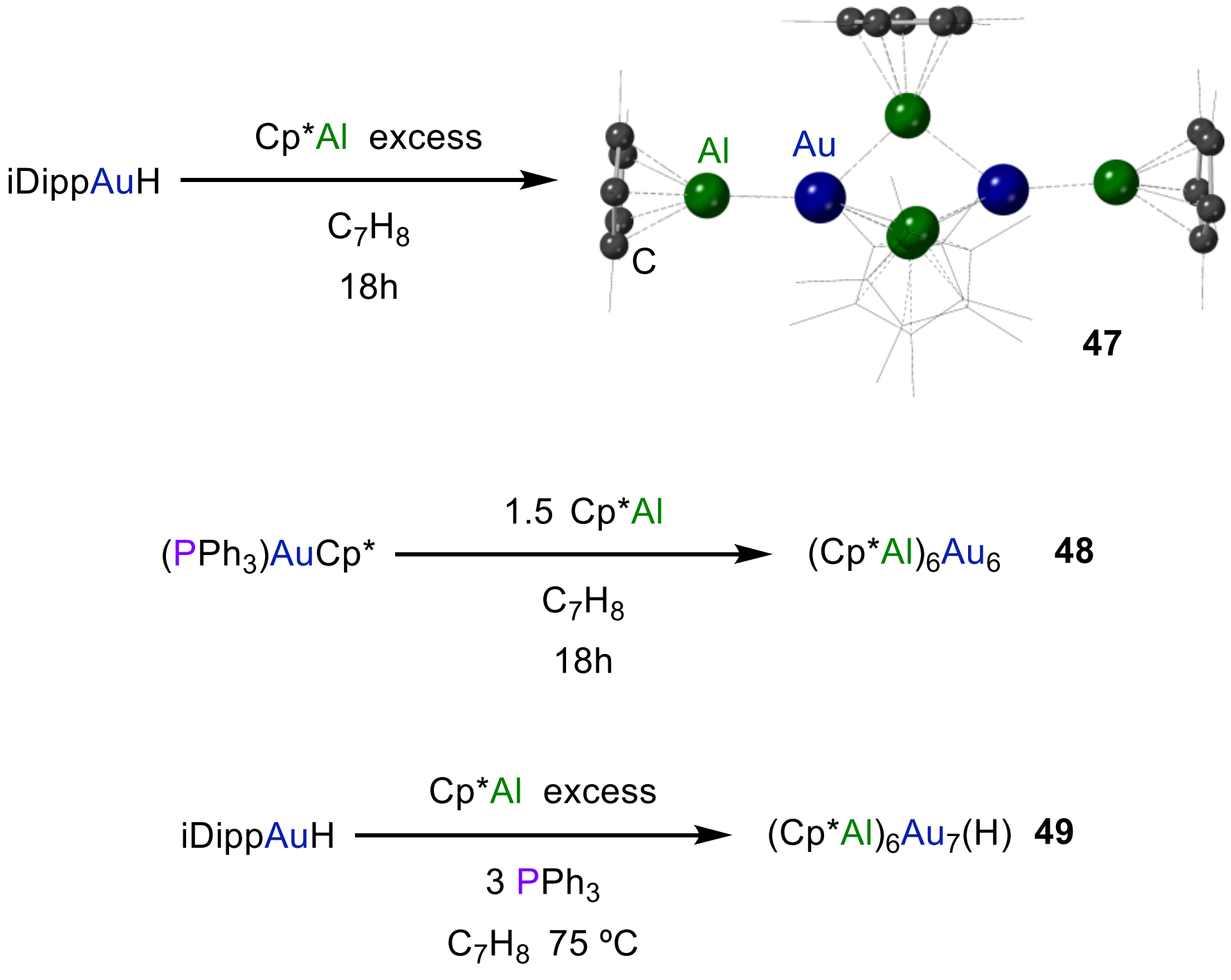

**Figure 13.** Al–Au clusters **47**–**49**.

Aluminylene L-type ligands Cp*Al have also demonstrated interesting reactivity with main-group reagents, exemplified by insertion reactions into polar bonds. One area that has seen rapid development is the chemistry of coordination of Al(I) with zinc complexes. Cp*Al can insert into Zn–N bonds of zinc amides. For example, treating $Zn(HMDS)_2$ (HMDS = bis-(trimethylsilyl)amide) with Cp*Al resulted in the heterometallic complex $Zn(\mu_2\text{-}Cp^*Al)_2(HMDS)_2$ **50** (Figure 14).[39] In **50**, two Cp*Al units have each inserted into the two Zn–N bonds of $Zn(HMDS)_2$, such that each Al is bridging between Zn and the nitrogen (forming Zn–Al–N three-center linkages). **50** can be viewed as a four-membered $Zn(\mu\text{-}NSiMe_3)_2Al_2$ ring, with each Al also bearing a Cp* ligand. This demonstrated a clear parallel between Al(I) and classical carbene chemistry, as the insertion of a carbene into an M–N bond is known in other systems; here the aluminylenes Cp*Al behaves similarly. More recently, Hevia expanded on this chemistry by isolating monoinsertion products and examining further reactivity.[40] By carefully controlling the stoichiometry (using only one equivalent of Cp*Al per Zn center), the complexes Zn($\mu_2$-Cp*Al)(HMDS)$_2$ **51** and Zn($\eta^2$-Cp*Al)(TMP)$_2$ **52** (TMP = 2,2,6,6-tetramethylpiperidide) are obtained (Figure 14). **51** and **52** features single a Cp*Al bridging one Zn–N bond in a dialkylamide (TMP = 2,2,6,6-tetramethylpiperidide). In **51**, for example, one HMDS ligand remains purely σ-bound to Zn, while the other HMDS is linked to Zn and Al via a η-N–AlCp* bridge. These species confirm that the insertion of Al(I) into Zn–N bonds can be selective and controlled, leading to stable two-metal, two-ligand frameworks. The presence of both Zn–N and Al–N interactions in such complexes imbues them with both Lewis acidic (Zn) and Lewis basic (Al lone pair) sites in proximity. Exploiting this dual reactivity, the team showed that **51** and **52** can further insert unsaturated small molecules. For instance, exposure of the Al–Zn amide to carbodiimides (RN=C=NR) results in facile insertion of the carbodiimide into the Zn–Al ensemble.[40] The product (HMDS)(Cp*Al)Zn(C(NR)$_2$)(HMDS) **53** (R = iPr or Cy) is formed by insertion of the N=C=N unit. One N–C bond of the carbodiimide is cleaved and a new C–N bond is formed to the Al center (yielding an amidinate-type C(NR)$_2$ ligand bridging Al and Zn. Similarly, complex **52** (with TMP ligands) was shown to undergo Cp*/TMP ligand exchange to give **54** coupled with carbodiimide insertion into one of the Al–TMP moieties yielding **55**. In this product **55**, $CO_2$ has inserted into an Al–TMP bond to form a Al–O(CO)O–TMP carbonate linkage resulting with **56**, suggesting a tandem reaction where $CO_2$ and the diisopropylcarbodiimide were both incorporated. These transformations highlight how the Al–Zn combination can activate small molecules such as $CO_2$. The Lewis acidic Zn can polarize the incoming substrate (carbodiimide or $CO_2$), while the aluminylenes can form a bond to a fragment of the substrate (e.g., the carbon of $CO_2$ or the carbon of a C=N bond), effectively cleaving and rearranging the molecule.

The result is the formation of new C–N or C–O bonds bridging the two metals. This sort of reactivity is reminiscent of bimetallic cooperative catalysis and suggests that Al(I) complexes could play a role in the activation of inert substrates like $CO_2$, potentially contributing to future developments in main-group mediated transformations.

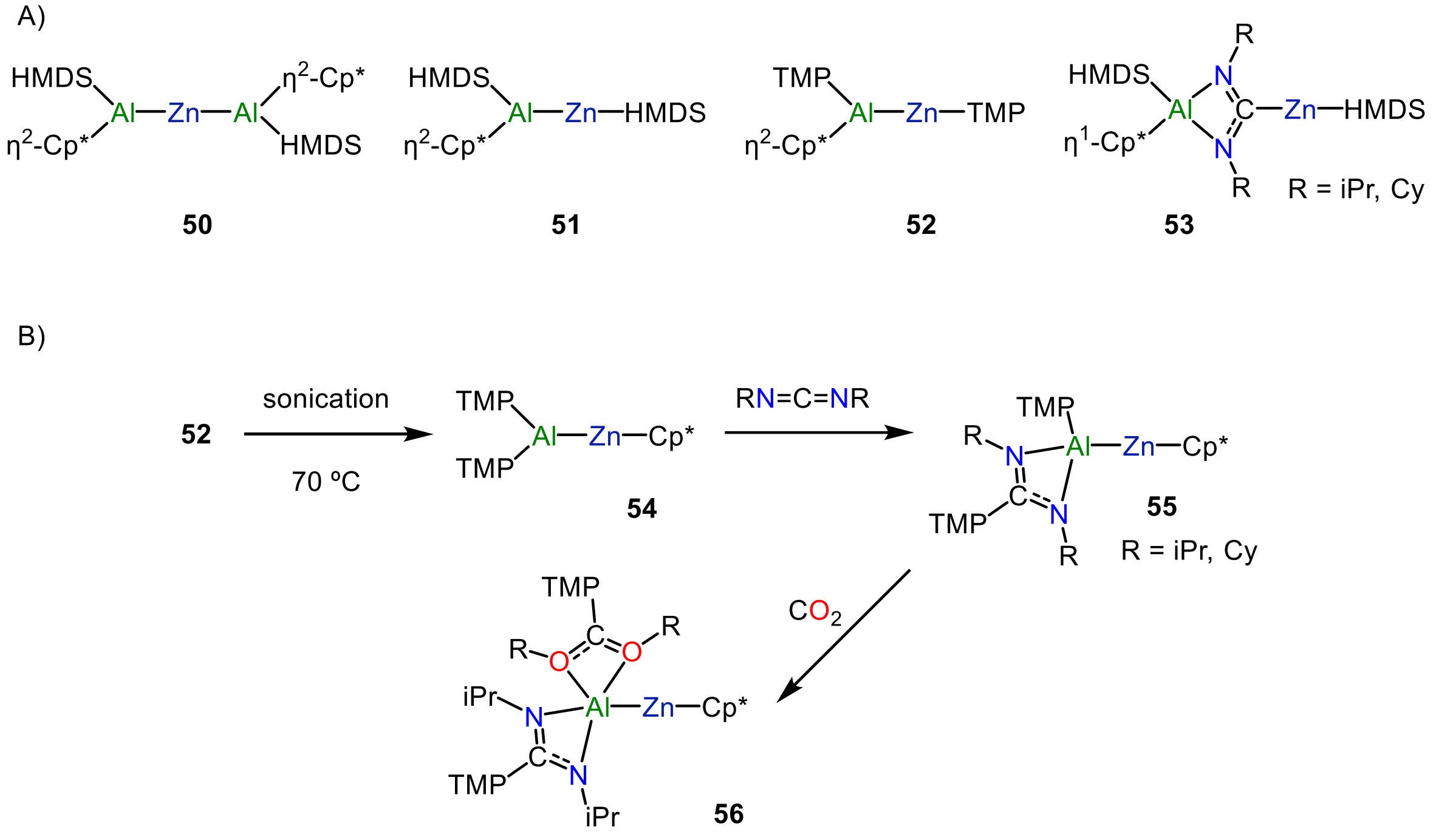

**Figure 14.** Representative Al–Zn compounds **50–52** and **54**, showcasing Cp*Al insertion into Zn–N bonds and subsequent carbodiimide/$CO_2$ insertion reactions **53**, **55**, **56**.

Overall, the chemistry of Cp*Al as an L-type (2-electron donor) metalloligand has rapidly expanded from early proof-of-concept complexes to a broad landscape of structures and reactivities. Cp*Al and related aluminylene species can coordinate to a wide range of transition metals, from early (Cr, W) to late (Ni, Pd, Pt, Cu, Au), in both terminal and bridging fashions. These Al(I) ligands often endow the complexes with unique features: they can stabilize high coordination numbers (as in Ni or W with up to 6 Al ligands), promote metal–metal bonding in clusters, and crucially, enable unusual bond activation processes (E–H and even C–H bond cleavage, alkyne coupling, etc.)

through synergistic interactions with the metal center. The ability of aluminylene Cp*Me ligands to act as both donor and acceptor (e.g., accepting a hydride or organic fragment) underlies much of this reactivity. Aluminum's unique balance of size, electron-donating ability, with respect to Ga, In and Tl (*vide infra*) and propensity to form multicenter bonds often makes its chemistry the most robust and varied. The continued exploration of Cp*Al–TM complexes is not merely a cataloging of new compounds but is leading to a deeper understanding of how highly electron-rich, low-valent environments can activate substrates and forge novel inorganic structures, a frontier that blurs the line between transition-metal and main-group chemistry.

## 2.3 Bulky Al(I) donors: Ambiphilicity, bridging and cooperative reactivity

Beyond the Cp* ligand, various bulky ligand frameworks have been used to stabilize monovalent Al(I) centers that act as neutral two-electron donors (L-type ligands) to transition metals. Such Al(I) species, termed as aluminylene ligands (metalloligands), exhibit coordination chemistry analogous to N-heterocyclic carbenes (NHCs).[41] Notably, β-diketiminate (BDI or NacNac) ligands and bulky aryl substituents have enabled the isolation of Al(I) complexes that coordinate to metals in this fashion. These bulky substituents provide steric protection and electronic stabilization for the low-valent Al center, preventing dimerization and allowing Al(I) to persist as a two-electron donor unit.

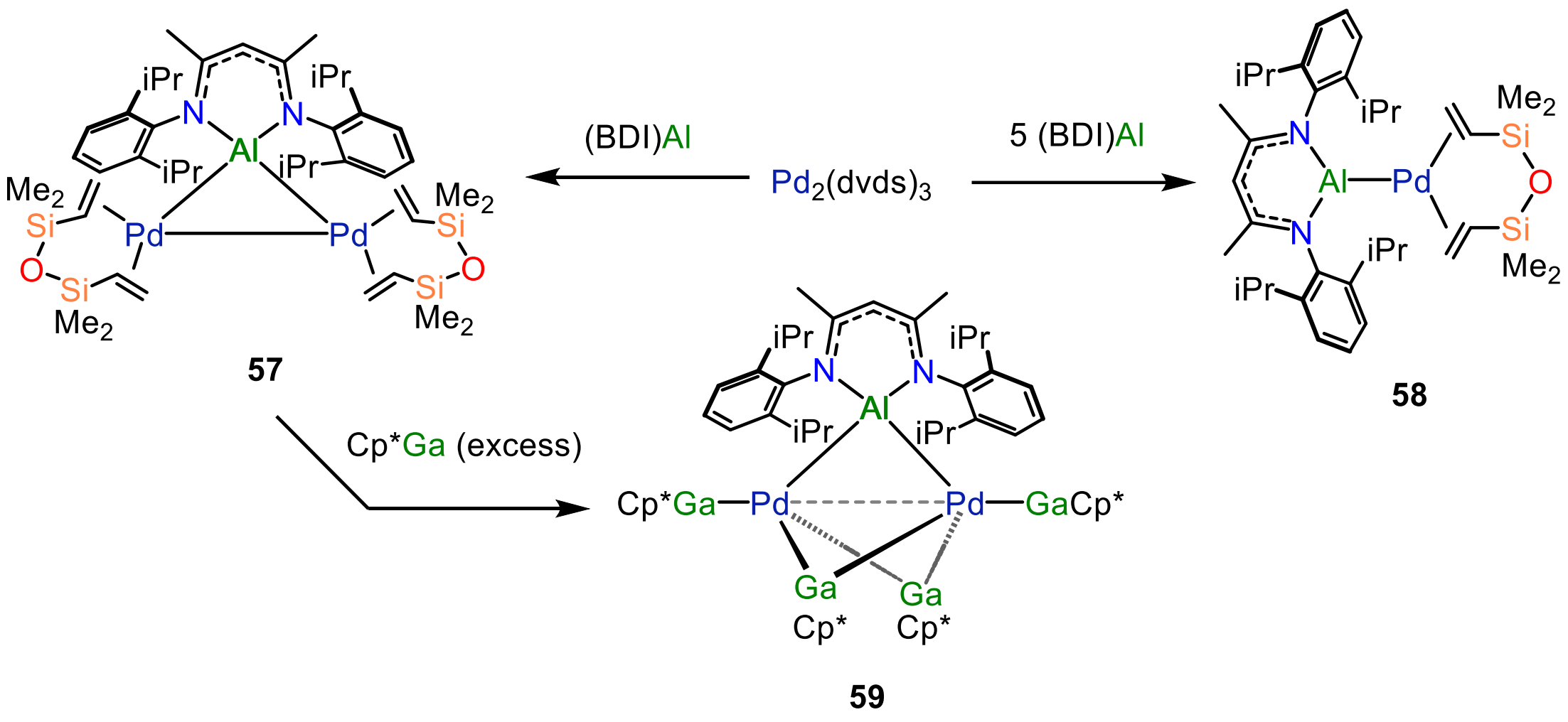

**Figure 15.** Formation of a bridging and terminal Al–Pd complexes **57**–**59**.

One of the earliest examples is the β-diketiminate-stabilized aluminylene (BDI)Al **6** (in Figure 3). Fischer showed that this Al(I) fragment, often denoted as (DDP)Al, can coordinate to Pd(0) centers, either in a bridging or terminal mode. For instance, reacting the $Pd_2(dvds)_3$ complex (dvds = tetramethyl-divinyldisiloxane) with (BDI)Al yielded a $AlPd_2$ cluster $(BDI)Al(Pd(dvds))_2$ **57** (Figure 15) in which the aluminylene bridges the two Pd centers.[42] However, under different conditions (e.g. higher Al:Pd ratios), the aluminylenes (BDI)Al could instead replace a dvds ligand to form a mononuclear complex (BDI)AlPd(dvds) **58**, featuring a terminal Al(I) ligand bound to a single $Pd^0$.[42] This demonstrated that (BDI)Al can serve as a discrete donor to metals, much like an NHC. Moreover, such Al–Pd species showed rich reactivity. Exposure of the Al-bridged $AlPd_2$ cluster to an excess of Cp*Ga (a Ga(I) source) resulted in ligand substitution and formation of a mixed-metal Al–Ga–Pd cluster $(\mu^2\text{-}(BDI)Al)Pd_2(\mu^2\text{-}Cp^*Ga)_2(Cp^*Ga)_2$ **59**,[41] containing a bridging Al alongside multiple Ga centers (with both terminal and bridging Ga positions), highlighting the potential for constructing multi-metal Group 13 metal-transition-metal assemblies.

Subsequent studies expanded the scope of bulky Al(I) metalloligands to a variety of transition metals. Power reported a landmark complex featuring an unsupported Al–Cu bond, formed by coordinating (BDI)Al to a Cu(I) center supported by its own β-diketiminate ligand in (BDI)Al–Cu(BDI') **60** (Figure 16A).[43] The Al–Cu bond length in this two-coordinate complex (~2.30 Å) is markedly shorter than the Al–Cu distances seen in earlier alumino-copper clusters (e.g. an $Al_6Cu_6$ aggregate with AlCp* had Al–Cu ~2.41 Å).[35] This significant contraction (~5% shorter) is attributed to the compact 3s-based lone pair on Al(I), which facilitates especially strong L-type σ-bonding with the soft Cu(I) center in **60**. In parallel, Crimmin prepared a whole series of first-row transition metal–aluminylene complexes featuring Al–Cr **61**, Al–Mn **62**, Al–Fe **63**, Al–Co **64**, Al–Cu **65** using the (BDI)Al ligand (Figure 16B).[2] These were obtained by simple addition of (BDI)Al to the metal precursors (such as carbonyl or arene complexes), causing displacement of labile ligands (e.g. CO or benzene), a testament to the potent L-type Lewis basicity of the (BDI)Al metalloligand. Once coordinated, the (BDI)Al metalloligand confer unique electronic properties to the corresponding bimetallic units with the transition metals. Computational analyses reveal that the (BDI)Al unit in these complexes is not only a strong L-type σ-donor but can also act as a π-acceptor via its vacant p-orbital.[2] This adaptively allows the L-type Al–M bonding to accommodate both electron-rich and electron-poor metals, unlike Cp*-stabilized aluminyls which behaved as almost pure σ-donors. For example, a (BDI)Al(I)–Fe(0) carbonyl complex **63** (Figure 16C) could undergo photolytic ligand substitution via irradiation of an Al–$Fe(CO)_4$ species in the presence of $PCy_3$ led to replacement of a CO by the phosphine, yielding a new Al–Fe complex **66** with $Fe(CO)_3(PCy_3)$.[2] Such reactivity underscores that (BDI)Al can function as a classical supporting L-type ligand, enabling typical organometallic transformations (e.g. ligand exchange) within these heterometallic transition metal frameworks.

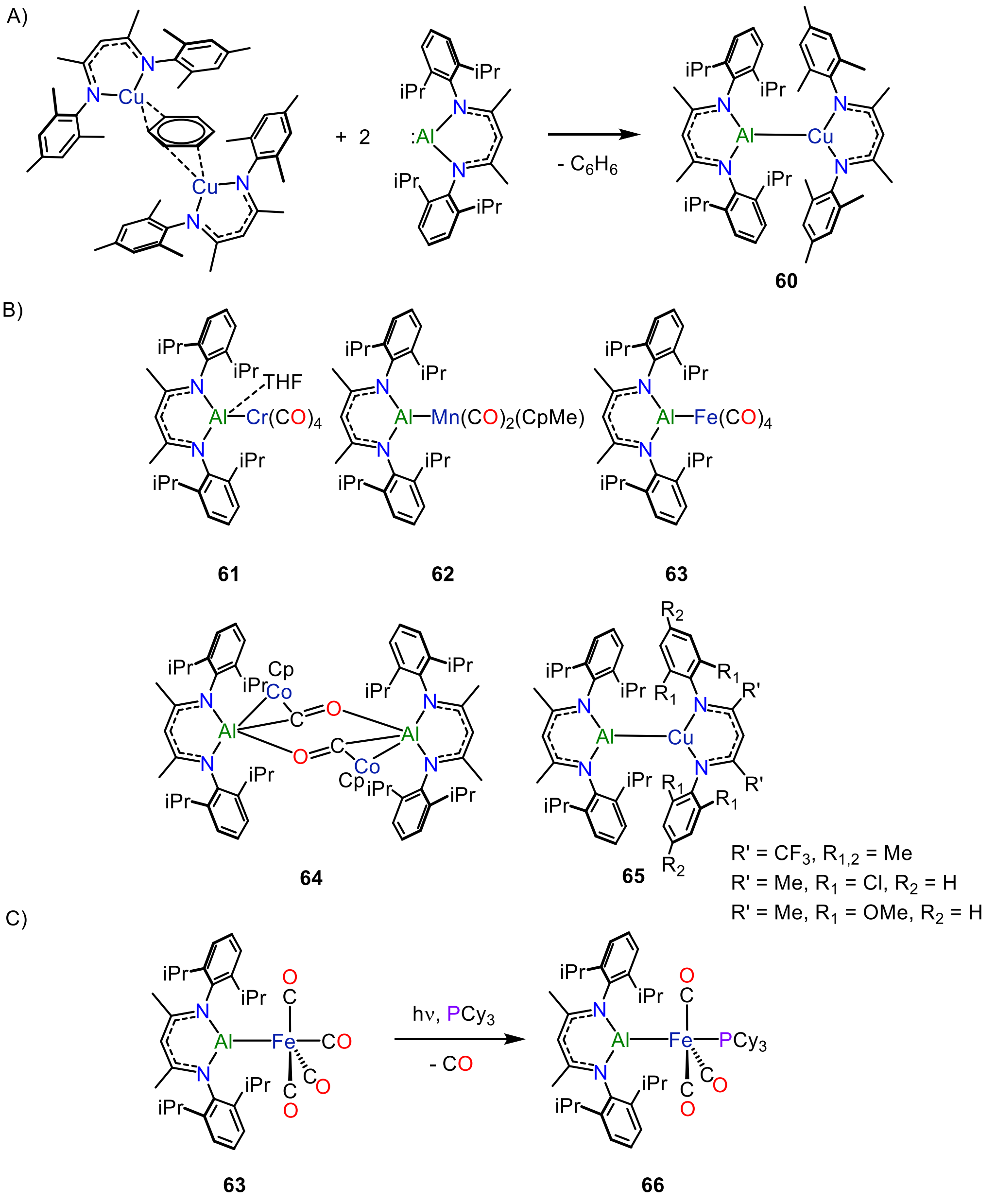

**Figure 16.** A) Molecular structure for two-coordinate Al–Cu complex (BDI)Al–Cu(BDI'), featuring an unsupported Al–Cu bond. B) Examples of first-row transition metal complexes bearing a (BDI)Al L-type ligand, including Al–Cr, Al–Mn, Al–Fe, Al–Co and Al–Cu adducts. C) Photolysis of (BDI)Al–Fe$(CO)_4$ in the presence of $PCy_3$ yielding (BDI)Al–Fe$(CO)_3$($PCy_3$) via ligand substitution analogous to classic 18e$^-$ Fe(0) chemistry.

Al–Zn complexes have also been targeted, extending the concept of (BDI)Al aluminylene donors to more electropositive metals. For instance, reacting (BDI)Al with a β-diketiminate–Zn alkyl complex leads to insertion of Al(I) into the Zn–C bond, yielding a heterometallic complex (BDI)(Et)Al–Zn(BDI’) **67** (Figure 17) with a direct Al–Zn bond and an alkyl group now bound to aluminum.[44] Similarly, (BDI)Al oxidatively adds across a Zn–H bond giving the species (BDI)Al(H)–Zn(BDI) **68**, featuring a terminal Al–H and an Al–Zn linkage.[45] Even insertion into Zn–halogen bonds is feasible. Treating (BDI)Al with $ZnBr_2$(tmeda) produces a bromide-bridged adduct, (BDI)(Br)Al–Zn(Br)(tmeda) **69**, in which Al and Zn are directly bonded.[46] Structurally, these Al–Zn bonds (approximately 2.45–2.50 Å) are slightly longer than the sum of Al–Zn covalent radii, reflecting a polarized, largely ionic character to the L-type bonding of the (BDI)Al–M unit. This trend is consistent with the higher electropositivity of Al, the Al–Zn L-type interaction is strong but has more electrostatic character compared to the more covalent Al–Cu bond. Nonetheless, these complexes firmly establish that low-valent (BDI)Al centers can form stable L-type bonds even with s-block or early p-block metals, vastly expanding the family of BDI aluminylene–metal complexes.

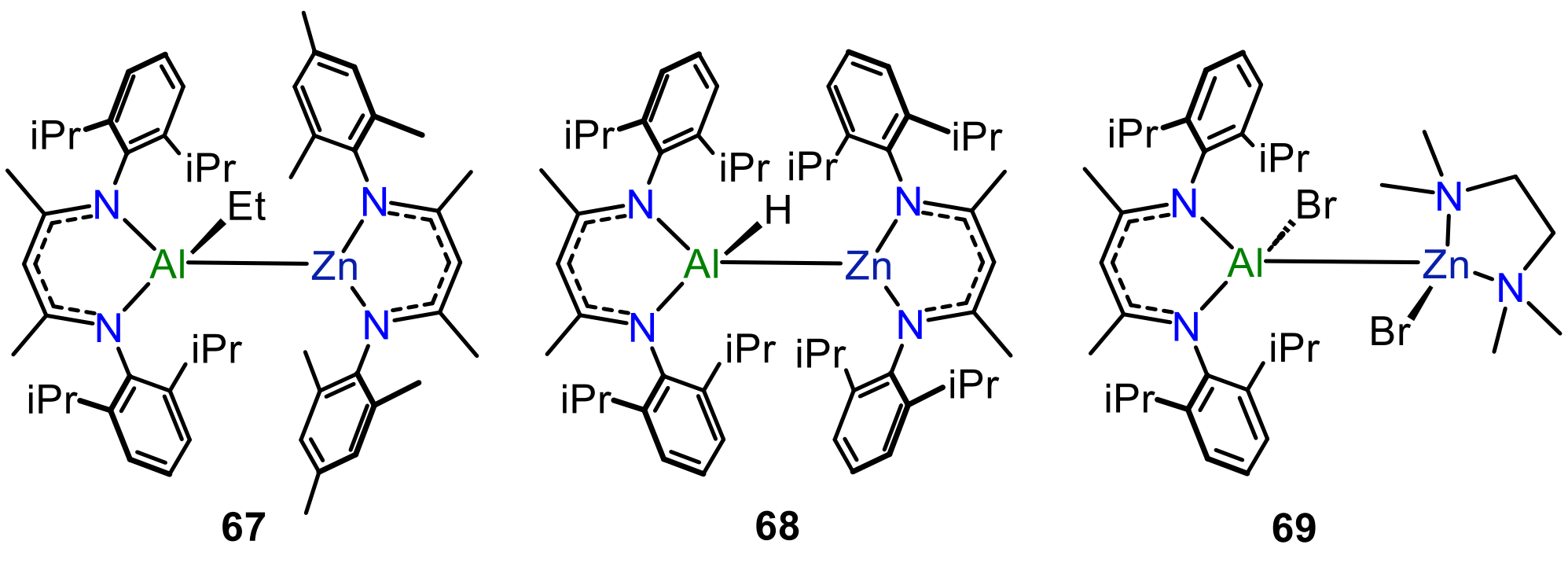

**Figure 17.** Insertion reactions of (BDI)Al into Zn–X bonds (X = alkyl, hydride, halide).

Another important class of non-Cp* aluminylene ligands involves bulky aryl-substituted ArAl fragments. Tokitoh synthesized the first terminal arylalumylene ArAl complexes where the Al(I) centers supported by extremely bulky aryl groups (e.g. Bbp = 2,6-$(CH(SiMe_3)_2)_2C_6H_3$ or Tbb = 2,6-$(CH(SiMe_3)_2)_2$-4-(tBu)$C_6H_2$) that coordinate in an L-type manner to a transition metal.[6a] In this seminal work, a dialumene (Al–Al double bond compound) was used to transfer an ArAl unit to Pt(0), yielding a discrete Al–Pt complex with the formula ArAl–Pt$(PCy_3)_2$ **70** (Figure 18, Ar = bulky Bbp aryl). Using 1,2-dibromodialumanes, $KC_8$ and Pt$(PCy_3)_2$ yielded the analogous (Ar')Al–Pt$(PCy_3)_2$ **71** (Ar' = Tbb = 2,6-$(CH(SiMe_3)_2)_2$-4-(tBu)$C_6H_2$). X-ray crystallography confirmed a short Al–Pt bond, and the Pt center exhibited a trigonal planar geometry (two $PCy_3$ and the Al ligand) as expected for a Pt(0) complex. Notably, bonding analysis of these Pt–Al complexes showed that the Al(I) ligand has a significant π-acceptor ability in addition to σ donation. NBO calculations indicated the Al–Pt bond is ~56% σ and 44% π in character. This degree of π-backbonding is remarkable for a main-group ligand and contrasts with the behavior of Cp*Al, which was found to bond to metals in a mostly ionic, purely σ-donating fashion.[6a] DFT studies on these aryl alumylene complexes **70** and **71** further suggested that the Al–Pt interaction is highly polar (dominated by electrostatic attraction), consistent with Al carrying a partial negative charge and accepting electron density from the metal. These findings underscore that the electronic profile of an Al(I) metalloligand can be tuned by its substituents: a strongly electron-releasing ligand like Cp* yields an Al donor that is essentially a two-electron σ-base, whereas a more π-acidic framework (bulky aryl or NacNac ligand) makes Al(I) into an ambiphilic donor with some carbene-like π accepting character.

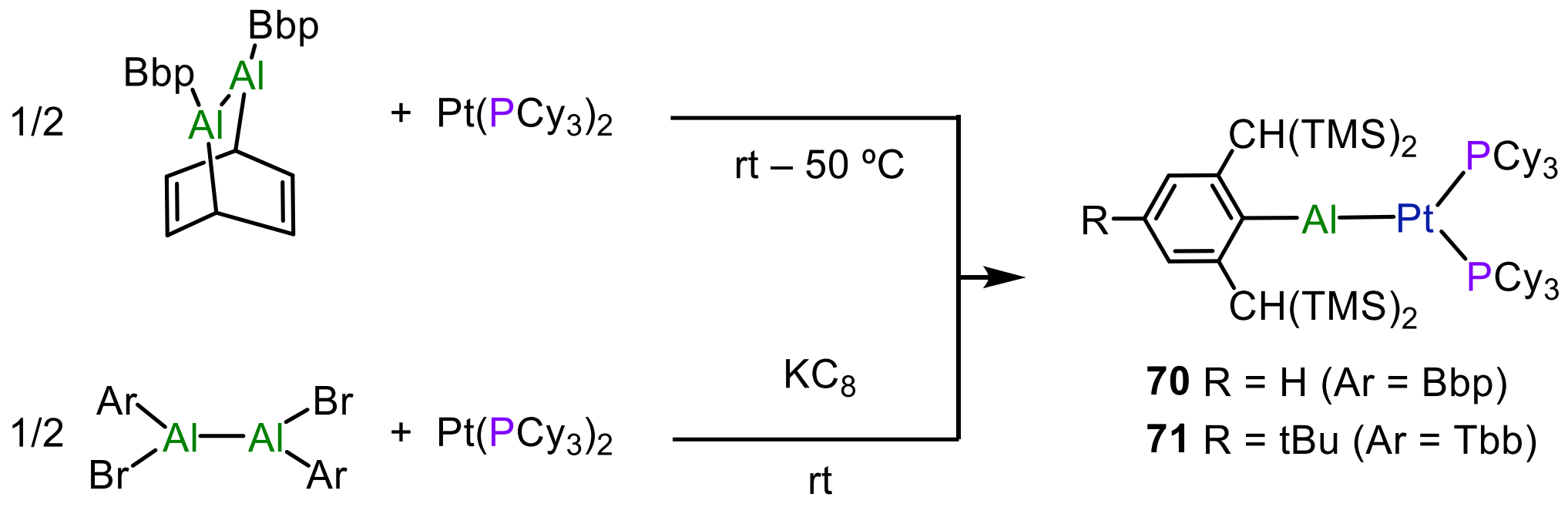

**Figure 18.** Terminal ArAl–Pt complexes.

These examples demonstrate that the development of bulky Al(I) ligands beyond Cp* has opened a broad landscape of heterometallic Al–transition metal complexes. These examples span nearly the entire periodic block of transition metals (from late TMs like Cu and Pd to mid/early metals like Fe, Cr and even Zn) demonstrating the versatility of Al(I) as a coordination partner. The Al(I) center in these complexes typically functions as a strongly donating L-type ligand, capable of stabilizing low-coordinate metal centers and even mimicking noble-metal behavior by facilitating two-electron processes at base metals.

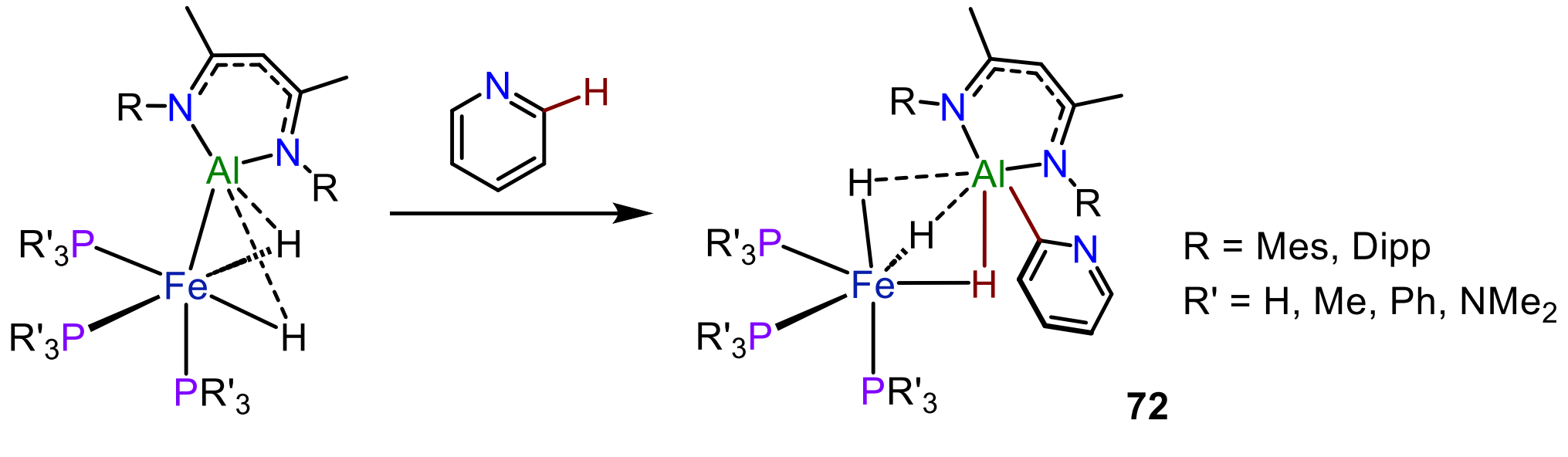

**Figure 19.** Cooperative C–H activation enabled by a bimetallic Al–Fe system.

At the same time, the ability of Al(I) ligands to engage in π-backbonding and polarization provides a unique lever to modulate reactivity at the metal center. Indeed, recent studies have shown that incorporating an aluminylene ligand can enable unusual cooperative reactions, for example, the Al–Fe **72** complex can activate C–H bonds that a monometallic iron could not, with the Al acting as an "electron reservoir" during the process (Figure 19).[47] As such, Al(I) metalloligands with robust, bulky frameworks are emerging as powerful tools to imbue base-metal complexes with new reactivity and stability, highlighting an exciting new frontier in organometallic chemistry.

# 3 Gallium as an L-type ligand in transition-metal complexes

## 3.1 Monomeric Neutral Ga(I) Species: Design and metrics

Interest in low-valent gallium chemistry surged in the 1990s, when bulky ligands made it possible to isolate monovalent Ga(I) species. A practical enabling synthon was the gallium monoiodide GaI prepared via ultrasonication of Ga metal with $I_2$ in toluene in 1990.[48] This simple GaI salt opened straightforward routes to low-valent gallium compounds. Shortly thereafter, pentamethylcyclopentadienyl gallium, Cp*Ga **73**, emerged as a useful Ga(I) source. Early access to Cp*Ga via metastable GaCl solutions and Cp*Li or $(Cp^*)_2Mg$ reagents was reported by Schnöckel in 1993,[49] and Jutzi established a convenient synthesis of Cp*Ga by reacting GaI with Cp*K, furnishing isolable Cp*Ga in useful yields in 2002 (Figure 20A).[50] In the solid state Cp*Ga forms an hexameric aggregate $[Cp^*Ga]_6$,[51] whereas under coordination conditions it behaves as a monomeric, neutral L-type donor, an important practical distinction. In parallel, the tris(trimethylsilyl)methyl ("trisyl", $C(SiMe_3)_3$) ligand delivered robust $(Me_3Si)_3CGa$ synthons **74**. Uhl reported (1992) the well-defined tetramer $[(Me_3Si)_3CGa]_4$,[52] through the disproportionation reaction of $Ga_2Br_4$ with $(Me_3Si)_3CLi$ (Figure 20B).[53] Now, this

trisyl-Ga(I) tetramer is a standard bench synthon for monomeric Ga(I) transfer in solution and an entry point to Ga(I)–TM complexes.[54]

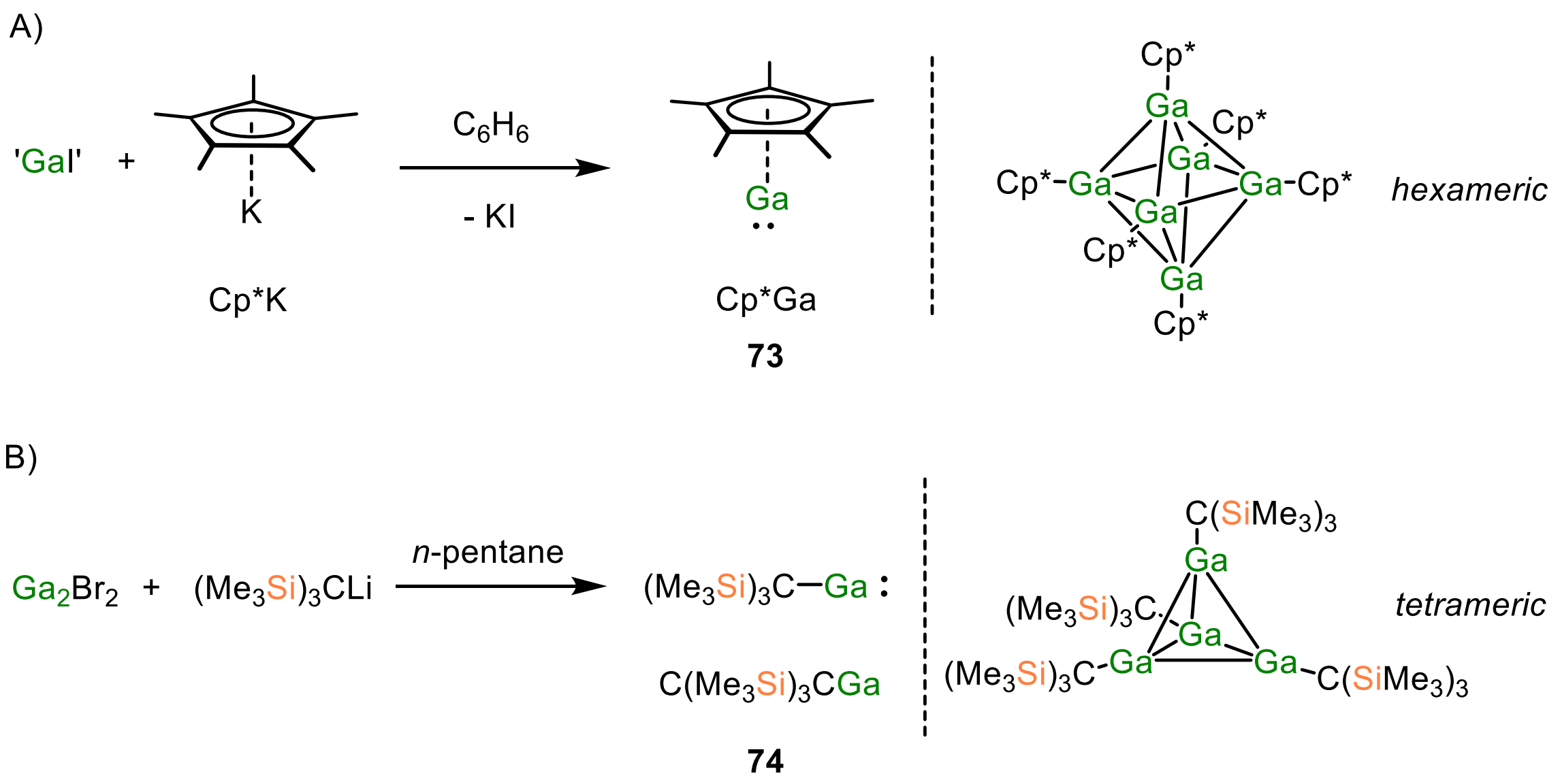

**Figure 20.** Preparation of Cp*Ga **73** and trisyl-gallium $(Me_3Si)_3CGa$ **74**.

Beyond Cp* and trisyl donors, several bulky ligand frameworks stabilize monomeric gallylenes (Ga(I) with a lone pair and an accessible p orbital) that can act as neutral L-type donors to transition metals. β-Diketiminate (BDI/NacNac) platforms[55] are especially versatile and tunable (Figure 21). They deliver well-characterized monomeric Ga(I) "carbene analogs" **75** and remain a mainstay for Ga(I) chemistry.[56] Other bulky stabilizing frameworks involve the use of tris(pyrazolyl)hydroborate[57] **76** (Tp) ligands and guanidinates[58] **77** are also good σ-donors and have been explicitly developed as L-type Ga(I) metalloligand systems to transition metals. More recently, xanthene-based phosphino-amide (PON) scaffolds **78** have been introduced by Aldridge[59], expanding the menu of strongly donating neutral Ga(I) L-ligands obtainable by salt-metathesis or reduction routes. Super-bulky aryl frameworks have also proved powerful in suppressing Ga–Ga aggregation and the formation of digallenes (RGa=GaR).[60] In particular, terphenyl-substituted ligands **79** were used by Power and Nagase[61] to access to monomeric aryl-Ga(I) derivatives, highlighting how steric protection steers monomer vs. dimer outcomes. These aryl systems have become archetypes for discrete, monomeric Ga(I) donors, and more recently, Kretschmer[62] showed that super bulky aryl skeletons, such as the s-hydrindacene skeleton **80**, can also stabilize monomeric Ga(I) species with unusual redox reactivity.

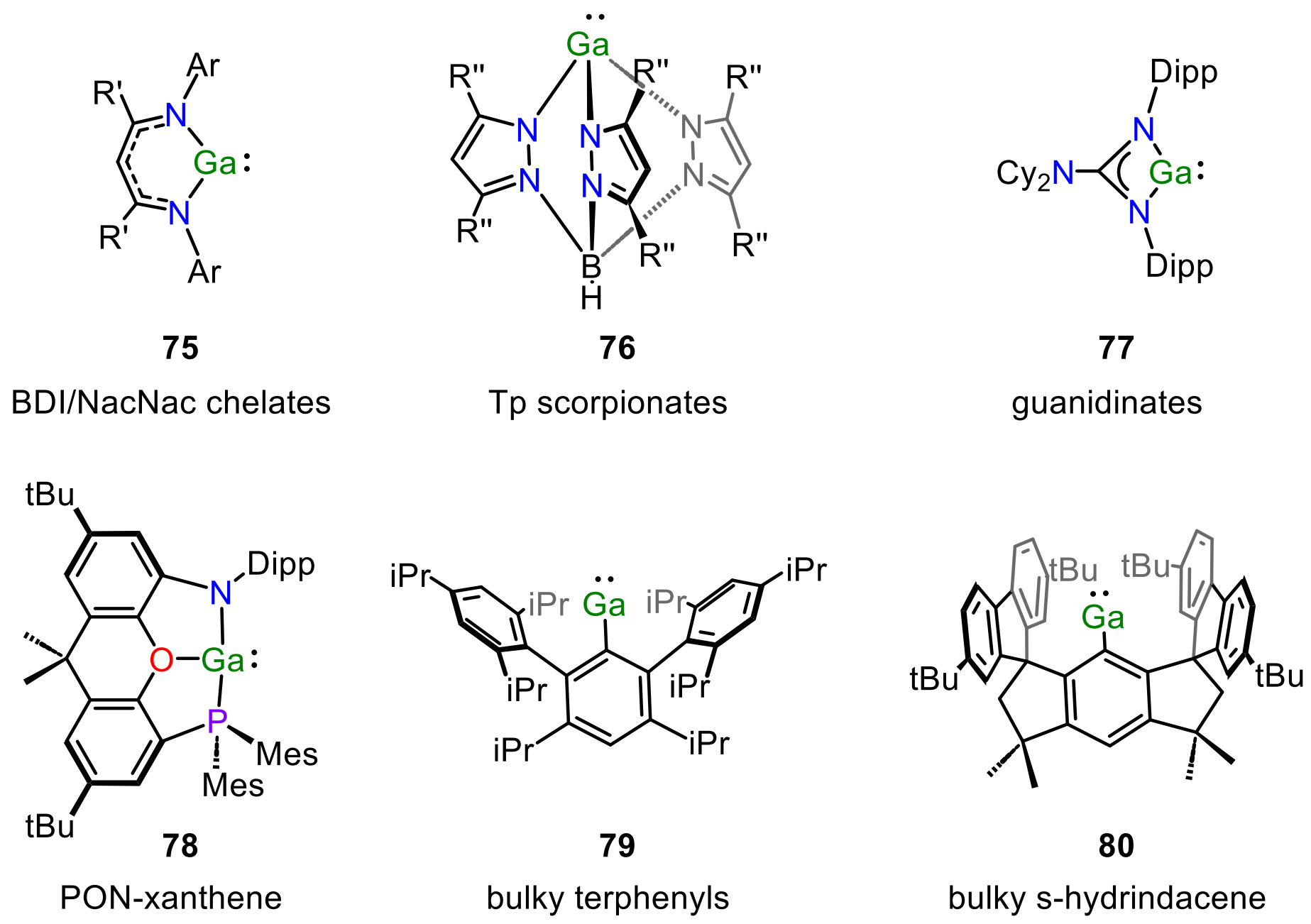

**Figure 21.** Monomeric Ga(I) compounds featuring bulky ligand skeletons **75–80**.

## 3.2 Cp*Ga as an L-type Donor: Cooperative manifolds

The earliest coordination studies treating Cp*Ga as a neutral (L-type) donor established that low-valent gallium can function as a bona fide supporting ligand to transition metals (TMs), modulating electron density and enabling substitution patterns not accessible with classical organic ligands. In direct analogy to CO/phosphine substitution chemistry, Cp*Ga inserts into the ligand sphere of electron-rich carbonyl fragments.[63] A benchmark demonstration is the clean replacement of cis-cyclooctene in $Cr(CO)_5(COE)$ (COE = *cis*-cyclooctene) and nitriles in $[fac\text{-}(RCN)_3M(CO)_3]^+$ (M = Mo, W) by Cp*Ga, affording $(Cp^*Ga)Cr(CO)_5$[64] **81** and *fac*-$(Cp^*Ga)_3M(CO)_3$[65] and **82** (Figure 22), respectively. In both series, systematic ν(CO) shifts report the σ-donor strength of the Ga(I) metalloligand, and these complexes have become the prototypical "carbonyl-analogue" Ga adducts at Cr, Mo and W. Additionally, this type of L-type Ga(I) ligation is also is also observed in related *cis*-$(Cp^*Ga)_2M(CO)_4$ (M = Cr, Mo) and $(Cp^*Ga)_4Ni$ species.[66] The tetrahedral homoleptic $(Cp^*Ga)_4Ni$ deserves special mention as an early, clear CpGa/CO analog at a zero-valent late metal, cis-$(Cp^*Ga)_2M(CO)_4$ (M = Cr, Mo) further underscores the isolobal analogy.[67] Consistent with a primarily σ-donor/weak π-acceptor profile, ν(CO) patterns show modest backbonding engagement at Ga relative to classical π-acid ligands. Complementarily, addition of Cp*Ga to the unsaturated dimers $[CpM(CO)_2]_2$ (M = Mo, W) furnishes μ-$\eta^1$-Cp*Ga bridging complexes **83** in which gallium engages two metal centers (formal 3c–2e L-donation), again accompanied by diagnostic ν(CO) changes that track the electronic perturbation at the metal core. Further L-type Ga examples include $CpGaCr(CO)_5$[68] **84** and the tetrahedral $((Me_3Si)_3CGa)_4Ni$ **85**,[69] the latter providing a clear isostructural/electronic analogy to $Ni(CO)_4$. The tetrahedral Ni complex with four neutral Ga(I) donors underscores the isolobal/"CO-analogue" behavior of L-type Ga(I).[68] These reactions established Cp*Ga as a transferable neutral donor for both terminal Ga→M and bridging μ-Ga motifs in Werner-type carbonyl chemistry.

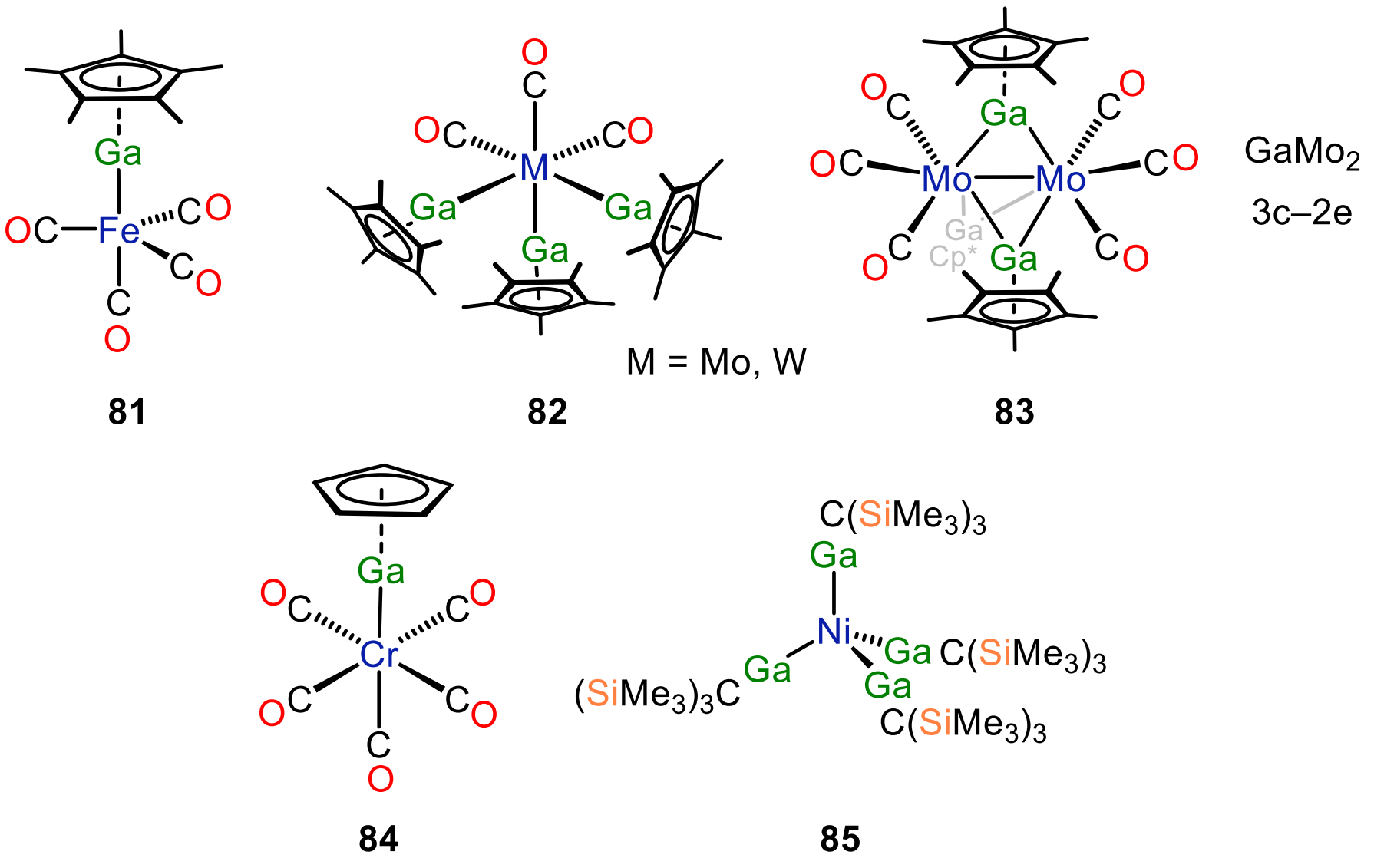

**Figure 22.** Representative L-type Cp*Ga→TM coordination complexes.

The extent of Cp*Ga substitution at TM centers depends sensitively on i) the electronic demand (basicity) of the metal fragment and ii) the co-ligands present. Thus, at $d^8$ Ru(0)/Rh(I) centers, full replacement of soft π-acceptors is uncommon. Ru($\eta^4$-butadiene)($PPh_3$)$_3$ gives (Cp*Ga)Ru($\eta^4$-butadiene)($PPh_3$)$_2$ **86** (Figure 23A), rather than homoleptic CpGa products, consistent with comparable binding energies for Ga vs. the incumbent olefin or $PR_3$ and with the electron-rich nature of the metal cores that disfavors exhaustive Ga-for-olefin/$PR_3$ exchange.[33] Comparable behavior is seen for Rh(I) in **87** (Figure 23B). [33] Here, the soft co-ligand sets (olefins) favor mixed Cp*Ga/olefin products, because complete substitution is disfavored at basic $d^8$ centers (Ru(0), Rh(I)), and an analogous outcome is observed when phosphines are present, reflecting similar ligand binding energies and competitive coordination equilibria.[10a, 67-68] In contrast, Lewis-acidic carbonyl cations (e.g., [*fac*-(RCN)$_3$M(CO)$_3$]$^+$, *vide supra*) readily accept three Cp*Ga donors to furnish *fac*-(Cp*Ga)$_3$M(CO)$_3$, underscoring substrate electrophilicity as a key handle to drive full L-ligation. Stoichiometric substitution at Ru($\eta^4$-diene)($PPh_3$)$_3$ and related fragments further illustrates that Cp*Ga competes effectively with π-ligands yet rarely displaces all soft donors at basic $d^8$ centers.[70] In practical terms, electrophilic substrates are the best entry to clean homoleptics.

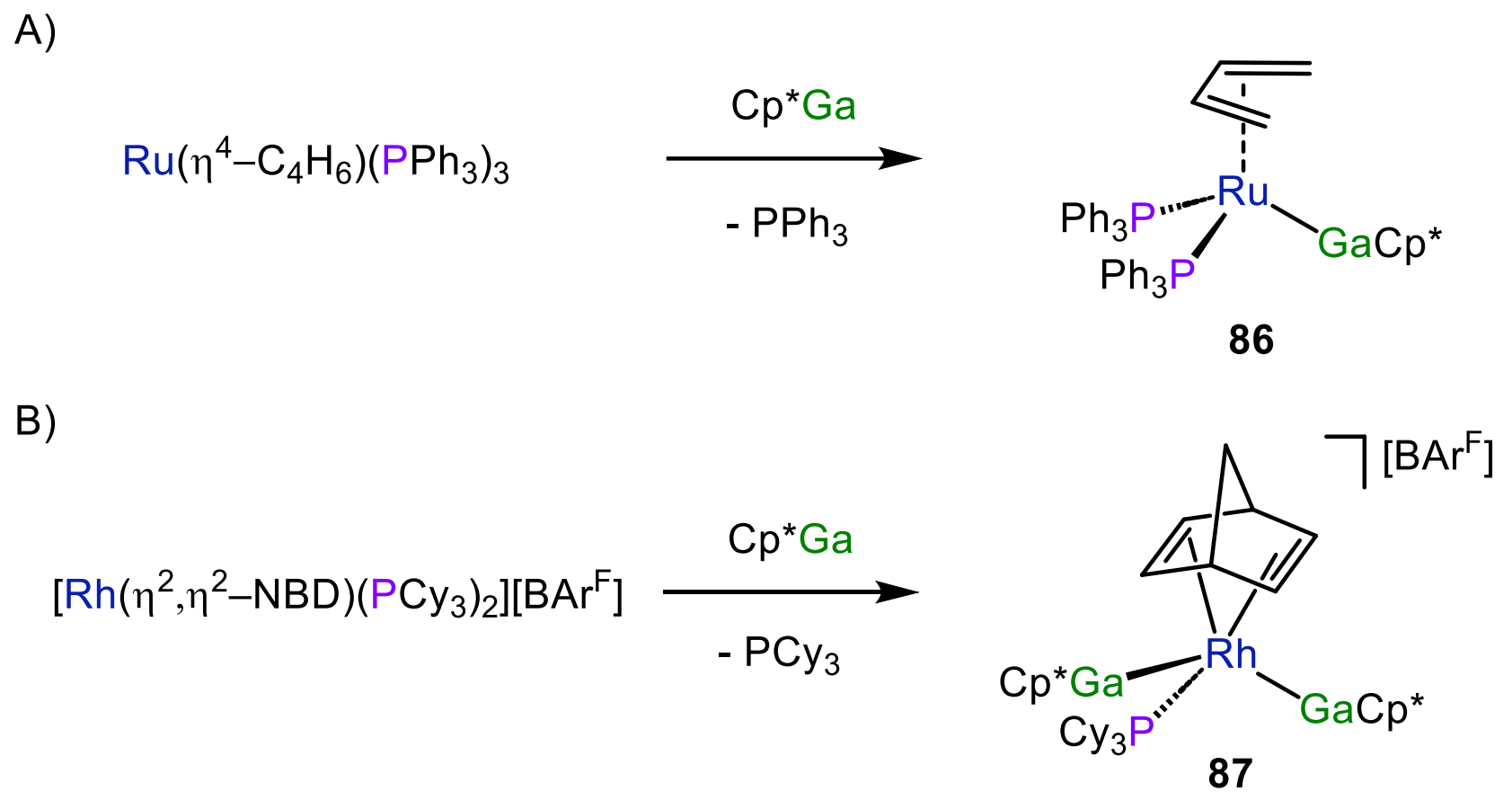

**Figure 23.** Ligand-displacement chemistry of Cp*Ga at $d^8$ Ru(0)/Rh(I) centers.

The ligand at gallium critically shapes the coordination outcome. Cp*Ga is a strong σ-donor with modest π-accepting capacity, because the Ga p orbital is partially engaged by π-interaction with the Cp* ring.[10a, 64] Even so, numerous Cp*Ga L-type metalloligands are known across the periodic table, including homoleptic $(Cp^*Ga)_4Cu$ **88**[71] and heteroleptic $(Cp^*Ga)_2Cu(CO)_4$ **89**[67] and $(Cp^*Ga)_2W(PMe_3)$ **90**[32] systems, Figure 24. Closely related "CO-analogue" chemistry is seen with trisyl-Ga(I) donors, where $((Me_3Si)_3CGa)_4Ni$ **85** was isolated as a tetrahedral Ni(0) complex directly isostructural with $Ni(CO)_4$, reinforcing the analogy between neutral Ga(I) donors and classical two-electron-ligands in the carbonyl series.[68] Many of these heterometallic complexes arise from partial CO displacement in $Ni(CO)_4$ or $Co_2(CO)_8$ and related carbonyls (Figure 24A).[64] However, a well-known limitation is the pronounced tendency of Ga(I) to aggregate (Ga–Ga nucleation/cluster growth) under certain conditions.[10a] For the $[CpM(CO)_2]_2$ (M = W or Mo) platform, Cp*Ga often binds in a labile manner[65] in **92** (Figure 24B), and η⁵→η¹ reorganization of the Cp* ring at Ga(I) can take place, promoting Ga–TM bridges. For comparison with phosphines, an estimated cone-angle proxy (~110–115°) derived from crystallographic overlays supports the steric argument, which helps rationalize competitive substitutions at crowded sites.[50, 72]

A) Cp*Ga L-type coordination resulting from CO substition at transition metal carbonyl complexes

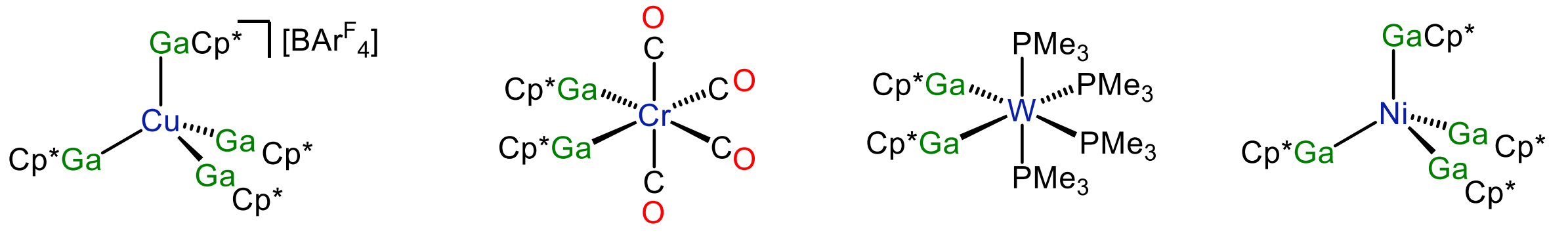

**88** *homoleptic Ga–Cu*
**89** *heteroleptic Ga–Cr*
**90** *heteroleptic Ga–W*
**91** *homoleptic Ga–Cu*

B) Addition Cp*Ga as an L-type bridging metalloligand in mutiple bonded complexes

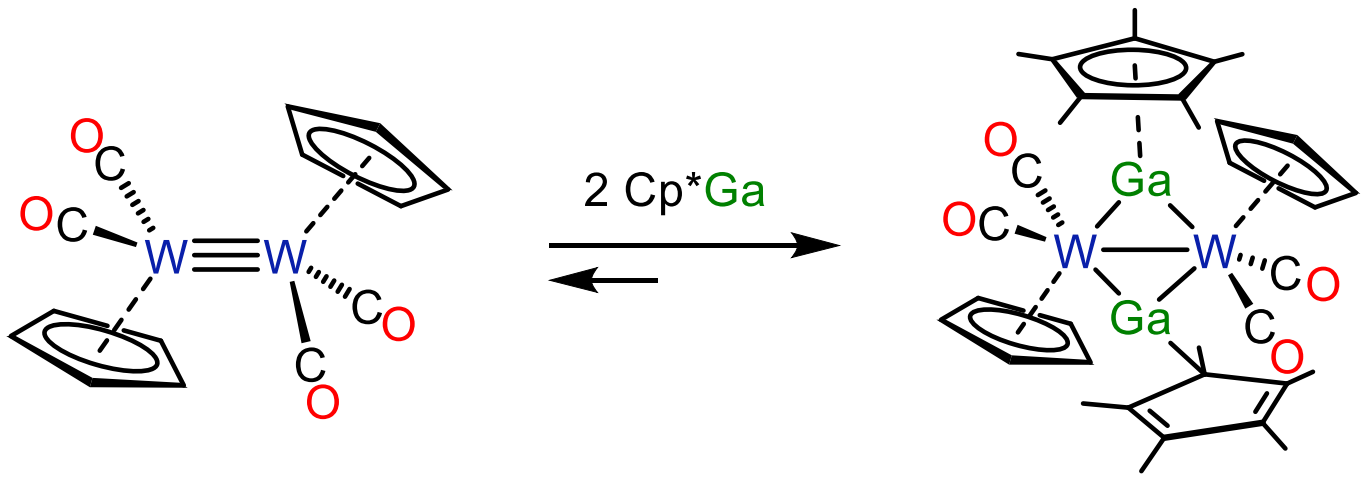

**92**

**Figure 24.** Homoleptic and heteroleptic Cp*Ga complexes with late metals.

Beyond mononuclear adducts, cluster formation is a hallmark outcome under thermodynamic control or with multinuclear precursors. A classic illustration is $Pd_2(dvds)_3$ (dvds = 1,3-divinyl-1,1,3,3-tetramethyldisiloxane), which reacts with Cp*Ga to give distinct products depending on temperature/solvent. At –30 °C in n-hexane the dinuclear complex $Pd_2(Cp^*Ga)_2(\mu_2\text{-}Cp^*Ga)_3$ **35** (Figure 25) is formed in high yield, whereas at room temperature in toluene the reaction furnishes the trinuclear species $Pd_3(Cp^*Ga)_4(\mu_2\text{-}Cp^*Ga)_4$ **93**.[34] These outcomes exemplify kinetic/solubility control in homoleptic $(Cp^*Ga)_nM_m$ clusters and rationalize why cluster growth often competes with discrete terminal Ga→TM formation in late-metal manifolds. The related family $(Cp^*Ga)_nM_m$ (M = Pd, Pt) exhibits analogous fluxionality and ligand-exchange behavior under mild conditions, rationalizing why cluster growth often competes with discrete terminal Cp*Ga→M formation in late-metal manifolds.[34]

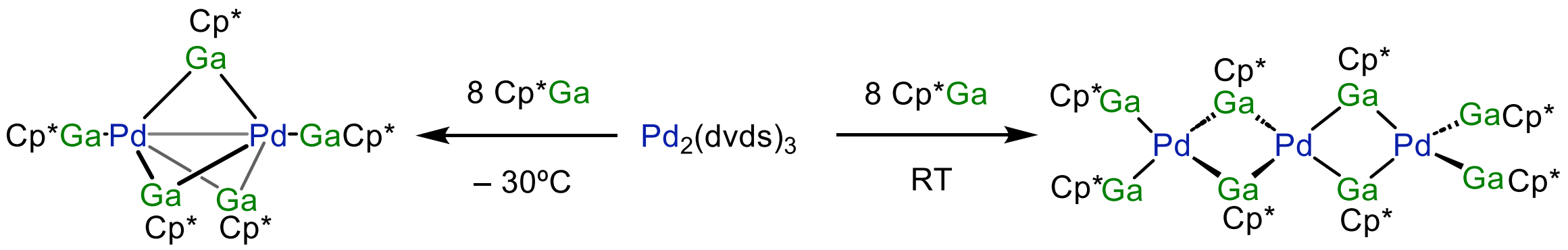

**35** **93**

*Ga3Pd2 cluster* *Ga8Pd3 cluster*

**Figure 25.** Temperature control in Ga/Pd cluster growth (dvds = 1,3-divinyl-1,1,3,3-tetramethyldisiloxane).

The formation of the different Ga→TM coordination compounds is also driven by the TM metal identity. For soft coinage ions, homoleptic all-gallium adducts are well documented. For example, reactions of $[Cu(MeCN)_4][BAr^F{}_4]$ or $Ag[BPh_4]$ with Cp*Ga give $[(Cp^*Ga)_4Cu][BAr^F{}_4]$ **94** and $[(Cp^*Ga)_4Ag][BPh_4]$ **95** (Figure 26A and B), and Ag[OTf] furnishes a dimeric $[(Cp^*Ga)_3Ag_2(\mu\text{-}Cp^*Ga)_2][OTf]_2$. These crystallographically characterized complexes demonstrate Ga–Cu and Ga–Ag bonding supported solely by L-type Ga(I) donors.[71] By contrast, more electrophilic Fe(II)/Co(II) substrates frequently undergo Cp* transfer/redox rather than simple L-type adduct formation (e.g., $[(Cp^*Ga)_4Co]^{2+}$ **96**), yielding half-sandwich $[Ga(Cp^*)Ga]^+$ products **97** (Figure 26C).[71] In other words, Cp*Ga can act as donor, reductant and Cp*-transfer reagent with sufficiently electrophilic partners, which is crucial when planning syntheses.

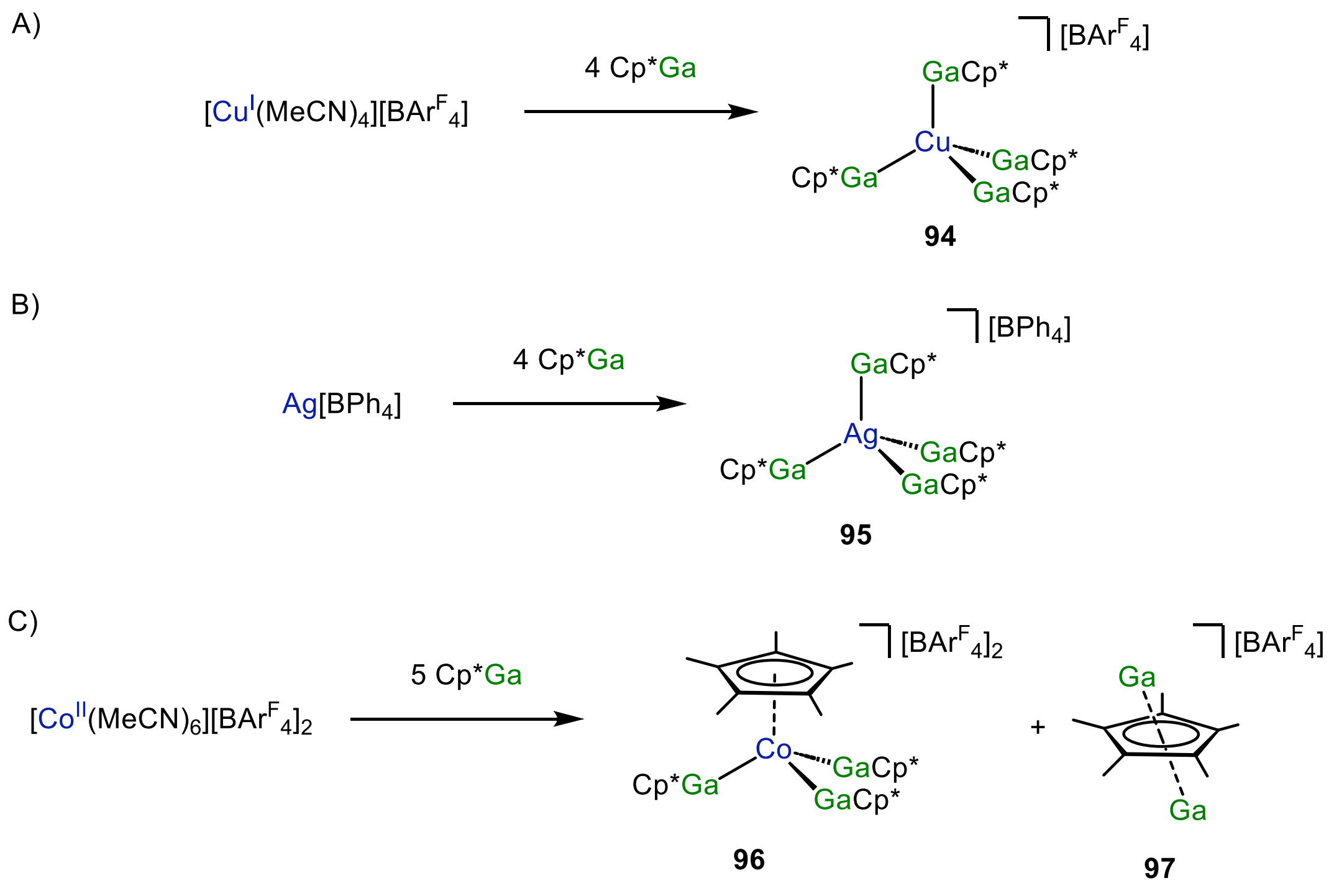

**Figure 26.** Divergent outcomes by TM identity (soft coinage *vs* electrophilic Co(II)).

Modern case studies show that L-type Ga(I) can cooperate with late metals to activate small molecules under mild conditions. For Ru, treatment of Ru(COD)(MeAllyl)$_2$ or Ru(COD)(COT) with Cp*Ga under $H_2$ generates intermediates such as (Cp*Ga)$_3$RuH$_2$, which activate Si–H ($HSiEt_3$) and arene C–H (toluene) to the well-defined products (Cp*Ga)$_3$Ru(SiEt$_3$)H$_3$ **98** and (Cp*Ga)$_3$Ru(C$_7$H$_7$)H$_3$ **99** (Figure 27A).[73] Energy decomposition and QTAIM analyses indicate predominantly σ-donation from Ga with limited π-backbonding, supporting the L-type description while rationalizing the observed cooperative bond activations. Irradiation (λ = 350 nm) of the well-defined silane adduct (Cp*Ga)$_3$Ru(SiEt$_3$)H$_3$ **98** triggers a cascading reductive elimination sequence ($H_2$ and $HSiEt_3$ loss) to generate the unsaturated (Cp*Ga)$_3$Ru(dppe) (dppe = 1,2-bis(difenilfosfino)etano) species **100** still supported solely by L-type Cp*Ga donors.[73-74] These reactive sites can be trapped with diphosphines to give **100**, providing direct structural evidence for a $(Cp^*Ga)_3Ru$ platform that remains intact after the photochemical event. In the same manifold, the photogenerated $(Cp^*Ga)_3Ru$ species shows catalytic activity in alkyne hydrogenation, although intermetallic Ru/Ga cluster growth emerges as a competing, activity-limiting side reaction.

Another example of the potential reactivity of L-type Ga–TM frameworks is the engagements of Cp*Ga with Ru to yield electron-rich GaRu polyhydrides. Treatment of $(Cp^*Ru)_2(\mu_2\text{-}H)_4$ or $(Cp^*Ru)_3(\mu\text{-}H)_3(\mu_3\text{-}H)_2$ with Cp*Ga furnishes $(\mu\text{-}Cp^*Ga)_2(Cp^*Ru(\mu_2\text{-}H)(H))_2$ **101** and $(\mu_3\text{-}Cp^*Ga)(Cp^*Ru)_3(\mu\text{-}H)_5$ **102** (Figure 27B), respectively.[75] These species explicitly retain L-type Cp*Ga donors and display Ru–H–Ga linkages and μ-Cp*Ga bridges, mapping a hydride-rich energy landscape in which Cp*Ga stabilizes high-hydrogen-content Ru cores without invoking X- or Z-type Ga chemistry. The same study shows that hydrazine-ligated Ru precursors undergo ligand substitution by Cp*Ga to give $[(Cp^*Ga)_3Ru(COD)(H)][BAr^F_4]$, highlighting that L-donation by Cp*Ga is compatible with both neutral and cationic Ru environments and serves as a gateway to hydride aggregation.

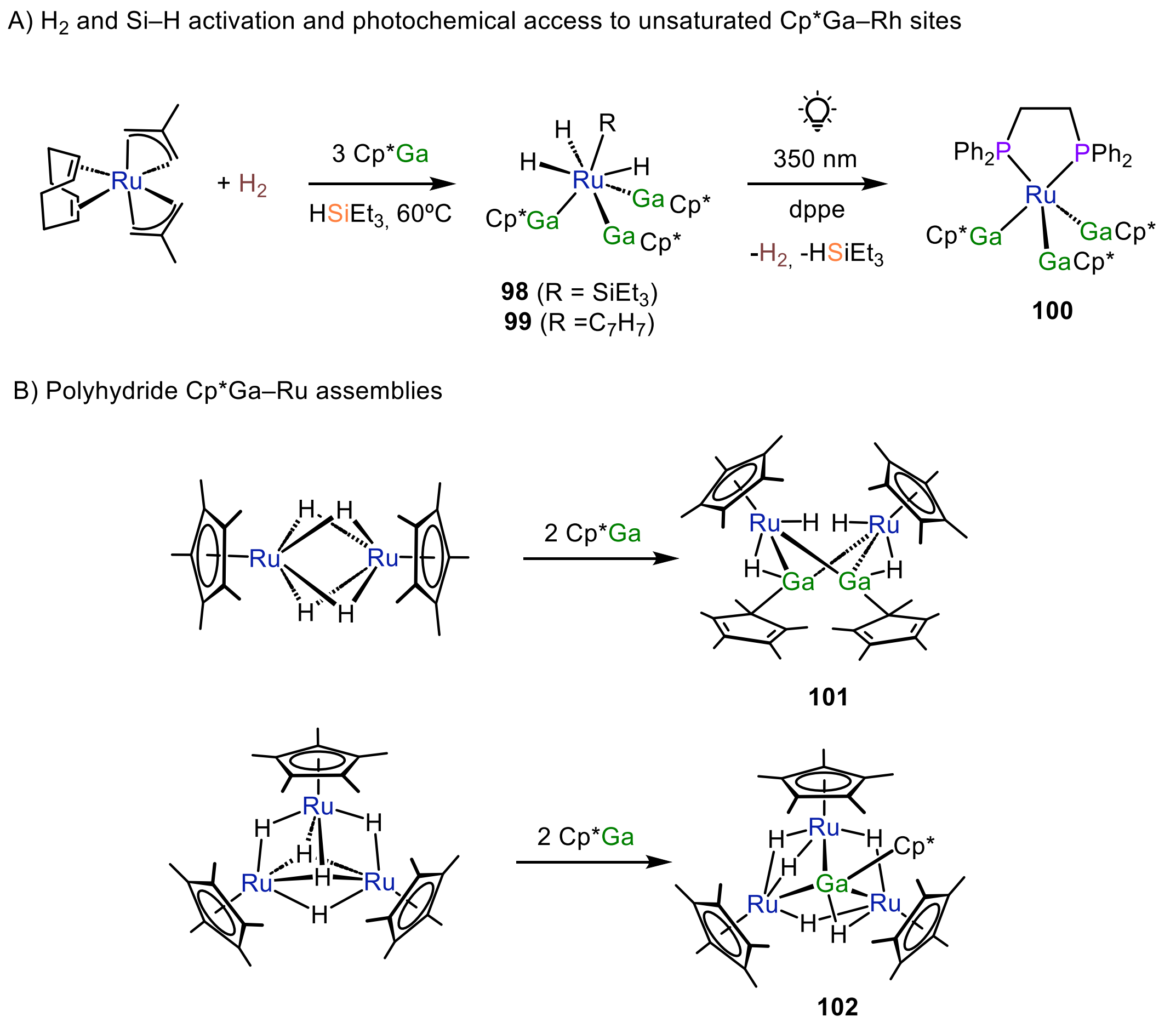

**Figure 27.** $H_2$/Si–H/C–H activation at a Cp*Ga–Ru L-type platforms and polyhydrides.

Beyond substitution chemistry, selected Cp*Ga-containing heterometallics display bond activation. In Cp*Ga–Rh systems intramolecular C–C activation of a Cp* ring has been observed, and the thermally unstable intermediate [(Cp*Ga)RhCp*(Me)$_2$] **103** evolves in solution to a zwitterionic rhodium species [Cp*Rh($C_5Me_4$–$GaMe_3$)] **104** (Figure 28A).[8b, 76] Mechanistic work supports a concerted C–C activation pathway centered at Ga, consistent with rapid Ga(I)→Ga(III) oxidation during the rearrangement.[8b, 10c] Although this transformation is not a small-molecule activation per se, it starts from a bona fide L-type Ga→Rh bond and exemplifies ligand non-innocence that can appear in borderline redox regimes while staying consistent with an L-type entry point. Separately, insertion of Cp*Ga into TM–halide bonds can dramatically increase solubility of otherwise intractable heterometallics (e.g., Fe–Cl insertions), providing highly soluble Ga(I) “carbenoid” species that stabilize reactive intermediates and reveal Cp*Ga’s dual role as donor, reductant, and Cp*-transfer reagent in **105** (Figure 28B).[77] This Fe platform illustrates how initial Cp*Ga→Fe L-coordination can evolve into bond insertion and ligand migration. The reaction of $FeCl_2$ (or $Cp^*(CO)_2FeCl$) with Cp*Ga first engages L-type binding, then effects Ga insertion into Fe–Cl, followed by Cp/Cl exchange, to give **105**.[72] These steps dramatically increase solubility, deliver highly accessible “Ga(I) carbenoid” fragments, and rationalize why Fe-group substrates often diverge from simple adduct formation despite the L-type starting point.[78] Notably, comparison of carbonyl substitution products obtained downstream (e.g., $Cp^*GaCr(CO)_5$, $Cp^*GaFe(CO)_4$) with ν(CO) data supports the view that Cp*Ga remains a strong σ-donor with limited π-acceptor ability across these families.

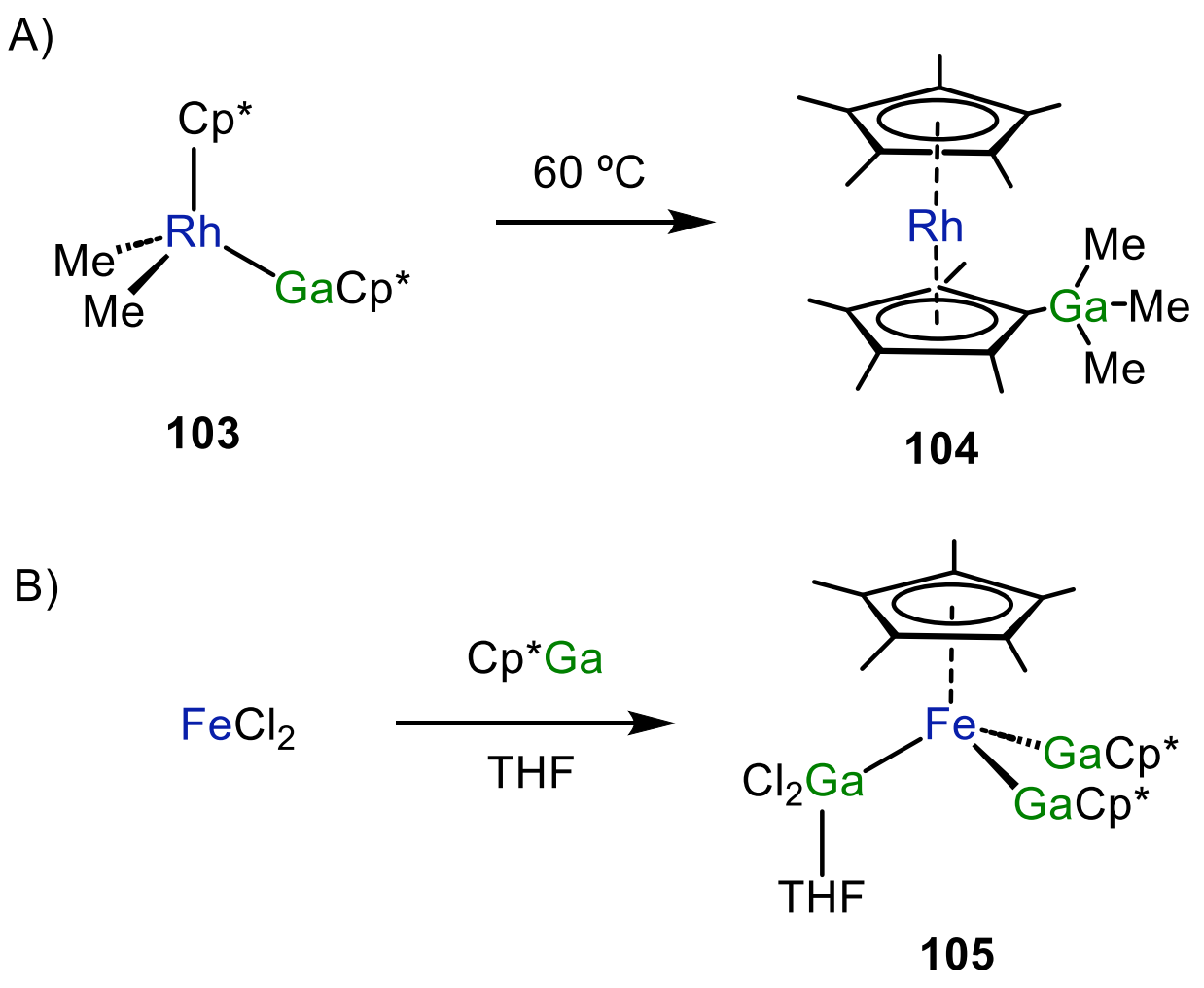

**Figure 28.** Cp*Ga originating bond-activation and bond-insertion manifolds.

Across these manifolds, a unifying mechanistic thread emerges. The Cp*Ga platform behaves as a robust, primarily σ-donating L-type ligand that survives along the reaction coordinate (including photochemical and hydrogenolytic manifolds) and cooperates with an electron-rich transition metal centers to lower barriers for bond-making/breaking. Where divergent outcomes occur (e.g., Fe–Cl insertion, Rh Cp* activation), they can be traced to substrate electrophilicity and accessible redox states rather than a change in the elemental L-type character at Ga. Computational comparisons[73] to $PR_3PRu$ analogs reinforce the donor-only profile of Cp*GaRu and provide a quantitative rationale for why Cp*Ga can enable H–H, Si–H, and C–H activation while resisting strong π-backbonding. These insights firmly position Cp*Ga as a reactivity-enabling L-donor, transforming the traditional view of "inert" main-group ligands into one of active participants in cooperative Ga–TM chemistry.

Two trends merit emphasis for Cp*Ga. First, while Cp*Ga is a powerful σ-donor yet a poor π-acceptor, a consensus supported by ν(CO) data across the Mo/W carbonyl series as well as modern computational analyses of Ru/Ga systems.[10c, 73] Second, metal identity strongly biases outcomes. Soft Cu(I)/Ag(I) stabilize homoleptic Cp*Ga adducts (and even dimeric cations with bridging Ga under halide abstraction), whereas more electrophilic Fe(II)/Co(II) frequently undergo Cp transfer/redox rather than simple adduct formation, underscoring the competing roles of Cp*Ga as donor, reductant, and Cp-transfer reagent in borderline cases. These nuances do not detract from the central message of this section: Cp*Ga behaves as a versatile CO-analog L-ligand, enabling predictable substitution at electrophilic carbonyl fragments, μ-Ga bridging at unsaturated dimers, and under cluster-forming conditions homoleptic $(Cp^*Ga)_nM_m$ aggregates, all of which can be leveraged for cooperative small-molecule activation in Cp*Ga–TM platforms.

### 3.3 Bulky Ga(I) donors: Steric/electronic divergence and cluster propensity

In parallel to Cp*Ga (discussed in Section 3.2), aryl-substituted “gallylenes” (Ar*Ga), β-diketiminate gallylenes (BDI)Ga, tris(pyrazolyl)borate derivatives (TpGa), four-membered NHC-analog Ga rings ($N_2Ga$), and newer multidentate N,N-frameworks such as phenalenyl, carbazolate pincers ($RN_2Ga$) all behave as strong σ-donors with measurable π-accepting capacity. In essence, these Ga(I) ligands mimic classical CO/$PR_3$/NHC donor manifolds while retaining distinctive main-group character.[43, 79] L-type two-electron donor bonding is the default assignment for these neutral Ga(I) donors; however, some borderline or debated cases are instructive and are highlighted below (e.g., putative multiple bonds, μ-Ga bridging, halide-insertion manifolds). Overall, bulky Ga(I) ligands afford an expanded coordination chemistry at transition metals, often allowing only one Ga per metal (monosubstitution due to sterics) and sometimes unlocking reaction pathways unavailable to purely organic ligands.

#### 3.3.1 Terphenyl Ga(I) frameworks

Early reports of multiple bonding in ArGa–TM complexes prompted seminal discussion of the true nature of Ga–TM bonding. The often-cited “Fe≡Ga triple bond” reported by Robinson in 1997 for **106** (Figure 29),[80] was re-evaluated shortly thereafter by Cotton and Feng in 1998. Their analysis reassigned the interaction as a donor–acceptor Ga→Fe bond (L-type) based on the trigonal planar geometry at Ga, the ν (CO) frequency pattern, and DFT calculations, all inconsistent with a true Fe≡Ga triple bond.[81] This re-interpretation has become a touchstone for the field. Even when some backbonding from the metal into the vacant p-orbital at Ga is detectable, the electron counting and spectroscopic signatures are best captured by treating Ga(I) as a neutral two-electron donor (L-type) rather than an X-type or multiply bonded anionic $Ga^{n-}$ fragment.

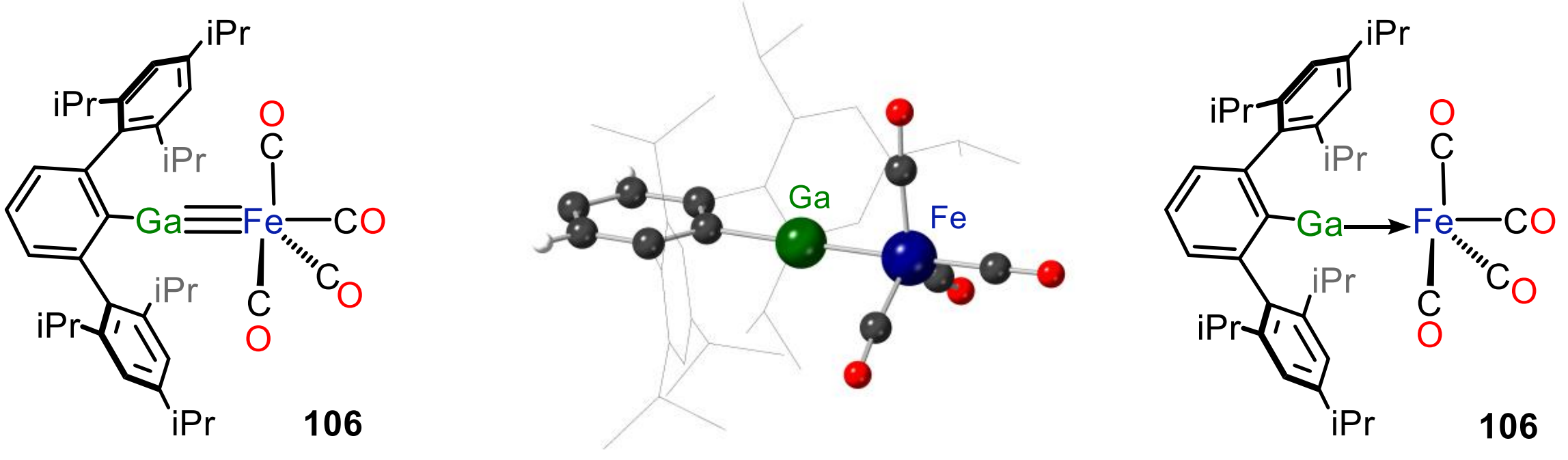

*multiple bond character* (Robinson, 1997) | crystal structure | *L-type donation* (Cotton and Feng,1998)

**Figure 29.** Ga–Fe bond: original triple bond claim (left) *vs* donor–acceptor L-type Ga→Fe reassignment (right).

Terphenyl-stabilized gallylenes extend this motif. Thus, treatment of $Ni(CO)_4$ with the bulky aryl gallylene initially gives the mononuclear adduct $(ArGa)Ni(CO)_3$ (Ar = $C_6H_3$-2,6-$(Dipp)_2$) via CO displacement by L-type Ga. Over time, however, cluster growth dominates and the carbonyl $(ArGa)_3Ni_4(CO)_4$ **107** is isolated (Figure 30).[82] In forming **107**, three Ga(I) centers cap a $Ni_4$ carbonyl core. IR ν(CO) trends for this system map the donor impact of Ga(I): the Ga→Ni donors attenuate the CO stretching frequencies, indicating strong σ-donation by Ga(I) but without the pronounced π-acidity of CO ligands. An important general theme emerges, the extreme steric encumbrance of bulky Ga ligands often limits the number of Ga donors bound to a single metal (favoring only one Ga per metal) and can divert the chemistry toward thermodynamically stabilized clusters under mild conditions, when multiple unsaturated fragments aggregate.

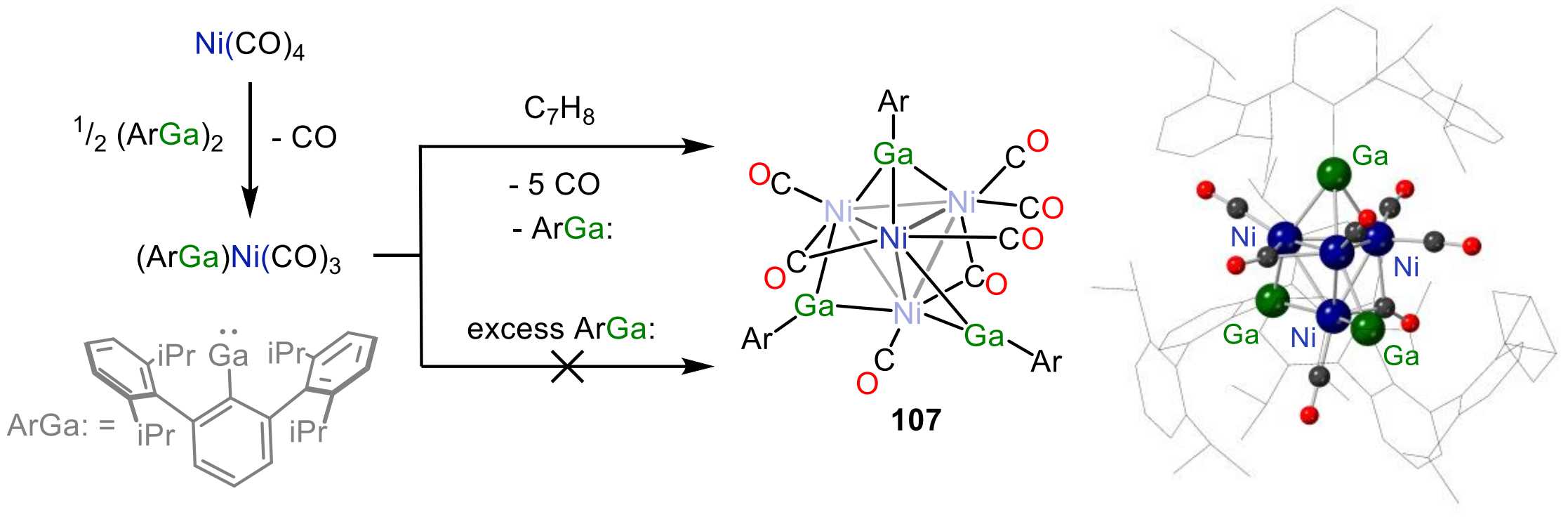

**Figure 30.** Formation of (ArGa)Ni(CO)$_3$ and evolution to (ArGa)$_3$Ni$_4$(CO)$_4$ (Ar = terphenyl).

Notably, bridging Ga(I) donors (either μ–ArGa or μ–Cp*Ga) can share one lone pair across two metal centers in a 3c–2e fashion. Such delocalized Ga–M–Ga interactions defy simple 2c–2e bonding labels yet are still best described as Ga(I) acting as a bridging L-type metalloligand (see the Ag–Ga bridging case in Section 3.2).[71] In these cases, Ga(I) donates to both metals, and the electron pair is spread over the Ga–M–Ga unit. This again underlines that even in unconventional geometries, the Ga is providing a neutral donor function rather than forming direct Ga=M multiple bonds.

Very recently, Kretschmer and co-workers showed that super-bulky aryl frameworks (e.g., s-hydrindacene) can stabilize truly monomeric, single-site Ga(I) donors that remain intact while displaying redox-variable and redox-inert reactivity modes, a stringent proof that aryl scaffolds alone can deliver robust L-type Ga(I) ligands poised for TM coordination developments in the future.[62]

### 3.3.2 Tris(pyrazolyl)borate (Tp) Ga(I) and related multidentate systems

Reger's early work demonstrated the use of a tris(pyrazolyl)borate (Tp) scaffold to stabilize a Ga(III) precursor, which could then be chemically reduced in the presence of a transition metal fragment to generate a Ga→TM complex. For example, starting from the tris(pyrazolyl)borate complex HB(3,5-$Me_2$pz)$_3$GaClMe, reduction with $Na_2$[Fe(CO)$_4$] afforded the iron carbonyl adduct [HB(3,5-$Me_2$pz)$_3$Ga]Fe(CO)$_4$ **108** (Figure 31).[83] Electron counting and spectroscopy indicate a Ga(I)→Fe interaction of bond order 1, i.e., Ga(I) is acting as a neutral L-type donor to Fe, rather than a multiply bonded Ga≡Fe unit. This synthetic umpolung strategy (Ga(III) → Ga(I) at the metal center) remains methodologically useful. The redox change at the metal fragment unmasks a coordinative Ga(I) ligand from a Ga(III) precursor, delivering a well-defined L-type Ga→Fe carbonyl complex in a single step.

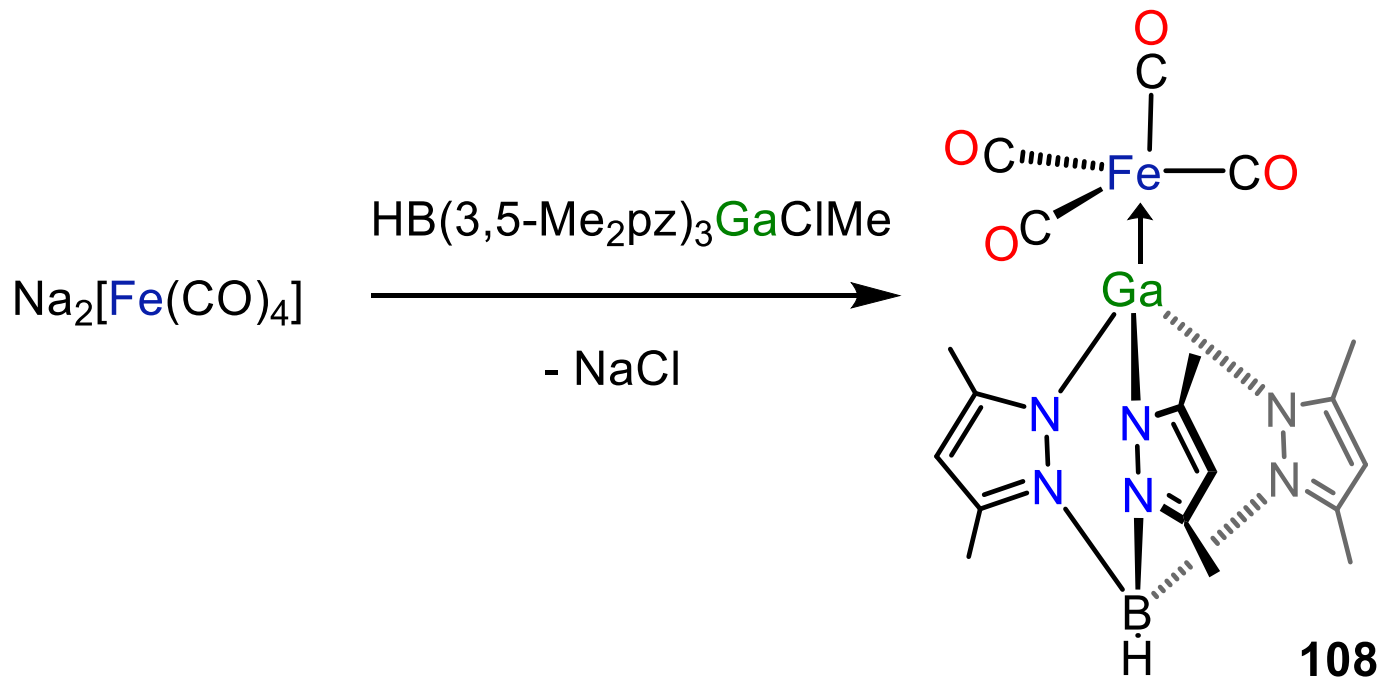

**108**

**Figure 31.** Synthesis of [HB(3,5-$Me_2$pz)$_3$]GaFe(CO)$_4$ via $Na_2$[Fe(CO)$_4$] reduction.

Newer multidentate N,N frameworks likewise stabilize Ga(I) donors as bona fide L-type ligands on electron-poor transition metal fragments. Kodama and Tobisu, for example, used a phenalenyl-based bidentate to isolate a monomeric Ga(I) that coordinates to W(CO)$_5$, giving a discrete Ga→W adduct **109**. The Ga–N bond lengthening in **109** relative to the free Ga(I) was noted, reflecting electron withdrawal by the electron-deficient $W(CO)_5$ fragment (Figure 32A).[84] Similarly, Tan employed a bis(imino)carbazolate pincer ligand to generate a gallylene that, under UV irradiation in the presence of $Cr(CO)_6$, cleanly coordinates to a $Cr(CO)_5$ fragment. The resulting complex **110** (Figure 32B) features Ga(I)→Cr bonding consistent with a neutral L-type donor interaction, and the observed ν(CO) shifts indicate the Ga(I) is a strong σ-donor (with relatively minimal π-backbonding from Cr).[85] In both cases, the multidentate ligand frameworks enforce low-coordinate Ga(I) centers that act as two-electron donors to the transition metal, analogously to how an NHC or phosphine would, yet with the added twist of the Ga's accessible empty p-orbital.

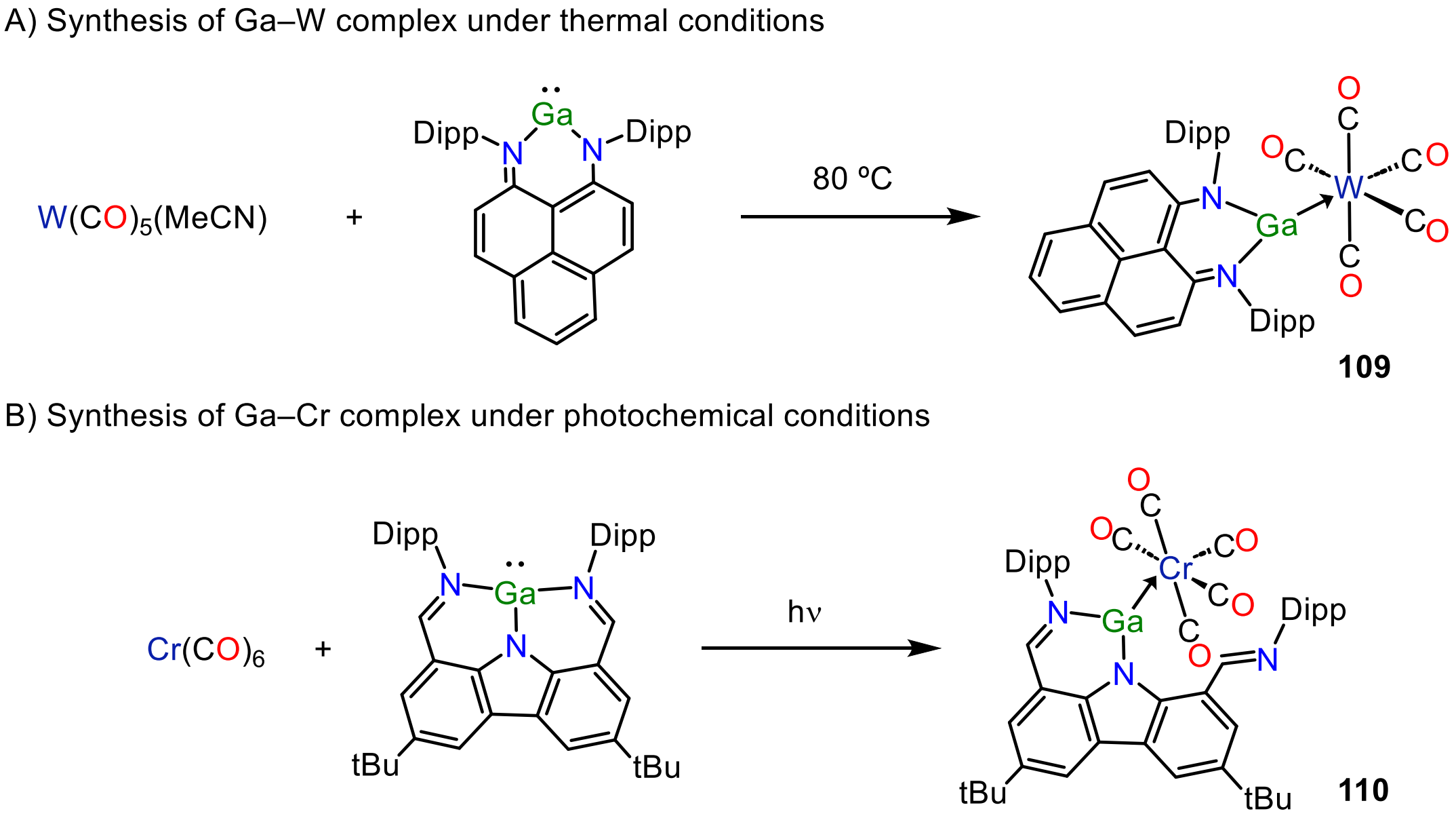

**Figure 32.** A) Phenalenyl-Ga(I)→[$W(CO)_5$] and B) carbazolate-Ga(I)→$Cr(CO)_5$ (Dipp = 2,6-diisopropylphenyl).

An especially intriguing extension involves redox-active ligand scaffolds. Fedushkin showed that a transition metal complex containing a diimine-stabilized Ga(I) (the dpp-bian ligand) can reversibly bind certain unsaturated substrates at the Ga center.[86] For example, the species (dpp-bian·)$GaCr(CO)_5$ **111** and (dpp-

bian)GaFeCp(CO)$_2$ **112** (Figure 33) will adduct tBuNC or PhNCS at the Ga, forming isolable complexes **113** and **114** in which the substrate is bound to Ga(I). Remarkably, some of these adducts can dissociate upon mild heating, regenerating the Ga→TM complex. This offers a means to regulate the transition metal's coordination environment and reactivity in a switchable fashion, effectively using Ga(I) as both a stabilizing donor and as a site for cooperative substrate activation. Such behavior underscores that certain Ga(I) ligands, especially when incorporated in redox-non-innocent frameworks, can partake in shuttling reactants to and from a metal center in a manner not achievable with inert donors.

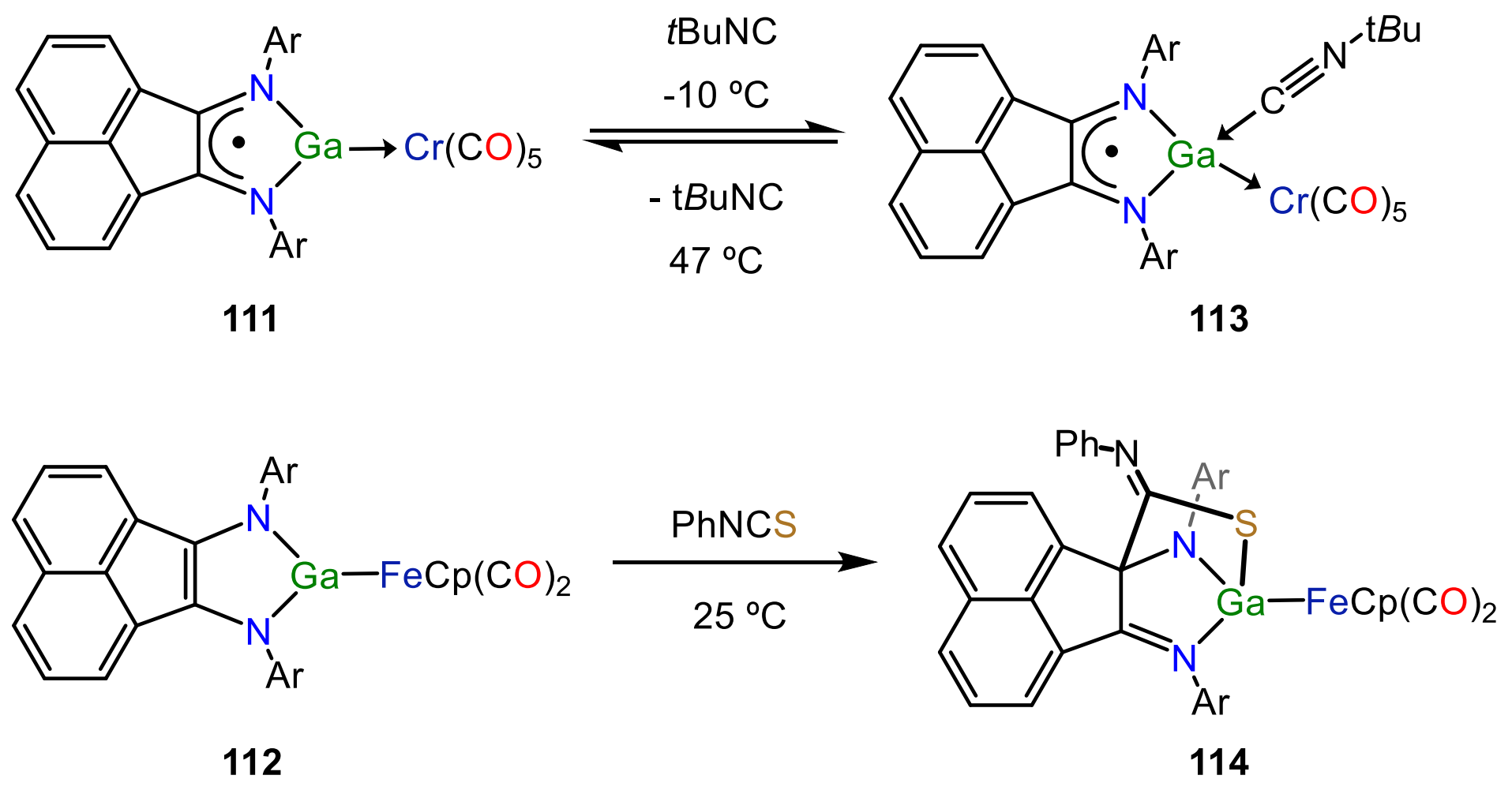

**Figure 33.** Reactivity of (dpp-bian)Ga–Cr and (dpp-bian)Ga–Fe species.

Importantly, ligand design continues to advance in this area. Very recently, Ghadwal reported an annulated carbocyclic diimine (ADC) pincer-type ligand that stabilizes a two-coordinate Ga(I) dimer **115** (Figure 34). This species can bind to $Fe_2(CO)_9$, forming a Ga(I)→Fe(CO)$_4$ adduct **116** in which each Ga donates a two-electron pair to an Fe(CO)$_4$ unit.[87] Remarkably, the same $Ga_2$ framework was shown to activate white phosphorus ($P_4$) and even effect C(sp$^2$)–F bond cleavage in aryl fluorides, underscoring that while these Ga(I) donors primarily act as L-type ligands in **116**, they can also engage in substrate activations reminiscent of transition-metal reactivity. This highlights the rich, expanding chemistry enabled by bulky Ga(I) metalloligands.

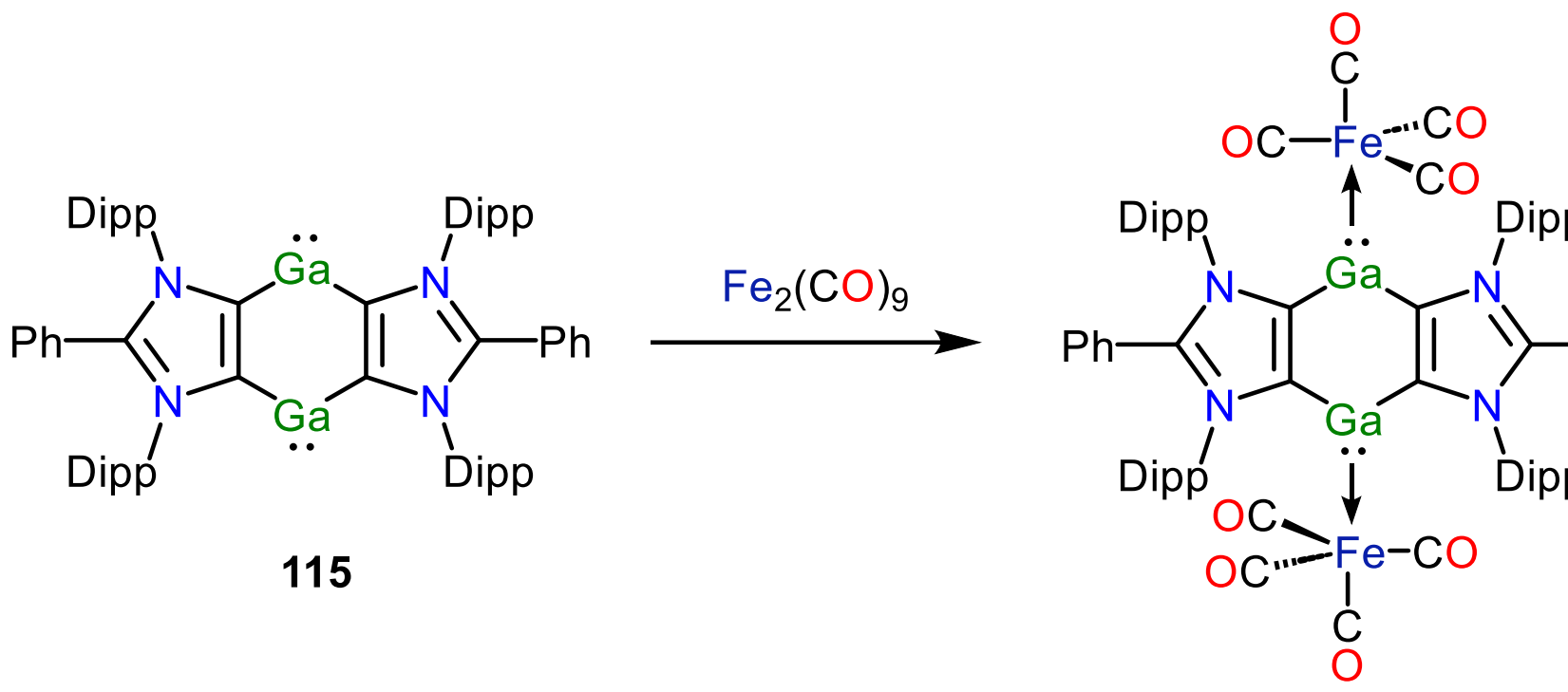

**Figure 34.** Carbocyclic diimine Ga(I) and L-type Ga→$Fe(CO)_4$ species.

### 3.3.3 Four-membered Ga(I) NHC-analogs and π-backbonding

In 2006, Jones and co-workers reported the first neutral four-membered Ga(I) heterocycle, an “NHC analogue” featuring a $C_2N_2Ga$ ring, from a diimine–carbene-like ligand.[58a] his compound, (DippNC($NCy_2$)NDipp)Ga (Dipp = 2,6-$iPr_2C_6H_3$, Cy = cyclohexyl), behaves as an ambiphilic Ga(I) ligand. Coordination to group-10 metal fragments was examined across different stoichiometries (Figure 35).[88]

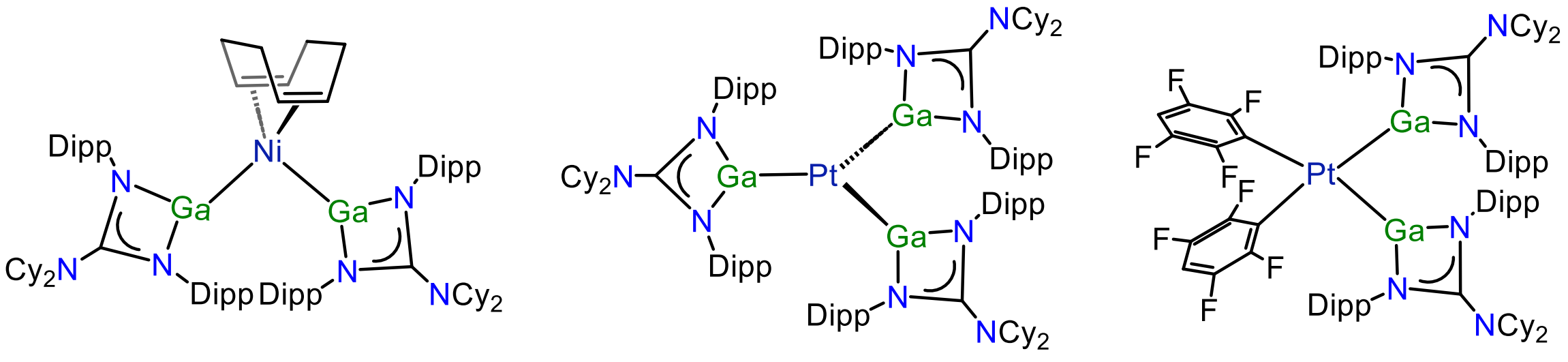

117 118 119

**Figure 35.** NHC-analog Ga(I)–Ni and Ga(I)–Pt systems.

Computational analysis shows a highly directional lone pair at Ga (strong σ-donor) and a low-lying empty p-orbital (π-acceptor) on this Ga(I) cyclic ligand. The HOMO–LUMO gap of the four-membered Ga(I) ring (~60 kcal/mol) is significantly smaller than that of a six-membered Ga(I) ring (~100 kcal/mol),[89] implying enhanced ambiphilicity (i.e. a greater readiness to both donate and accept electron density). In practical terms, these NHC-analog Ga(I) donors tend to be slightly less nucleophilic (weaker σ-donors) than Cp*Ga, owing to the electron-withdrawing effect of the chelating amido-carbene framework. However, crystallographic and computational studies of Pt complexes with this Ga(I) ligand reveal significant d(π)→p(π) backbonding from the metal into the Ga's vacant orbital.[88a] This π-acceptor character is more pronounced than in Cp*Ga complexes, because the Cp* ligand in Cp*Ga partly delocalizes electron density into Ga (dampening Ga's π acidity, cf. Section 3.2). Thus, the four-membered "carbene-analogue" Ga(I) combines strong σ-donation with a notable capacity for π-backbonding acceptance, a balance that parallels the behavior of N-heterocyclic carbenes but with Ga's different energetic profile. These findings highlight how altering the Ga(I) ligand's ring size and ligand bite can tune the donor/acceptor properties of Ga dramatically.

### 3.3.4 β-Diketiminate Ga(I) vs Cp*Ga: Steric and electronic divergence

Among non-Cp* Ga(I) frameworks, β-diketiminate-supported Ga(I), commonly denoted as (BDI)Ga, is arguably the most widely employed L-type Ga(I) donor in transition-metal coordination chemistry. There are stark electronic and steric differences between (BDI)Ga and Cp*Ga. In Cp*Ga, the Cp* ($C_5Me_5$) ligand is anionic and π-bound, giving the Ga a formal sp configuration with two vacant p-orbitals that are partially stabilized by Cp*→Ga π donation.[42] This renders Ga in Cp*Ga more Lewis acidic (and often a stronger σ-donor but a weaker π-acceptor) compared to (BDI)Ga. In (BDI)Ga, by contrast, the Ga center is $sp^2$ (trigonal planar) and is supported by a monoanionic, rigid N,N ligand, the electron-rich β-diketiminate π-system locks some electron density at Ga and makes it less Lewis acidic than Cp*Ga. Consequently, (BDI)Ga is less prone to oligomerize or form Ga-rich clusters. It tends to favor clean, heteroleptic Ga–TM complexes at a single metal center, whereas Cp*Ga often exhibits a greater tendency toward multi-Ga aggregation under similar conditions.

Indeed, alternative bulky monoanionic scaffolds can achieve even greater steric protection. For instance, Aldridge's xanthene-based PON (phosphino-oxide-amido) ligand was shown to support a monomeric Ga(I) center (PON–Ga), with steric bulk exceeding that of BDI. The (PON)Ga(I) complex exhibits no tendency to oligomerize, highlighting how extreme ligand sterics can thwart Ga–Ga aggregation and enforce a strictly mononuclear Ga(I) donor.[59] Such findings reinforce the principle that both electronic saturation and steric hindrance at Ga(I) are key to stabilizing isolated Ga→TM units.

These differences are evident in simple substitution reactions. For example, treatment of $Pd_2(dvds)_3$ (dvds = divinyldisiloxane) with (BDI)Ga cleanly affords a mono-substituted complex, (BDI)GaPd(dvds) **120** (Figure 36) as the major product.[42] In contrast, reacting $Pd_2(dvds)_3$ with Cp*Ga leads rapidly to Ga-rich Pd/Ga clusters (see Section 3.2).[34] Likewise, $Pt(1,5\text{-}COD)_2$ undergoes displacement of one COD ligand by (BDI)Ga, with concomitant isomerization of the remaining 1,5-COD to 1,3-COD, yielding $((BDI)Ga)_2Pt(\eta^2\text{-}1,3\text{-}COD)$ **121** or, under different stoichiometry, $((BDI)Ga)Pt(\eta^2\text{-}1,3\text{-}COD)$ **122**.[42] This rearrangement underscores the flexible, non-innocent behavior of the olefin ligand in presence of the Ga donor. Moreover, ligand exchange at the vacant site is facile, introduction of CO or tBuNC quantitatively replaces the bound 1,3-COD to afford carbonyl or isocyanide adducts **123**, attesting to the purely L-type (two-electron donor) character of the (BDI)Ga ligand and the kinetic lability typical of carbonyl-like ligand sets.

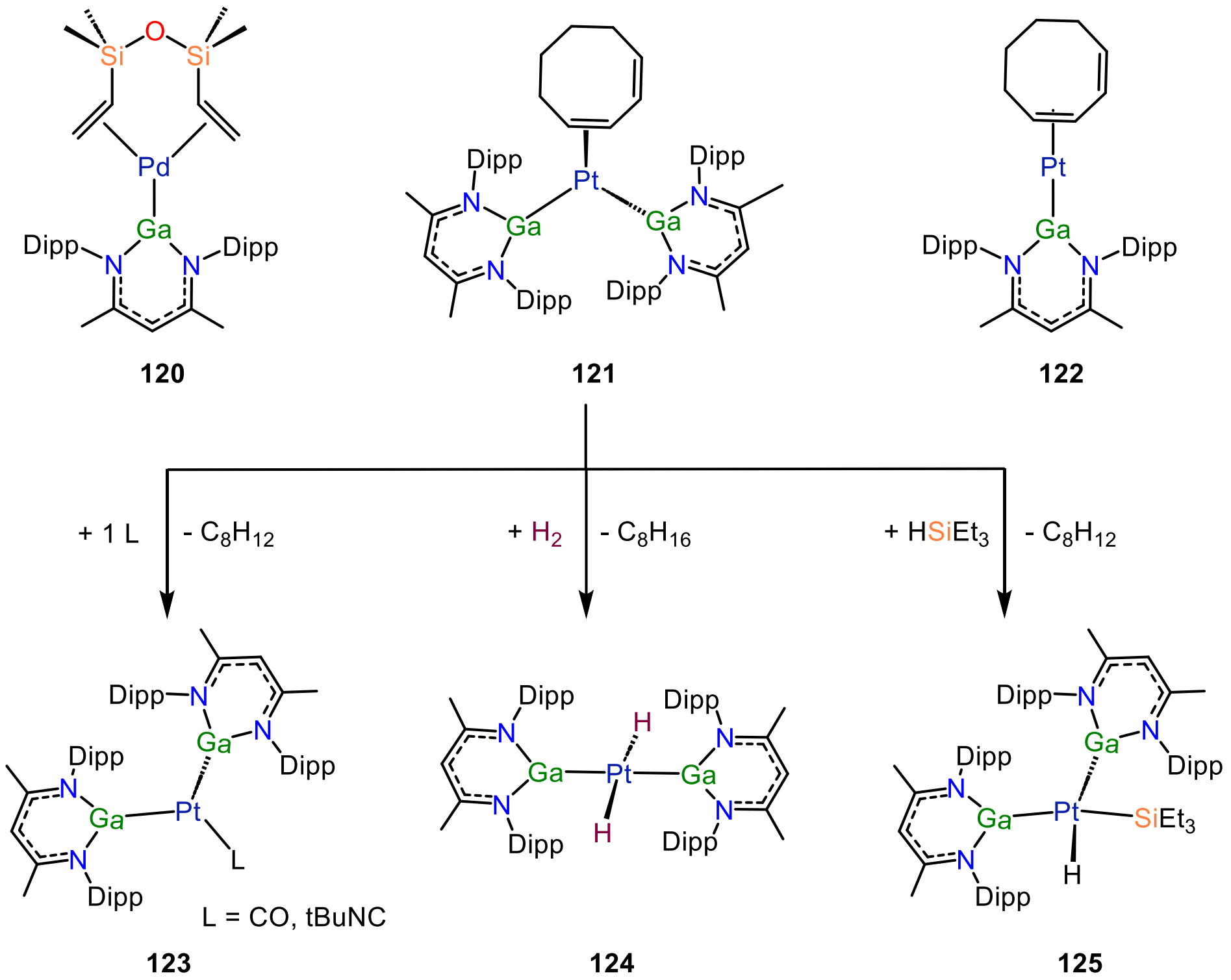

**Figure 36.** Diverse (BDI)Ga–Pd and (BDI)Ga–Pt platforms, highlighting mono-substitution and subsequent ligand exchange or small-molecule activation.

Notably, the (BDI)Ga–Pt fragment in **121** can engage in oxidative addition and insertion reactions reminiscent of those seen with Cp*Ga. The neutral Ga(I) donor in **121** can split $H_2$ under mild conditions, forming the Pt(II) dihydride (BDI)Ga–Pt(H)$_2$ **124.** Similarly, treatment of **121** with $HSiEt_3$ results in oxidative addition of the Si–H bond to yield a Pt(II) silyl hydride **125** with the (BDI)Ga ligand still bound. This chemistry closely parallels the small-molecule activation behavior described for the Cp*Ga–Ru platform in Section 3.2.[73] It confirms that bulky Ga(I) donors beyond Cp* can likewise support bond activation at late transition metals while Ga remains in an L-type role. In these reactions Ga(I) acts as an innocent spectator ligand that nonetheless can influence the electronic environment to facilitate oxidative addition at the metal.

Other studies further highlight the distinctive behavior of (BDI)Ga in transition-metal binding. For example, systems with Mo, Ru, Rh, Ni and Cu have been reported using (BDI)Ga donors.[70, 82, 90] In most cases, the high steric demand of the (BDI)Ga framework prevents the formation of kinetically inert multi-Ga assemblies, yielding discrete heteroleptic species instead. These trends underscore the importance of both electronic and steric tuning at Ga(I) in dictating reactivity: the softer, less Lewis acidic Ga(I) center in (BDI)Ga tends to engage in straightforward substitution at a single metal site, whereas the more Lewis acidic Ga(I) in Cp*Ga can lead to cluster growth or bridging interactions that alter the course of reactions.

When halide-bearing transition metal precursors are used, (BDI)Ga can also exhibit unique bonding modes. It may either form a μ-X bridge between Ga and the metal or insert into the M–X bond. For example, reacting a gold(I) halide with (BDI)Ga can yield a species like ((BDI)Ga)Au((BDI)GaCl) **126**. In this adduct, one Ga(I) is terminally bound to Au(I) (donor–acceptor) while the other Ga is bonded via a bridging chloride. Treatment of **126** with $NaBAr^F_4$ ($BAr^F_4$ = tetra-((3,5-bis-triflourmethyl)phenyl)borate) abstracts the chloride to produce a linear two-coordinate Au complex (BDI)Ga–Au–Ga(BDI) **127** (Figure 37), featuring a Au(I) center symmetrically bonded by two neutral (BDI)Ga donors.[91] This reveals both the coordination flexibility of Ga(I) (able to support a μ-Cl bridge or a terminal dative bond) and the utility of the Ga–X functionality as a handle for further reactivity (e.g. halide abstraction or σ-bond metathesis steps that lead to new products).[91b]

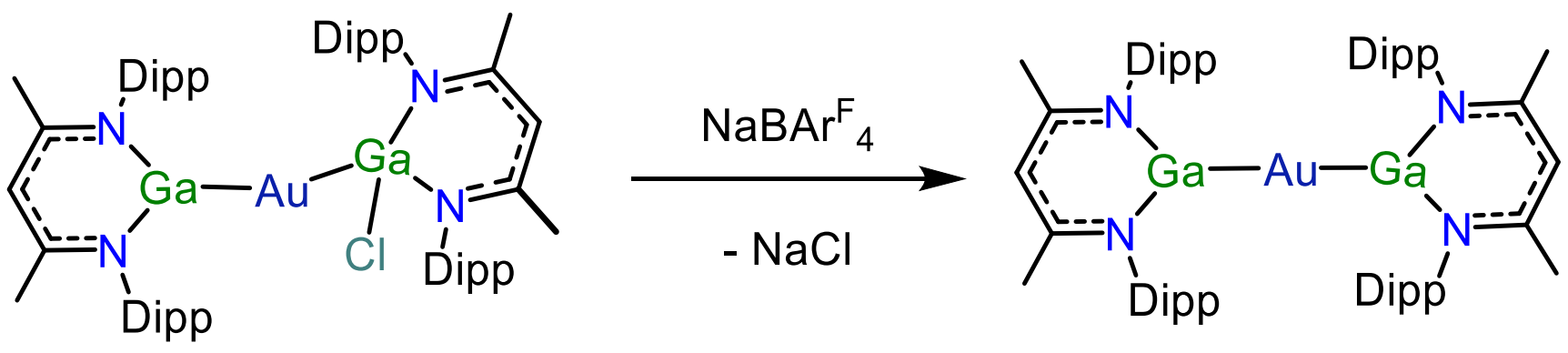

**126** **127**

**Figure 37.** Chloride abstraction from ((BDI)Ga)Au((BDI)GaCl) with NaBAr$^F_4$ to form a linear (BDI)Ga–Au–GaBDI species.

Finally, "soft metal" adducts illustrate the strength of Ga(I) σ-donation to more electrophilic partners. Recent studies by Power (on Ga–Cu **128**)[43] and Harder's group (on Ga–Zn **129**)[92] showed that electron-rich (BDI)Ga can form surprisingly strong unsupported Ga–TM bonds to soft, Lewis-acidic metal centers, Figure 38. For instance, the reaction of a Cu(I) β-diketiminate dimer with the (BDI)Ga monomer yielded the molecular complex **128** with a short Ga–Cu distance of ~2.29 Å.[43] DFT calculations revealed that roughly half of the Ga–Cu bond association energy in this system arises from London dispersion forces, but the remainder is a true covalent interaction, a testament to the potent donor ability of the Ga(I) center. In parallel, Harder's heterometallic complexes containing Ga(I) and Zn(II) showed that the Ga→Zn bond is both short and highly covalent in character.[92] In fact, the Ga–Zn bond in **129** is much more covalent than the analogous Ga–Mg bond, consistent with hard/soft considerations (Zn(II) being a softer Lewis acid matches well with the soft Ga(I) donor). Thus, the analogous Al–Zn combination (hard Al(I) donor with soft Zn(II)) proved so reactive that it could not be isolated, it spontaneously cleaved the C–F bond of a fluorobenzene solvent molecule during attempted synthesis. By contrast, the Ga–Zn adduct is thermally robust, indicating that Ga(I) hits a "softer spot" of high donor strength without tipping into uncontrolled reactivity.

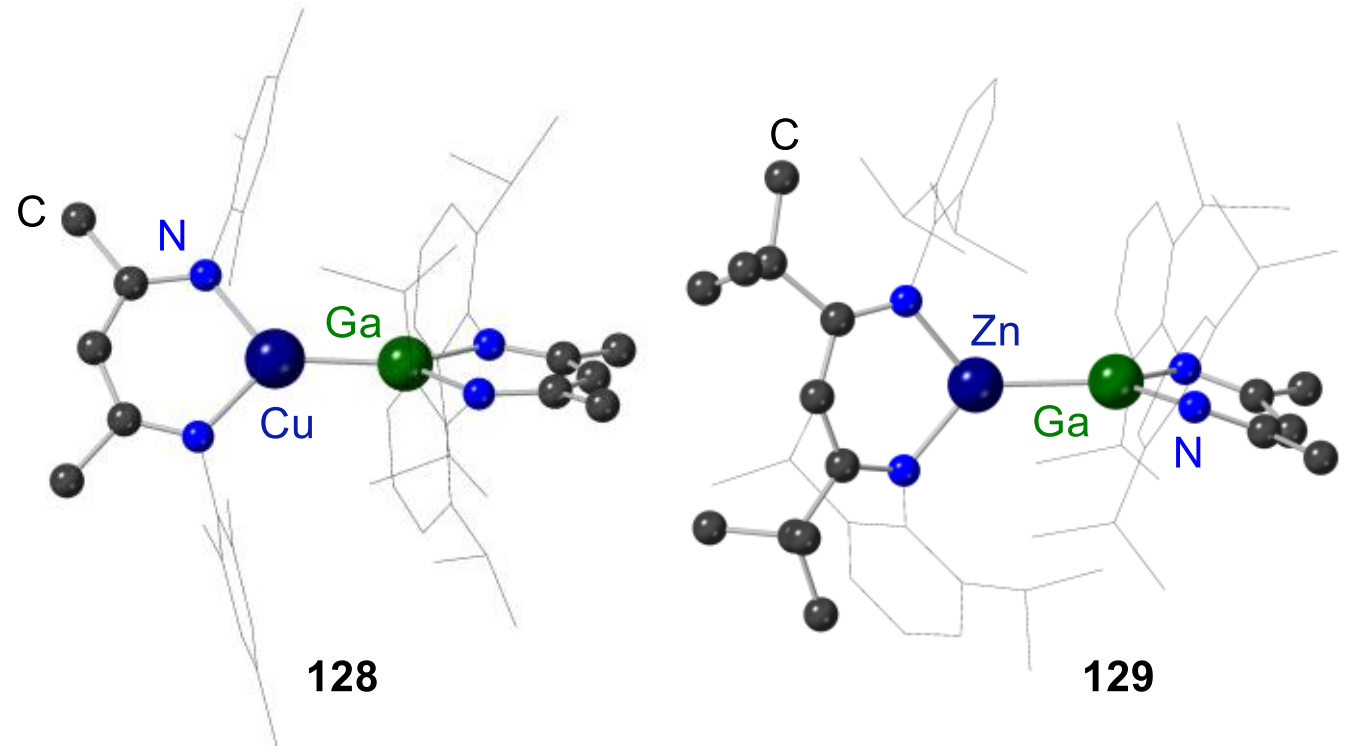

**128** **129**

**Figure 38**. Molecular structures of species **128** and **129** featuring unsupported (BDI)Ga–TM bonds.

The continued development of Ga(I) metalloligands[87, 93] thus offers a unique bridge between classical ligand chemistry and main-group reactivity. Indeed, the combination of low-valent Ga ligands with transition metals is beginning to realize the long-suspected potential for cooperative small-molecule activation. A striking recent example is the work of Hadlington, who showed that a geometrically constrained gallylene–Ni(0) complex **130** (Figure 39) can readily cleave $H_2$, yielding a formal [Ga(III)–H][Ni(II)–H] dihydride species **131**.[94] This $H_2$ splitting at the Ga–Ni interface is reversible and was leveraged in the catalytic semi-hydrogenation of alkynes, with Ga and Ni working in concert in the mechanism. Notably, the Ga(I) ligand in this system is non-innocent. It accepts a hydride (becoming Ga(III)–H) and then transfers that hydride to the alkyne substrate in the key step, while Ni delivers the other hydrogen to the alkyne. This dual-center pathway, involving an initial hydride transfer from Ga to the substrate, is unprecedented in homogeneous catalysis and underscores how Ga(I) can act as more than a static donor, instead actively participating in bond-breaking and making.

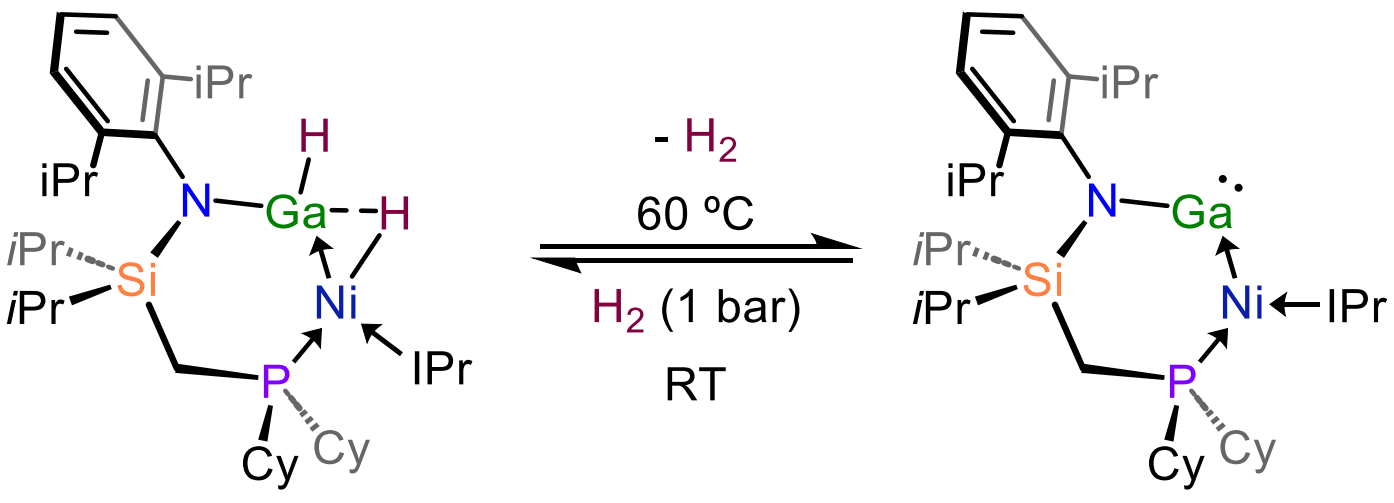

**Figure 39.** Reversible activation of $H_2$ by a Ga–Ni species.

These advancements highlight the untapped potential of Ga(I) metalloligands to serve as non-innocent partners in catalysis. By leveraging both the lone-pair donation and the accessible oxidation state changes at Ga, chemists are beginning to devise new mechanisms for small-molecule transformations that engage Ga and transition metals in truly cooperative roles. This emerging paradigm solidifies the view that new bulky Ga(I) ligands are not only effective spectators stabilizing unusual complexes but can also bring unique reactivity to the metal's doorstep, a "main-group meets transition-metal" synergy that is defining new frontiers in organometallic chemistry. As a forward look, the emergence of super-bulky aryl scaffolds that stabilize mono-coordinated Ga(I) donors with orthogonal redox behavior[62] expands the non-Cp toolbox and suggests that "designed aryl shells" will be a general entry point to innocent-by-default, non-innocent-on-demand Ga(I) ligands for cooperative TM chemistry.

In addition, a family of pincer-type Group 13 Ga–Rh complexes built from 6,6''bis(diphenylphosphino)-2,2':6',2''-terpyridine, abbreviated as $(PPh_2)_2$-terpyridine has also been reported. The Ga(I) dichloride precursor, upon reaction with 0.5 equiv of $[RhCl(coe)_2]_2$ (coe = cyclooctene), furnishes a Ga(I)→Rh adduct formulated $[(PPh_2)_2$-terpyridine)(Cl)GaRh(Cl)] **132** (Figure 40),[95] alongside the Al(I) and In(I) congeners prepared analogously. The Ga–Rh complex operates as a chemoselective hydrosilylation catalyst for nitriles to the imine oxidation level, thereby enabling a streamlined one-pot access to oximes; notably, under the reported conditions the Ga–Rh system favors oxime formation over amine reduction, highlighting how a Ga(I) L-type donor embedded in a pincer can bias multi-electron manifolds at rhodium.

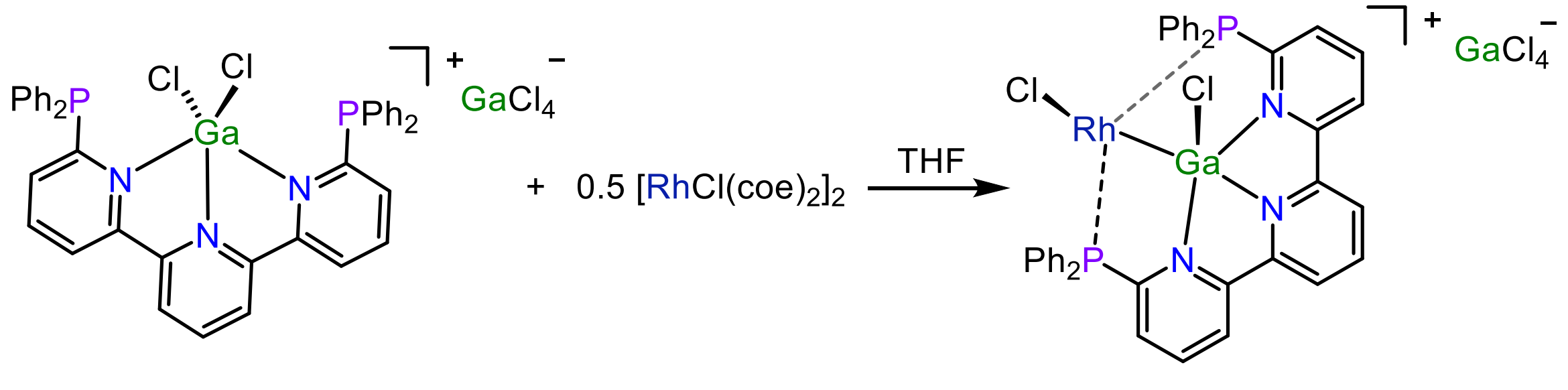

**132**

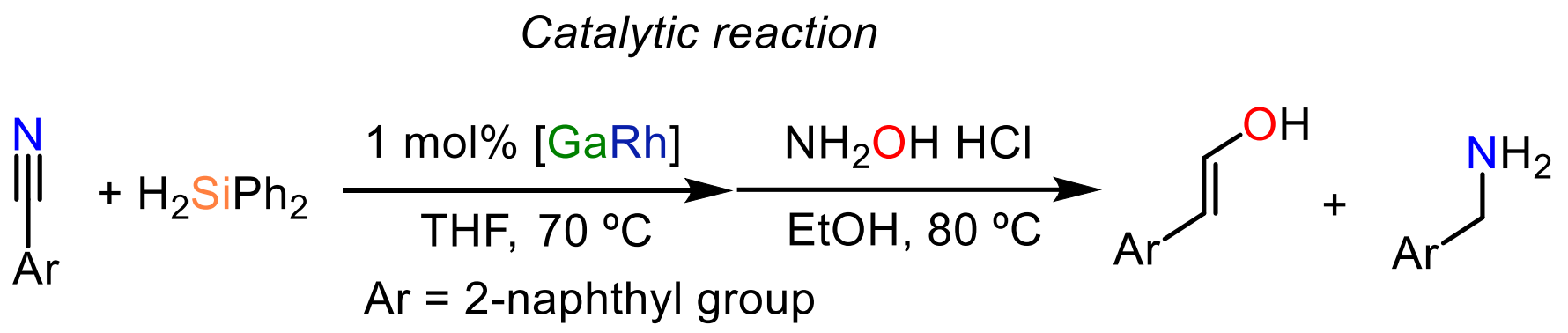

**Figure 40.** Ga–Rh system featuring a N,N,N-pincer-type ligand and catalytic reactivity.

## 4 Indium as an L-type ligand in transition-metal complexes

The chemistry of monovalent indium developed almost 30 years earlier than that of Al(I) and Ga(I). The first low-valent organoindium compound, cyclopentadienylindium(I) CpIn, was reported in 1957 by Fischer and Hofmann via thermolysis of $Cp_3In(III)$ to CpIn precursor.[96] Nearly two decades after the initial discovery, Peppe (1981) demonstrated an improved synthesis of CpIn via metathesis (InX + CpLi → CpIn + LiX).[97] Additionally, considering the advantages of Cp ligand such as the steric protection but also the fluxionality, other derivatives where prepared following a similar methodology.[98] By the mid-1980s, Beachley isolated the pentamethylcyclopentadienyl analogue, Cp*In **133**, as a thermally stable solid.[99] These cyclopentadienyl In(I) complexes provided early examples of L-type (two-electron donor) indium synthons, though in the solid state CpIn itself adopts polymeric or oligomeric structures (chains of CpIn units) due to In–In bridging interactions.[100]

Beyond Cp ligands, a variety of bulky anionic ligands have been employed to stabilize neutral In(I) centers (Figure 41). Atwood showed that the sterically encumbered neopentyl-like ligand $(Me_3Si)_3C$ can support In(I) in a cluster.[101] The compound $[((Me_3Si)_3C)In]_4$ **134** features a tetrahedral $In_4$ core, with each indium bridged by the bulky alkyl substituents. This $In_4$ cluster is robust in solution, tending to dissociate only partially to monomers, and can serve as a synthon for larger heterometallic clusters. Likewise, very bulky aryl ligands (such as *meta*-terphenyls) have enabled mono-coordinated terphenyl ArIn **135** monomers.[102] For instance, reaction of InCl with a lithium terphenyl reagent (ArLi, where Ar is a highly hindered m-terphenyl) yields a bright-orange ArIn species that remains monomeric in the solid state.[102b] This represents a rare example of a discrete base-free aryl In(I) compound, made possible by extreme steric protection of the In center. In general, using sufficiently bulky substituents (whether alkyl, aryl, or Cp-based) prevents extensive In–In aggregation and allows the isolation of monomeric or oligomeric In(I) compounds, analogous to the strategies employed for Al(I) and Ga(I).[103]

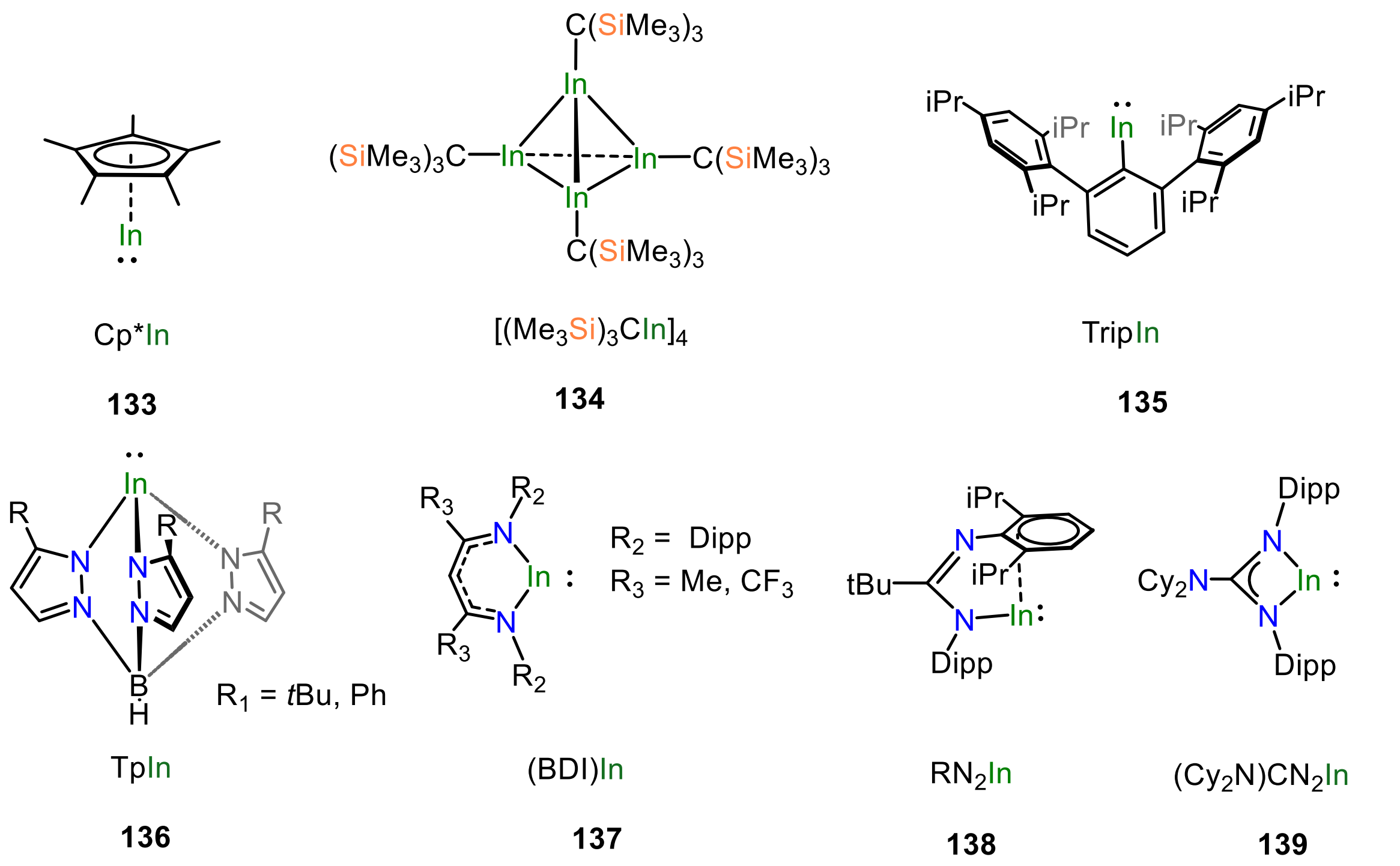

**Figure 41.** Sterically encumbered In(I) synthons/precursors.

Another successful approach to stabilize neutral In(I) species is through chelating mono-anionic ligands that mimic the coordination environment of N-heterocyclic carbenes. Hydridotris(pyrazolyl)borate (Tp) ligands can envelop the indium center and donate an electron pair from each pyrazolyl nitrogen furnishing monomeric TpIn **136** species (Figure 41).[104] The stability of such TpIn species with bulky functions underscores the effectiveness of multi-dentate ligation in preventing indium from disproportionating. Notably, some of these TpIn species **136** can undergo oxidative addition reactions without inducing In(I) disproportionation.[103] These examples highlight that (pyrazolyl)borate confer both steric bulk and a multidentate grip, yielding monovalent indium species that are stable enough to be handled and even exhibit reactivity.

Bulky β-diketiminates (NacNac ligands) represent another class of chelators that stabilize neutral monovalent In species. Inspired by the landmark work on Al(I) and Ga(I) β-diketiminate complexes,[16, 55a] Hill obtained the first two-coordinate In(I) β-diketiminato complex in 2004.[105] By a one-pot reduction of an In(III) precursor with potassium hexamethyldisilylamide $KN(SiMe_3)_2$ in the presence of the β-diketiminate ligand precursor, they isolated the neutral monomeric species (BDI)In **137** bearing a $CH(C(Me)NDipp)_2$ (Dipp = 2,6-$iPr_2C_6H_3$) ligand. This (BDI)In compound, is better described as an indium analog of a singlet carbene, featuring a two-coordinate In center with a lone pair and an accessible vacant orbital (formal oxidation state +1). In fact, its electronic structure (a filled np orbital and an empty (n+1)p orbital) mirrors the ambiphilic nature of NHC carbenes.[106] Around the same time, Jones investigated amidinate and guanidinate ligands on indium. When a bulky aryl-substituted amidinate K[ArNC(*t*Bu)NAr] (Ar = 2,6-$iPr_2C_6H_3$) was reacted with InCl, a monomeric In(I) amidinate was obtained.[107] Unexpectedly, the indium center in this complex coordinates not only to the N,N′-chelate but also to the aromatic ring of the ligand ($\eta^3$ bonding), giving a folded In–N–C aryl metallacycle **138**.[107] This mode, an N,$\eta^3$-arene coordination, can be viewed as a tethered arene adduct of a would-be four-membered $InN_2C$ ring in **138**. By employing an even bulkier guanidinate ligand (incorporating a cyclohexyl or diisopropyl substituent on the central carbon of the NCN fragment), the successful isolation of an In(I) complex **139** with true N,N′-chelating geometry (a four-membered $InN_2C$ ring with no arene interaction) was achieved.[58a, 108]

These neutral In(I) complexes **133-139** (Figure 41) represent key examples of potent two-electron donors, and they can serve as L-type ligands to transition metals in a manner analogous to NHCs or phosphines. However, in general, heavier group-13 diyls (Ga(I), In(I), Tl(I)) are somewhat less reactive for L-type ligation than the lighter Al(I) analogues, but they still exhibit remarkable reactivity patterns. Many In(I) compounds are thermally sensitive, tending to disproportionate to In(III) and indium metal if not sufficiently protected.[103] Nevertheless, when stabilized, they can engage in L-type coordination and even in bond activation chemistry.

## 4.1 Trisyl ($(Me_3Si)_3C$)In and Cp*In donor ligands

Low-valent monovalent indium species RIn can serve as 2-electron donor ligands in transition-metal complexes, acting as neutral L-type (Lewis base) ligands analogous to CO or phosphines. The nature of the substituent R strongly influences the electronic and steric properties of these ligands. In particular, bulky ligands (e.g. R = $(Me_3Si)_3C$ or Cp*) help stabilize discrete RIn units and tune their donor ability. Early studies (in parallel with Al(I) and Ga(I) analogues) showed that RIn fragments are isolobal with CO and can replace carbonyl ligands on metal centers.[109] However, indium's heavier character often necessitates more forcing conditions (e.g. heating) and leads to different outcomes (bridging modes, cluster formation) compared to lighter Group 13 analogues.

Initial explorations by Uhl and co-workers focused on reactions of the robust tetraindium(I) compound $[RIn]_4$ (R = $(Me_3Si)_3C$), a tetramer **134** (Figure 42), with transition-metal carbonyl complexes, to probe its ability to displace CO ligands.[110] Notably, **134** is more thermally stable and less readily dissociated into monomers than the Ga(I) analogue,[111] so higher temperatures were required to induce reactions. For example, heating the tetramer **134** with $Mn_2(CO)_{10}$ (which has only terminal CO ligands) in refluxing hexane led to replacement of two CO ligands by two RIn fragments in **140** (Figure 42).[110] The product is a dinuclear Mn complex wherein two $\mu_2$-RIn ligands bridge the Mn–Mn bond. This demonstrated that RIn can partially and selectively displace CO from a metal carbonyl, bonding in a bridging fashion between two metal centers A related experiment with the dimeric cobalt carbonyl $[Co(CO)_8(\mu_2\text{-}CO)]_2$, where two CO are acting as bridging ligands shows **134** substitutes one of the bridging carbonyls in **141** and an excess of **134** furnished the mixed Co–In–Co cluster **142** with doubly bridged by RIn ligands, analogous to **140**. In these reactions, In(I) displays a preference for bridging coordination, in contrast to typical terminal CO binding.[112]

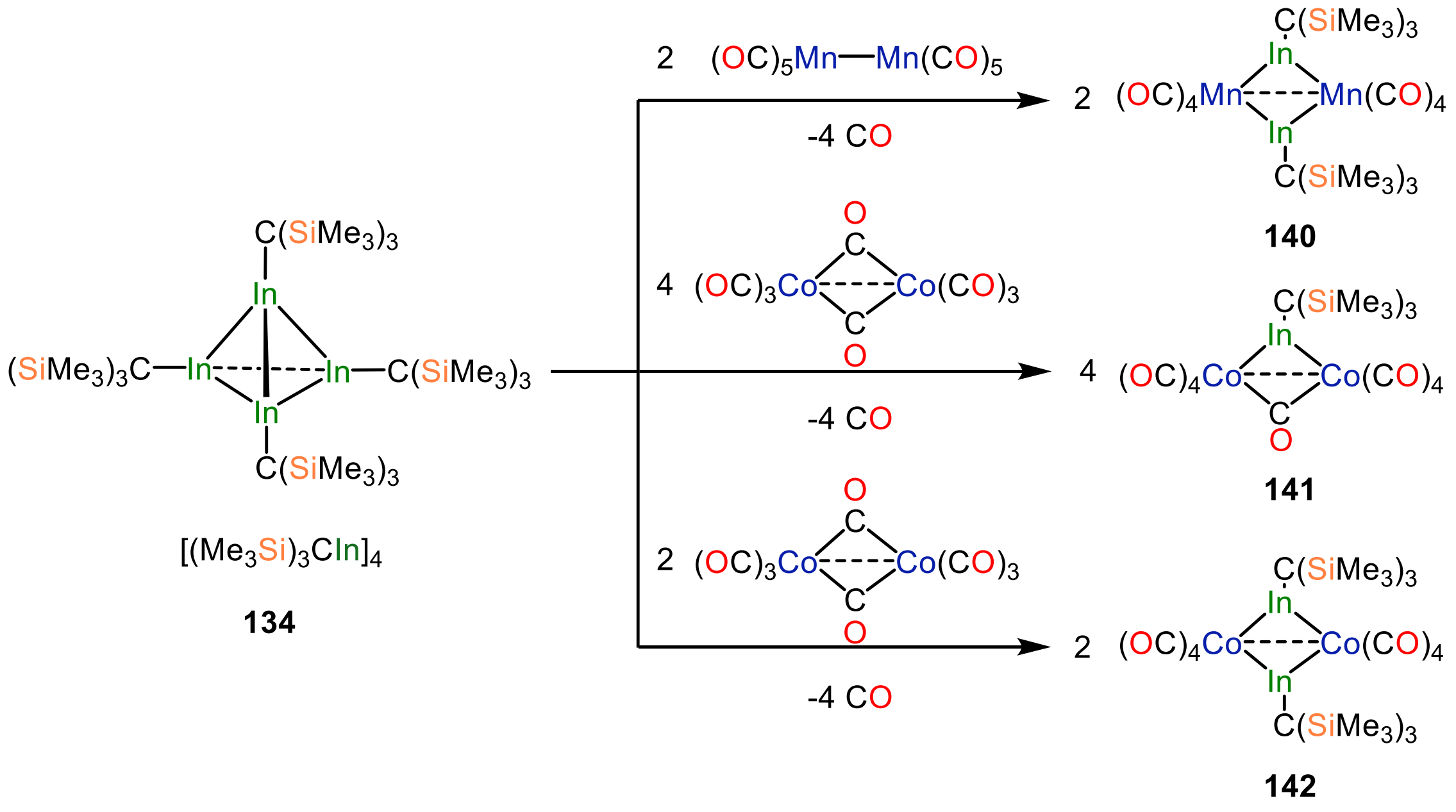

**Figure 42.** Substitution of CO ligand by $[((Me_3Si)_3C)In]_4$.

Uhl's group further tested the reactivity of the tetrameric $[RIn]_4$ compound **134** with higher nuclearity carbonyl clusters of iron (Figure 43). For example, $Fe_3(CO)_{12}$ which contains two bridging CO ligands reacted with **134** to afford a product consistent with two CO ligands being replaced by two RIn fragments. The crystallographic structure of this product showed an $Fe_3$ cluster where two edges of the $Fe_3$ triangle are each spanned by an RIn ligand and the third edge still bridged by a CO, resulting in $[Fe_3(CO)_{10}(\mu_2\text{-}InC(SiMe_3)_3)]_2$ **143**.[113] Similarly, $Fe_2(CO)_9$ gave a mixture of products upon reaction with **134**, one where the single bridging CO was replaced by an RIn ligand $Fe_2(CO)_8(\mu_2\text{-RIn})$ **144**, and another where a mixed bridging mode persisted $Fe_2(CO)_6(\mu_2\text{-}CO)(\mu_2\text{-RIn}$ **145** (Figure 43). These reactions were conveniently monitored by IR spectroscopy (shifts in ν CO bands indicated loss of bridging CO ligands) and by NMR of the $(Me_3Si)_3C$ group (the α-carbon resonance shifted in ways suggesting some π-backbonding from the metal to the indium ligand). Notably, despite multiple attempts, terminal carbonyl ligands on mononuclear complexes like $Fe(CO)_5$ could not be cleanly replaced by RIn to give a simple In–Fe bonded monocarbonyl species, the outcomes were invariably clustered species or no reaction. This parallels observations in Al(I) and Ga(I) chemistry, and highlights that bridging or cluster formation is often favored for In(I) unless special strategies are used to generate discrete monometallic complexes.

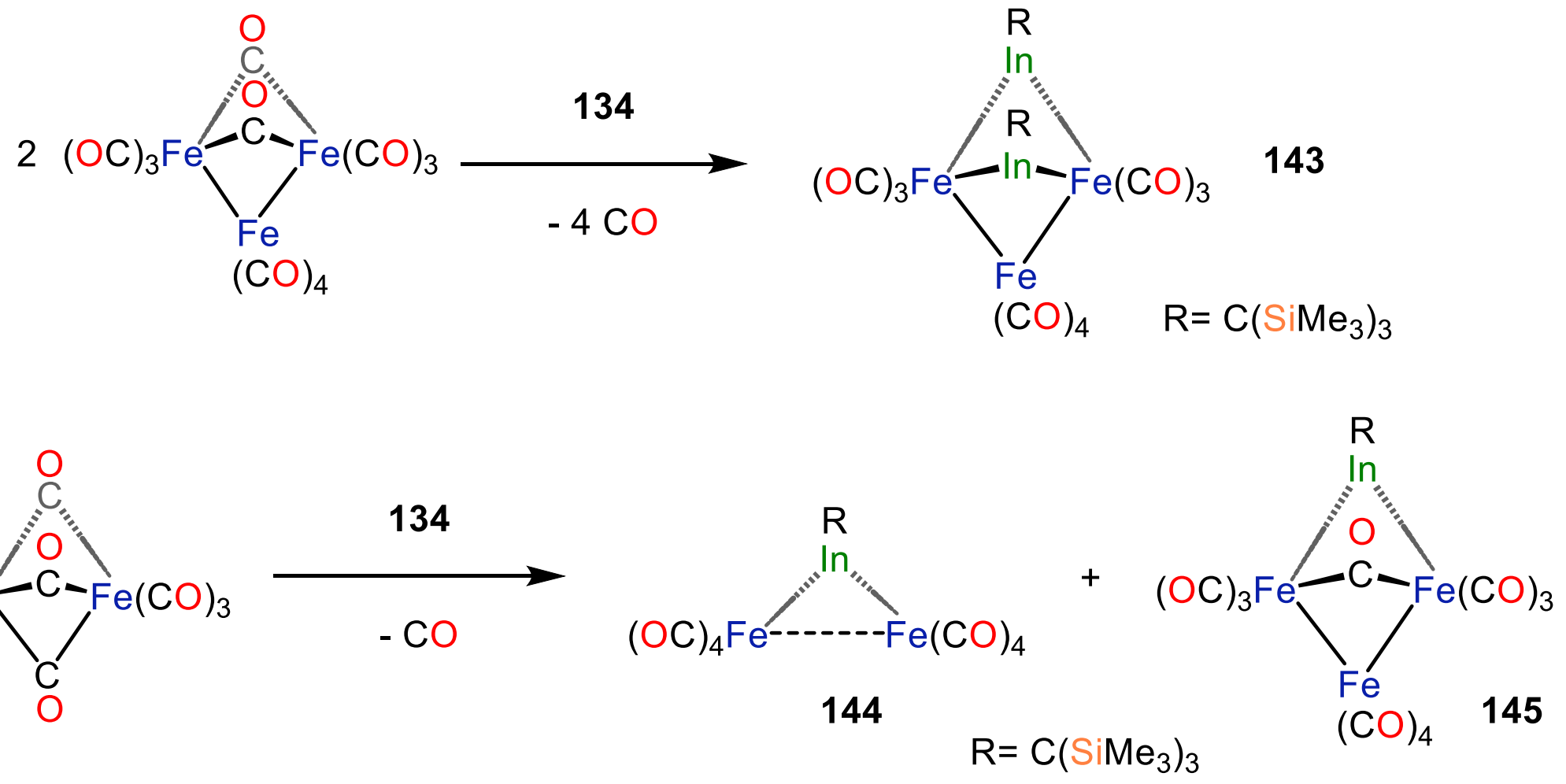

**Figure 43.** Substitution of CO bridging ligands by $[((Me_3Si)_3C)In]_4$.

One effective strategy to obtain mononuclear carbonyl analogs is to start from complexes with more labile ligands (such as olefins) instead of CO. Along these lines, $Fe(CO)_3(COT)$, where COT is cyclooctatetraene labile ligand, has been used as substate. Mixing $Fe(CO)_3(COT)$ with **134** in a 3:8 ratio (Fe:In) and heating led to replacement of the COT ligand by RIn, yielding **146** (Figure 44). The IR spectrum of **146** showed only terminal CO bands (no bridging CO signals), consistent with a mononuclear $Fe(CO)_3$ unit, meanwhile, the NMR exhibited a single set of $SiMe_3$ signals, indicating a symmetrical environment for the RIn ligand. X-ray crystallography revealed that the indium is indeed bound to the iron center and interestingly in a bridging fashion between two $Fe(CO)_3$ fragments in the solid state, suggesting an Fe–In–Fe motif, though in solution it behaves as a single Fe center with terminal CO ligands, possibly due to fluxionality or dissociation of a dimeric assembly. This result provided one of the first indications that an RIn fragment can coordinate in a terminal fashion to a single Fe metal center under the right conditions, although the solid-state structure still showed a dimeric clustering.

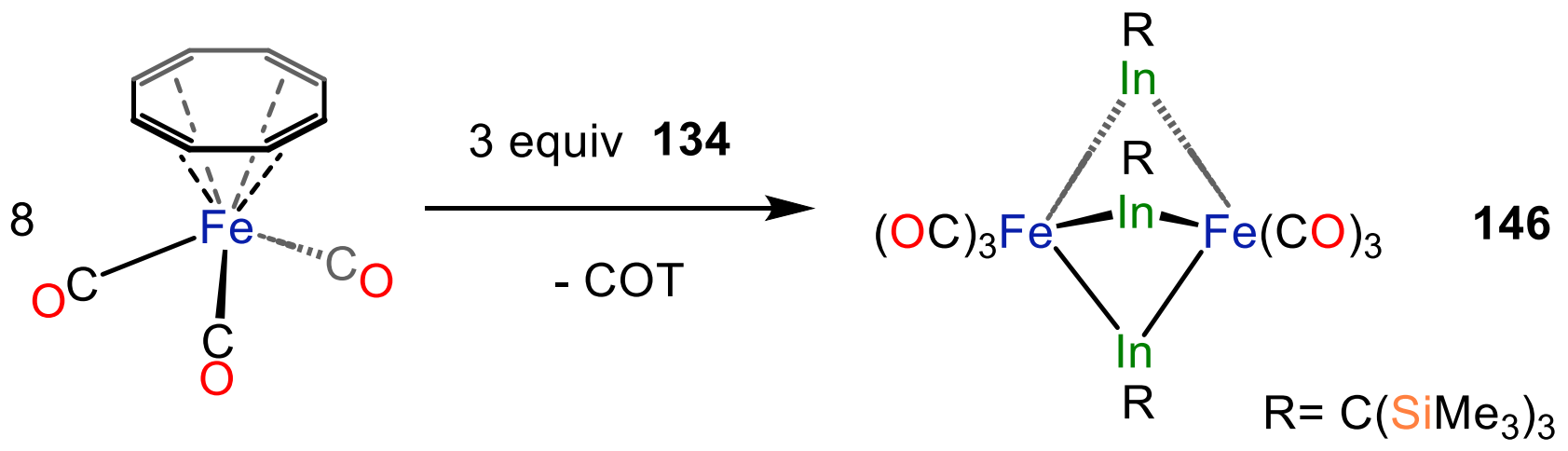

**Figure 44.** Substitution of olefin ligand by $[((Me_3Si)_3C)In]_4$.

A breakthrough was the isolation of discrete homoleptic complexes where a transition metal is coordinated solely by RIn ligands, akin to classic tetracarbonyl complexes. By choosing transition-metal precursors lacking CO and with labile ligands (like COD = 1,5-cyclooctadiene), Ni(0) and Pt(0) centers have been trapper with multiple RIn ligands. Thus, reacting bis(cyclooctadiene)nickel(0) $Ni(COD)_2$ with RIn **134** at slightly elevated temperature (~50 °C) produced the $(RIn)_4Ni$ complex **147** (Figure 45).[109] This remarkable complex features a Ni atom in a tetrahedral coordination environment defined by four RIn ligands, directly paralleling the structure of $Ni(CO)_4$. In fact, **147** is explicitly described as an organometallic In–Ni compound analogous to $Ni(CO)_4$, with the ligand RIn behaving isolobal to CO.[109] No significant In–In interactions are present in **147** (the In···In distances are long, indicating the indium atoms act as independent donors rather than clustering together). **147** is thermally stable, highlighting that RIn ligands can stabilize zero-valent metals M(0) effectively. A similar strategy to Pt(0) has also been reported. The reaction of bis(cyclooctadiene)platinum(0) $Pt(COD)_2$ with RIn **134** afforded $(RIn)_4Pt$ **148**, isostructural to $(RIn)_4Ni$ **147** (Figure 45).[111] This In–Pt compound is especially notable as an analog of $Pt(CO)_4$, a species that is thermally unstable and hypothetically exists only under matrix isolation. By contrast, $(RIn)_4Pt$ **148** is isolable under ambient conditions, underscoring the stabilizing power of the bulky L-donor RIn ligands. In addition, Fischer completed the Group 10 triad by synthesizing the palladium analogue. Using a Pd(II) precursor $(tmeda)PdMe_2$ (tmeda = tetramethylethylenediamine), that can generate a Pd(0) center *in-situ*, in the presence of RIn **134** under gentle heating. Thus, the Pd(II) complex undergoes reductive elimination of ethane, and the resulting Pd(0) is trapped by RIn ligands acting as stabilizing L-type metalloligands. This gave the palladium complex $(RIn)_4Pd$ **149**, which is the Pd counterpart to $Ni(CO)_4$.[114] With Ni, Pd, and Pt all able to support four RIn L-type metalloligands in a tetrahedral geometry, a full series of isolobal carbonyl analogs was thus established. Density functional calculations on these complexes indicated significant π-backbonding from the $d^{10}$ metal centers to the RIn L-type ligands, similar to the TM→CO back-donation in metal carbonyls.[109] This π-back-donation is evidenced experimentally by characteristic shifts in NMR and IR (for instance, the $^{13}C$ NMR resonance of the In-bound $(Me_3Si)_3C$ group shifts compared to free $[RIn]_4$ **134**, consistent with increased electron density at indium) and was further supported by computational charge analysis.

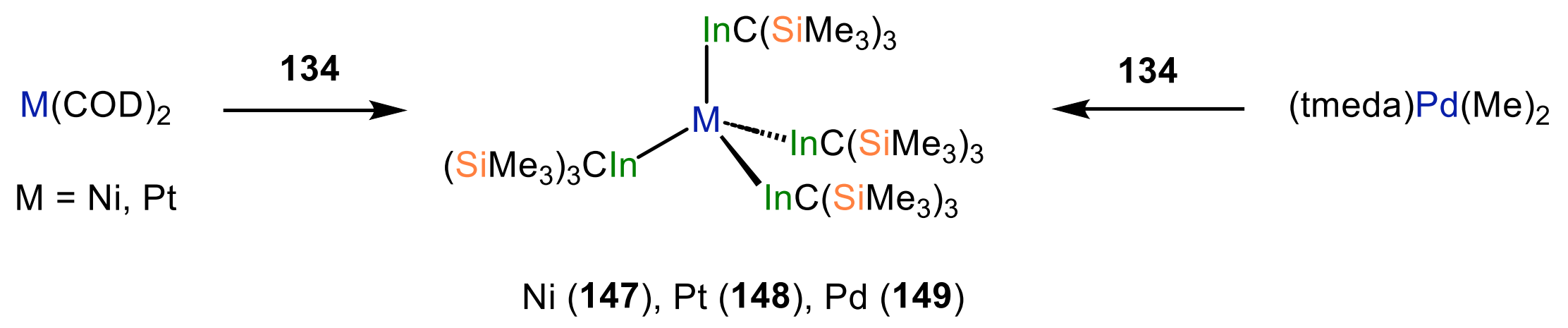

**Figure 45.** Synthesis of homoleptic complexes of Ni, Pd, and Pt with monovalent indium moieties.

After isolating these homoleptic complexes (which are structural analogs of $Ni(CO)_4$, $Pd(CO)_4$, $Pt(CO)_4$, Uhl revisited carbonyl substitution reactions to see if terminal CO ligands on dinuclear complexes could be replaced by RIn **134**. Direct substitution of terminal (non-bridging) CO ligands on metal carbonyls proved difficult, thus, no reaction was observed with simple mononuclear carbonyls. However, a notable success was achieved with the butterfly-shaped $Ni_2Cp_2(\mu\text{-}CO)_2$ complex (Cp = $C_5H_5$). This bimetallic species has two bridging carbonyls across a Ni–Ni core. Treating $Ni_2Cp_2(\mu\text{-}CO)_2$ with RIn **134** gave a stepwise substitution: initially, one RIn inserts or bridges, followed by the second (Figure 46).[115] The final product sees both bridging CO ligands replaced by two $\mu_2$-InR ligands, yielding a $Ni_2Cp_2(\mu\text{-}RIn)_2$ complex **150**, structurally isostructural to the starting carbonyl, with a "butterfly" $Ni_2Cp_2(\mu\text{-}CO)_2$ core. Interestingly, the reaction can also form an intermediate when an excess of $[RIn]_4$ **134** is used. An "insertion" of an RIn fragment into the Ni–Ni bond was observed under certain conditions. In that case, a monomeric RIn fragment cleaves the Ni–Ni bond, giving a product where an indium is coordinated to two Ni centers and carries its alkyl substituent, effectively a Ni–In–Ni bridging unit with Ni–Ni single bond broken. This oxidative addition-like insertion was seen only with indium and a high In:Ni ratio (4:1).[115] By contrast, the lighter Ga(I) analog $[RGa]_4$ did not show any Ni–Ni insertion, and only the bridging Ga substitution (analogous to the final indium product) was obtained regardless of stoichiometry. Quantum-chemical calculations confirmed that in the $Ni_2$(μ-RIn)$_2$ product there is no significant In–In bonding (and likewise no Ga–Ga bond in the Ga analogue). This reinforces that the RIn units function as true isolobal CO analogues, bonding separately to each metal center rather than aggregating or bonding to each other. The ability of In(I) to insert into metal–metal bonds observed here for Ni vs. merely bridge existing bonds (as for Ga) underscores a subtle distinction. The indium's lower tendencies to form strong In–In bonds and its relativistic effects can enable unique reaction pathways (insertion, cluster formation) not seen for Ga(I).

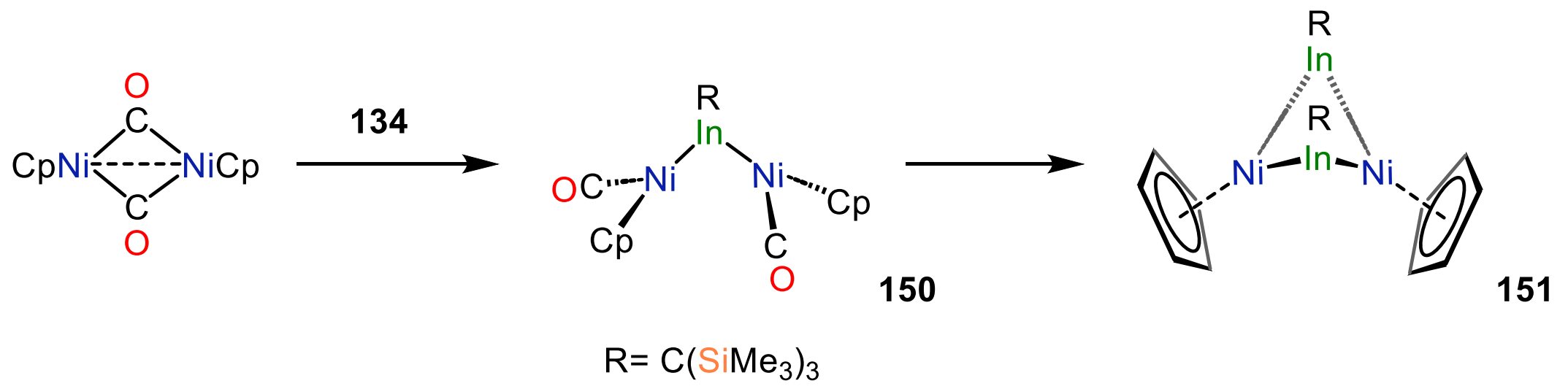

**Figure 46.** Reaction of $Ni_2$-$Cp_2$(μ-CO)$_2$ with [(($Me_3Si)_3C$)In]$_4$.

Overall, these demonstrations on RIn featuring L-type donor coordination as carbonyl analogs established that monovalent indium can coordinate to transition metals in a manner comparable to CO, especially in bridging modes on metal clusters and in forming homoleptic (RIn)$_n$TM complexes. These studies also highlighted practical challenges. For instance, In(I) reagents like (RIn)$_4$ **134** often require thermal activation to dissociate into reactive monomers, and the outcomes can be clusters rather than simple mononuclear adducts. Nevertheless, they opened the door to a new class of organometallic compounds where metals are stabilized by neutral group-13 ligands.

After Uhl's pioneering work with the bulky ($Me_3Si)_3C$ ligand, which confers exceptional thermal stability to RIn, attention turned to other In(I) species. For example,

Cp*In (R = $C_5Me_5$) can also act as an L-type ligand to transition metals.[116] Heating Cp*In (which is monomeric in the solid state) with a chromium complex bearing both CO and an olefin ligand, $Cr(CO)_5(COE)$ (COE = cyclooctene), resulted in displacement of the olefin by Cp*In. The product was a Cr(0) complex $(Cp^*In)Cr(CO)_5$ **152** (Figure 47) with In–Cr coordination and all carbonyls remaining intact. This reaction mirrors the Cp*Al and Cp*Ga chemistry, as they also had been shown to substitute olefins on $Cr(CO)_5$, filling an apparent gap in the triad. It demonstrated that even less bulky indyls like Cp*In can coordinate as a single-site ligand to a metal center, provided a sufficiently labile ligand is available for substitution.

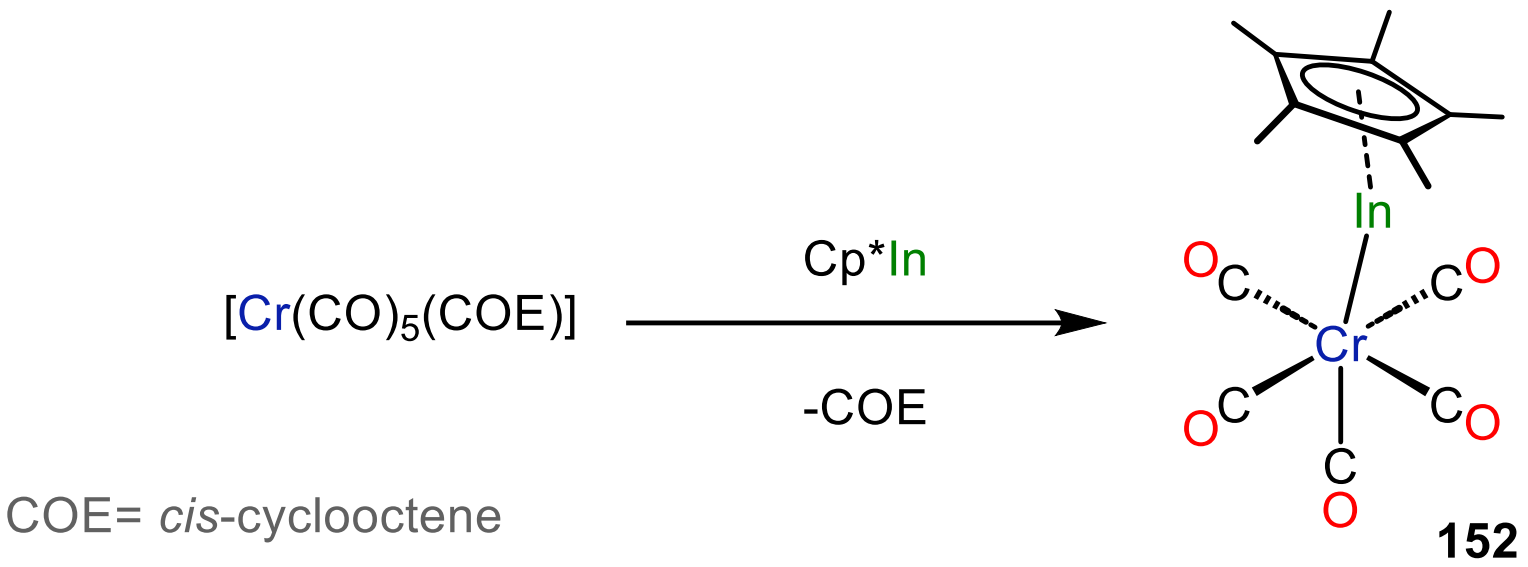

**Figure 47.** Reaction of $Cr(CO)_5(COE)$ with Cp*In.

Building on this, Fischer and coworkers expanded the chemistry to new transition-metal combinations, often comparing homologous Al, Ga, In systems. For instance, they prepared In–Pd clusters analogous to the gallium cluster **93**. Using a slightly different route, mild thermolysis of $(tmeda)PdMe_2$, in the presence of Cp*In, they obtained a triangular $Pd_3$ cluster complex formulated as $Pd_3(Cp^*In)_4(\mu_2\text{-}Cp^*In)_4$ **153** (Figure 48).[117] **153** contains four terminal Cp*In ligands, each bound to one Pd and four bridging Cp*In ligands each spanning a Pd–Pd edge, a geometry very similar to the previously reported Ga analog **93**. Interestingly, NMR studies indicate the Cp*In L-type ligands (terminal vs bridging) undergo exchange in solution, unlike the gallium version which was static on the NMR timescale. Attempts to push this Pd/In system to a mononuclear $(Cp^*In)_4Pd$ complex, by adding excess Cp*In, were unsuccessful. This highlights that indium's larger size and different kinetics can favor cluster formation over monomeric products in Cp*In–TM systems. Furthermore, by varying reaction conditions and the starting Pd precursors, one can isolate different In–Pd cluster nuclearities that are not interconvertible, for instance the $In_3Pd_3$ **154** and **155** clusters (Figure 48).[34] This suggests that kinetic factors, such as how quickly In inserts or how the cluster assembles, strongly influence the final products. Once formed, these clusters do not simply exchange indium ligands or convert into each other upon adding more Cp*In ligand. Such observations underscore that the assembly of heterometallic clusters with Cp*In is a controlled but complex process.

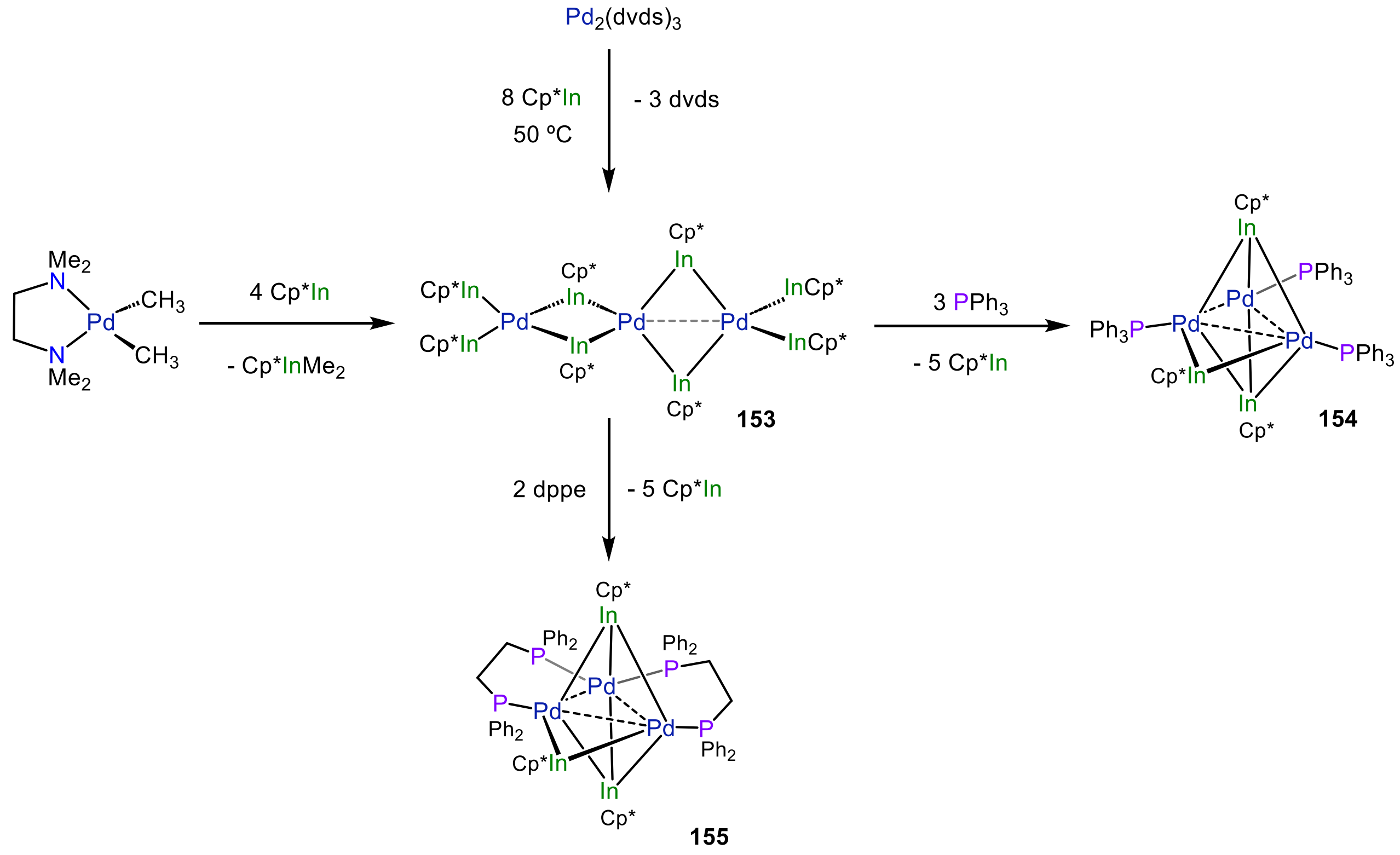

**Figure 48.** Preparation of In–Pd clusters.

Bimetallic complexes with single In–M bonds using are also known. One illustrative example is the In–Pt complex **156** (Figure 49),[118] which is directly analogous to the Al complex **16** discussed earlier.[25] The low-coordinate platinum(0) fragment can be generated by reductive elimination from a platinum(II) hydride precursor (dcpe)Pt(H)($CH_2$tBu) (dcpe= bis(dicyclohexyl-phosphino)ethane). The transient 14-electron Pt(0) species, bearing the bulky dcpe ligand, was trapped by Cp*In to form a stable In–Pt bonded complex $(Cp^*In)_2$Pt(dcpe), in analogy to the Al–Pt and Ga–Pt complexes with two Cp*E ligands.[25] This route provided a straightforward entry to a base-stabilized In–Pt complex, avoiding the need for CO ligands. Notably, computational analysis of this series, Al–Pt, Ga–Pt, In–Pt with dcpe supporting ligand, revealed that the nature of the M–Pt bond is influenced not only by the Group 13 fragment but also significantly by the ancillary ligands on Pt. The spectator ligands and overall (dcpe)Pt metal fragment modulate the polarity and donor–acceptor characteristics of the M–Pt bond. In general, In–Pt bonds were found to be somewhat less polarized than Al–Pt, reflecting indium's lower electronegativity and the relativistic expansion of its orbitals, which can accept electron density (π-backbonding) more readily.

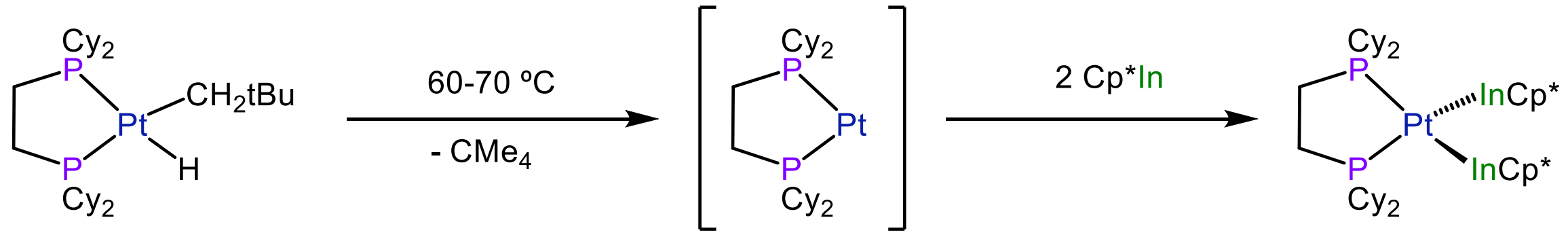

**156**

**Figure 49.** Preparation of In–Pt complex.

A particularly intriguing set of reactions was reported with half-sandwich Rh(III) and Ru(II) complexes, where the Cp*In inserts into metal–halide bonds. The reaction of the chloride-bridged dimer $[Cp^*RhCl_2]_2$ with either $[RIn]_4$ **134** and Cp*In **133** undergoes the insertion of of In(I) into the Rh–Cl bonds, yielding octahedral complexes of the type $(RIn)_3RhCl_2$ (**157**, R = Cp*; **158**, R = $(Me_3Si)_3C$, Figure 50).[119] Structurally, these compounds feature a single Rh center bonded to three In L-type ligands and two chloride ligands. One chloride remains bound to Rh, while the other is attached to one of the indium centers. The arrangement can be viewed as a “trapped intermediate” of a complete halide transfer from Rh to In. In essence, it’s as if an In(I) inserted into Rh–Cl, forming In–Rh and In–Cl bonds, but the resulting In(III) chloride is still coordinated to Rh. By varying the stoichiometry of Cp*In, different products could be isolated. With a 3:1 ratio of In:Rh, one obtains the insertion product above. Using only ~1 equivalent of Cp*In led to reduction of Rh(III) to a Rh(II) complex $[Cp^*RhCl]_2$ **159** with no indium bound.[120] **159** is essentially an electron transfer reaction where In(I) was oxidized, likely to In(III) chloride, and Rh(III) was reduced to Rh(II). On the other hand, an intermediate amount of indium (e.g. 3 eq on a dimer, giving 1.5 per Rh) yielded a mixed-valence product **160**.[120] In this complex, the anion contains an $In_3Rh$ unit in which one Cp* ligand actually bridges between an In and Rh (μ-Cp*), and an $In_2Cl_4$ unit suggests two $In(III)Cl_2$ moieties joined by a Cp* ligand. The rich chemistry observed with the rhodium system emphasizes how In(I) can induce oxidative addition and bridging outcomes, complicating the simple view of “RIn as just a L-donor”.

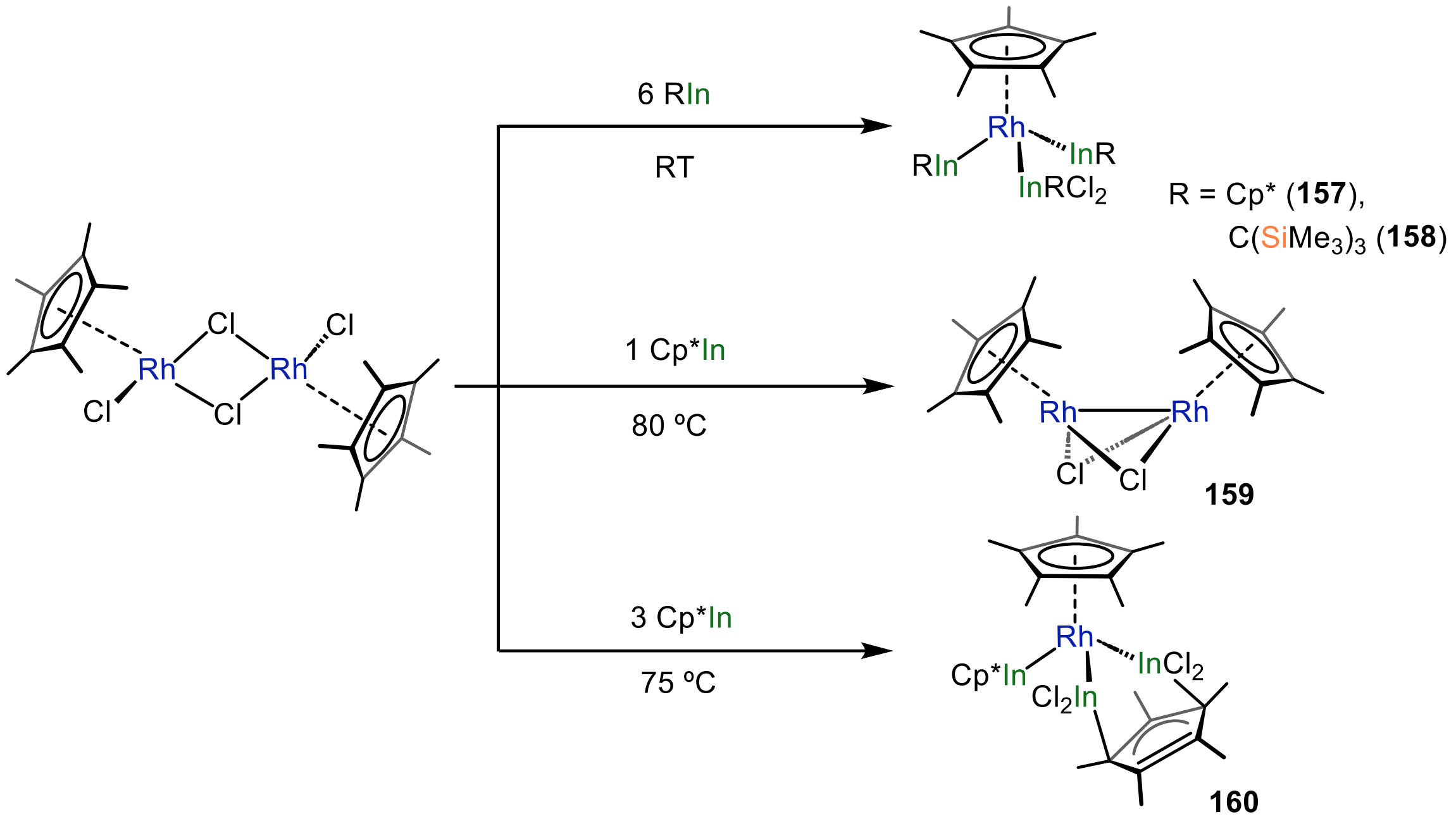

**Figure 50.** Preparation of In–Rh complexes.

The closely related ruthenium system $[Cp^*RuCl]_4$ behaved analogously. Reaction with $[RIn]_4$ or Cp*In gave the $(RIn)_3Ru(Cp^*)Cl$ species **161** and **162** (R = $C(SiMe_3)_3$ and Cp*, Figure 51).[120] These are "piano-stool" Ru(II) complexes with three In L-type ligands and one chloride completing the coordination sphere. In the solid state, some μ-Cl bridging between Ru and In is observed, whereas in solution the complex appears symmetric. The Rh and Ru results illustrate that In(I) can participate in insertion (oxidative addition) chemistry, blurring the line between a simple 2-electron donor ligand and a reactive metalloid that can form multicenter bonds. By choosing a more Lewis-basic metal center, like Cp*Rh or Cp*Ru fragments, it is possible to stabilize intermediate species that would otherwise break down. For instance, an In(III) byproduct coordinating to the metal instead of dissociating). This approach provided deeper insight into the "classic" insertion reaction mechanism for M(I) species into M–X bonds, showing it can stall at the half-transferred state when conditions are right.

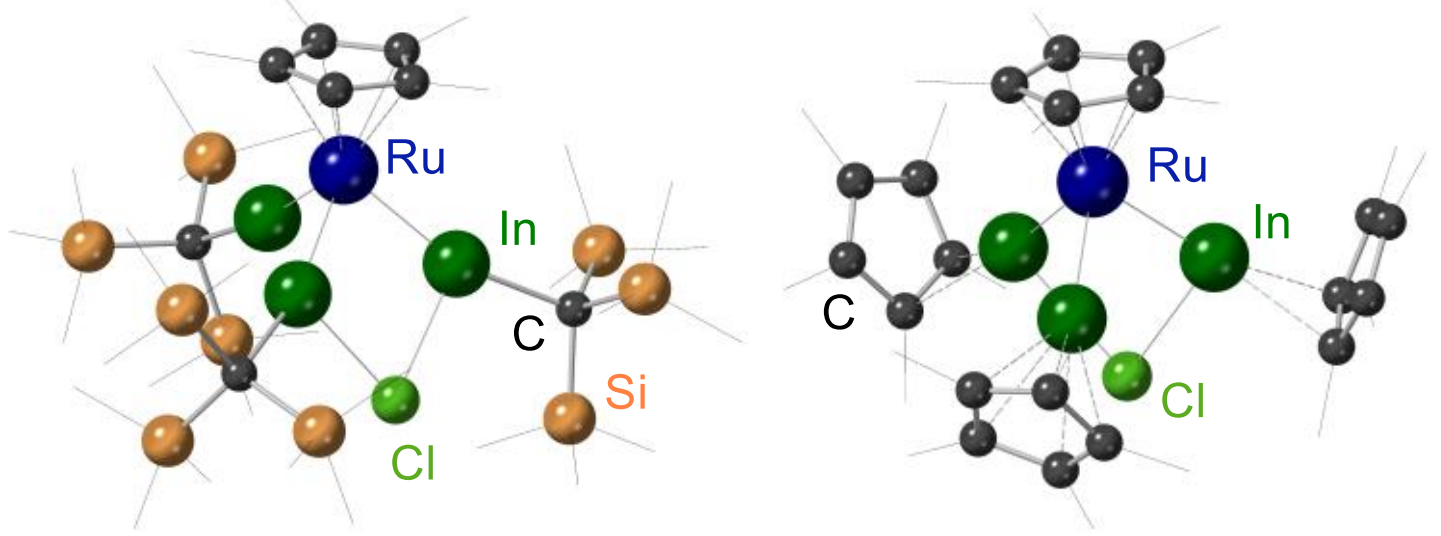

161 162

**Figure 51.** Molecular structures of In–Ru **161** and **162**.

## 4.2 Bulky In(I) systems: Design rules, bridging preference and photophysical leverage

While most of the examples featuring In–TM metals largely relied on the specific R = $C(SiMe_3)_3$ or Cp* ligands, a broader range of bulky aryl and chelate-supported In(I) compounds have been employed to coordinate transition metals. Monomeric In(I) species are inherently unstable. In absence of steric protection or stabilization, they tend to disproportionate (2 In(I) → In(0) + In(III)). Nevertheless, synthetic chemists have developed robust In(I) reagents with bulky or donor-functionalized ligands that resist disproportionation and act as neutral two-electron donors. These have enabled a variety of novel In–transition metal complexes to be isolated and studied in depth.

Robinson demonstrated one of the first examples of In–metal bonds supported by extremely bulky aryl ligands. They prepared heteronuclear complexes of the form $Cp_2M(ER)_2$ (**163** M = Ti, **164** Zr, **165** Hf), containing In–TM–In trimetallic linkages (Figure 52).[121] The complexes **163**–**165** were prepared by sodium or magnesium metal reduction reaction of $Cp_2MCl_2$ (M = Ti, Zr and Hf) with the reagent ($C_6H_3$-2,6-$Trip_2$)$InCl_2$ (Trip = $C_6H_2$-2,4,6-$iPr_3$). Each In(I) bridges between the group 4 metal and an R group, so structurally, they resemble $Cp_2M$ with two RIn units bound, one on each side. The syntheses were accomplished by reducing $Cp_2MCl_2$ with sodium or magnesium in the presence of the In(III) precursor (Trip)$InCl_2$. During the reduction, In–M bonds form with concomitant formation of NaCl/$MgCl_2$. This strategy, alkaline metal reduction in the presence of $RInCl_2$, provided access to the first group 4 complexes containing direct In–M bonds. Notably, DFT studies on **163** and **164** revealed that the In–Ti and In–Zr bonds exhibit π-backbonding from the $d^2$ metal centers into the empty p-orbitals of the indium ligands. In other words, the indium in these complexes is not a simple σ-donor; it also accepts electron density from the metal, analogous to CO ligands. This back-donation is facilitated by the relatively low energy acceptor orbitals on indium and is evidence that In(I) can function as a π-acid to some extent. The presence of M→In π bonding was reflected in the structural data (short In–M distances) and was supported by computational orbital analysis (populating indium p-orbitals).

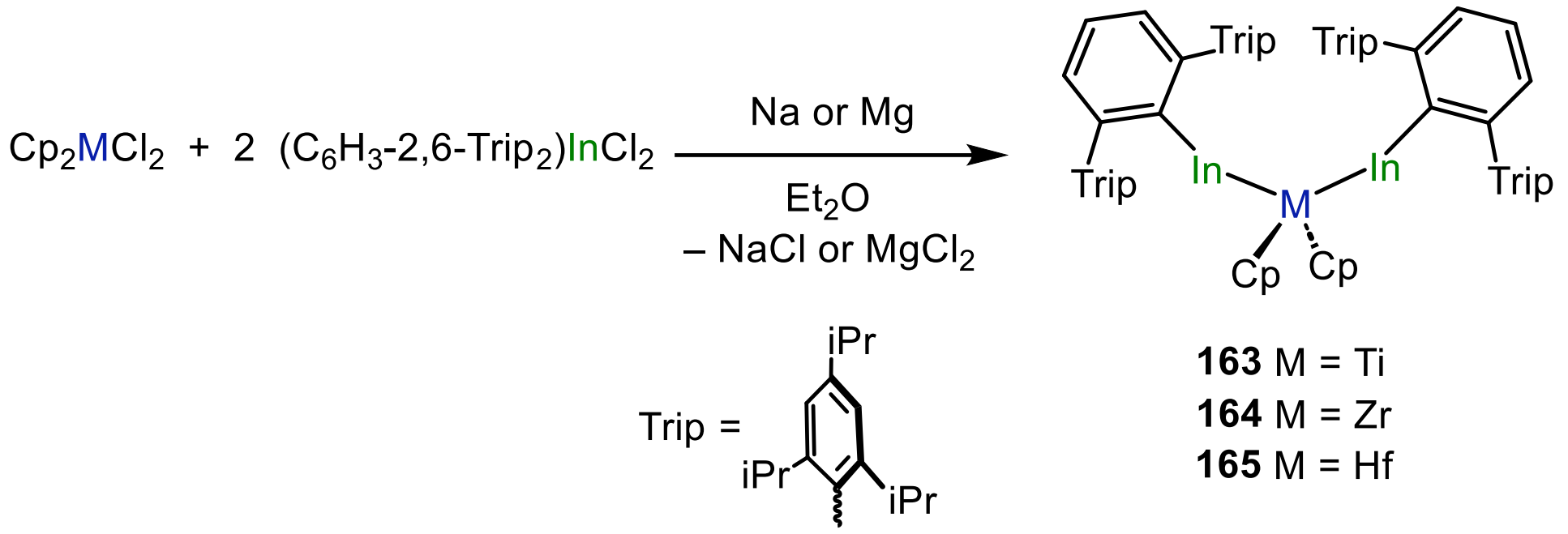

**Figure 52.** Synthesis of heteronuclear organometallic compounds containing In–TM group-4 bonds.

A variety of other transition metals across the periodic table have since been bonded to In(I) ligands, particularly using bulky aryl or chelating ligand frameworks. Power and co-workers isolated a neutral In–Mn complex using an extremely bulky o-terphenyl indium ligand. The indium reagent in this case was the diaryl-In(I) ($C_6H_3$-2,6-$Trip_2$) which is a monomeric ArIn species stabilized by the large terphenyl substituent. Reaction of this In(I), essentially a terphenyl-indium(I) “indyl”, with the complex (Cp)Mn(CO)$_2$Mn(THF), a CpMn(I) dicarbonyl with a THF ligand, resulted in THF loss and formation of the adduct allowed the formation of the product ($C_6H_3$-2,6-Trip$_2$)InMn(CO$_2$)(Cp) **166** (Figure 53).[102b] In this complex, the indium is bound to the Mn center, likely in a near-linear Mn–In–C(Ar) arrangement. Structurally, it resembles a manganese carbonyl with an In(I) ligand occupying a site, after THF departure. Notably, the In–Mn distance is short and the Mn–In–C(terphenyl) angle is almost linear, suggesting a strong donor bond from In to Mn with minimal bending, consistent with ArIn behaving like a pseudohalide or CO-like ligand. The isolation of this complex provided direct evidence that very bulky aryl In(I) compounds can coordinate to mid-row transition metals in low oxidation states, expanding the scope beyond the carbonyl substitution chemistry.

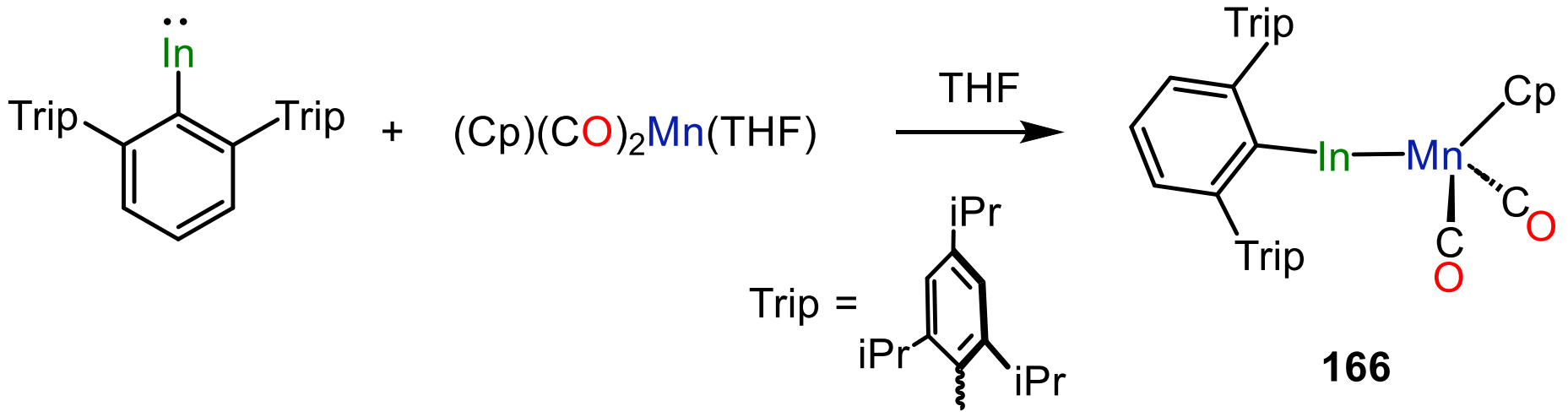

**Figure 53.** Synthesis of In–Mn complex based on bulky ligand [($C_6H_3$-2,6-Trip$_2$).

Reger reported that In(I) supported by a tris(pyrazolyl)borate scorpionate ligand could coordinate to both Fe(0) and W(0) carbonyl anions. Starting from the indium(III) dichloride (HB(3,5-Me$_2$pz)$_3$)InCl$_2$ (pz =pyrazolyl group), an In(I) species was generated *in situ* and reacted it with $K_2$[W(CO)$_5$] and Na$_2$Fe(CO)$_4$.[122] This yielded the complexes (HB(3,5-Me$_2$pz)$_3$)InW(CO)$_5$ **167** and (HB(3,5-Me$_2$pz)$_3$)InFe(CO)$_4$ **168** (pz = pyrazolyl group), respectively (Figure 54). In these molecules, often written as TpIn–M(CO)$_n$ (Tp = hydrotris(3,5-dimethylpyrazolyl)borate), In(I) donates to the electron-rich carbonyl metalate fragments. X-ray structures confirmed a single In–M bond with the supporting

Tp* ligand remaining bound to In(I). Indium is formally in +1 oxidation, bonded to the borate and the transition metal. The In–M distances were notably short, and the In–M–CO geometries indicate In is a terminal ligand on the metal, similar to a phosphine or CO. These represent a "heterometallic linkage" where a dative In→M bond is present. Interestingly, the In–Fe and In–W bond lengths were quite small, indicating strong interaction, and the authors described these as a new class of intermetallic complexes featuring short metal–metal contacts. The ability to form such adducts stems from the stabilizing influence of the polydentate Tp* ligand, which keeps indium monomeric and electron-rich.

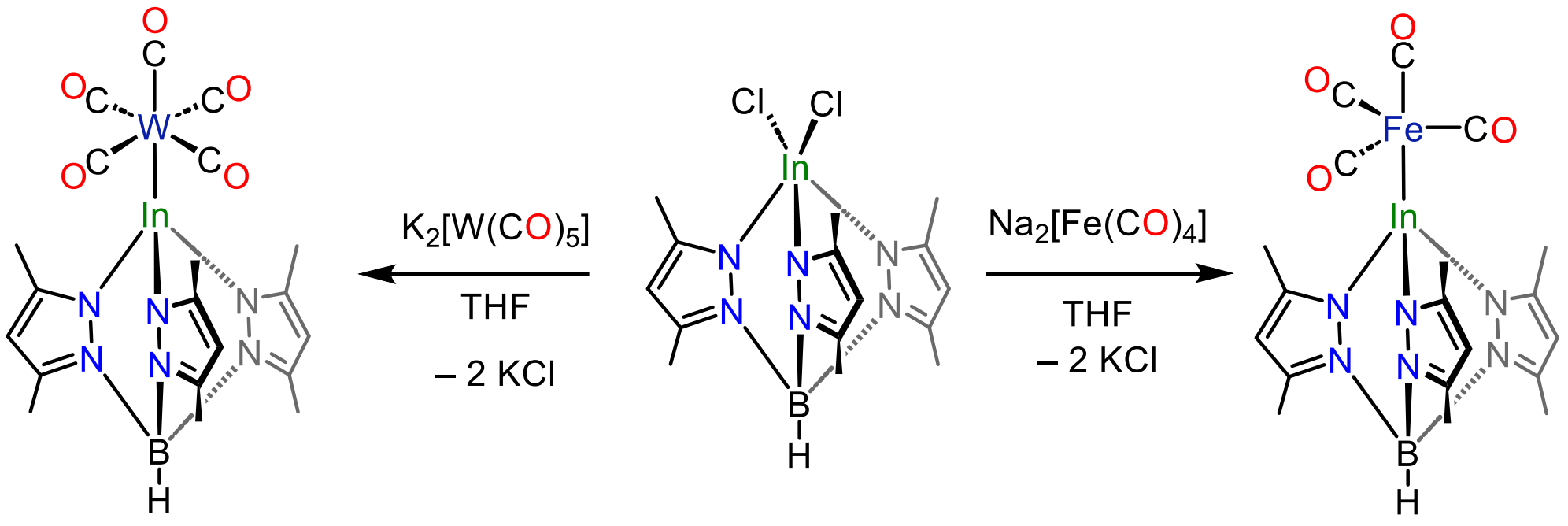

**Figure 54.** Synthesis and solid-state structures of $(HB(3,5\text{-}Me_2pz)_3)InW(CO)_5$ and $(HB(3,5\text{-}Me_2pz)_3)InFe(CO)_4$.

Hill synthesized In(I) analogs of N-heterocyclic carbenes, using β-diketiminate (BDI) ligands, (BDI)In. Treatment of such (BDI)In species with a transition metal halide can result in oxidative addition of the In(I) into the M–X bond.[123] For example, reacting (BDI)In with the 18-$e^-$ Fe(II) complex $CpFe(CO)_2I$ led to insertion of In into the Fe–I bond, yielding the $(BDI)InFe(I)(Cp)(CO)_2$ species **169** ($^{Dipp}$BDI) and **170** ($^{Mes}$BDI) (Figure 55).[123] In these complexes, the indium is formally oxidized to In(III), now bound to the BDI as well as an iodide, and the iron is reduced to Fe(I). So, it's an oxidative addition outcome, In(I) + Fe(II)–I → In(III)–Fe(I). The formulation (BDI)InFe(I)(Cp)$(CO)_2$ species **169** and **170** feature an In–Fe single bond and an Fe–I bond. One can also view it as an In(III) coordinated by BDI, Cp, two CO, and an iodide bridging to Fe(I). While these are not simple In(I)→Fe(0) donor complexes, since redox occurred, they demonstrate the rich reactivity of (BDI)In species as they can act as both donor and acceptor, inserting into polar bonds. Subsequent reactivity tests on **169** and **170** showed that abstracting the iodide with $NaBAr^F_4$ led to complicated mixtures, indicating that without the iodide bridge the In–Fe bond can be unstable or undergo further reactions. Nonetheless, these demonstrations are pivotal in showing that In(I) compounds can perform oxidative addition analogously to how a Pd(0) or Ni(0) might, expanding the notion of "In(I) as a L-type ligand" to include bond activation processes.

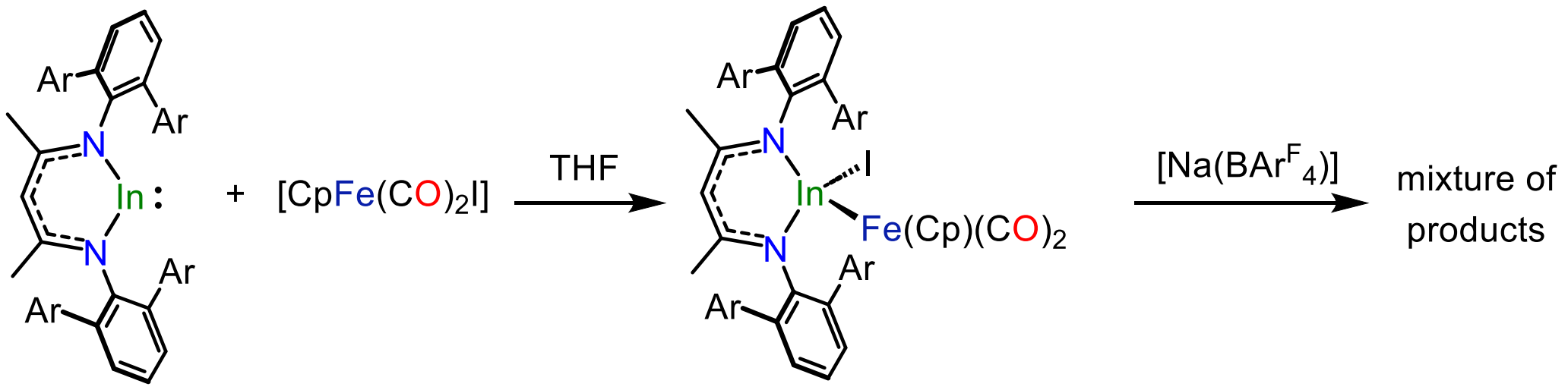

**169** Ar = Dipp
**170** Ar = Mes

**Figure 55.** β-diketiminate scaffolds for the stabilization of In–Fe complexes.

Futhermore, Jones extensively investigated four-membered ring NHC-analog ligands on indium, notably the ligand "Giso" = $[N(Ar)]_2CN(C_6H_{11})_2)^-$, an anionic guanidinate-type ligand with Ar = $C_6H_3$-2,6-i$Pr_2$. The species (Giso)In, a cyclic alkyl amino carbene analog of In(I), is a monomeric, highly nucleophilic In(I) center. (Giso)In reacts with various transition-metal complexes to form novel adducts. With $Ru(CO)_2(PPh_3)_3$ it yields the (Giso)InRu$(CO_2)(PPh_3)_2$ species **171** (Figure 56 A).[124] Here, a (Giso)In unit coordinates to a Ru(0) center that still holds two CO and two $PPh_3$ ligands. Structurally, it can be viewed as $Ru(PPh_3)_2(CO)_2$ with an added L-type indium ligand (Giso)In. This was one of the first examples of a mononuclear $d^8$ metal, Ru(0), bound by an NHC-supported In(I). The In–Ru bond length was indicative of a single bond, and spectroscopic data (e.g. IR shifts of CO) suggested some degree of Ru→In back-donation. With a Pt(0) complex bearing a chelating diphosphine, $Pt(dppe)(C_2H_4)$, (Giso)In it gives a bimetallic $((Giso)In)_2Pt(dppe)$ species **172** (Figure 56 B).[125] In this structure, two (Giso)In units coordinate to a Pt(dppe) fragment. The geometry at Pt is approximately square-planar, as expected for a Pt(II) with two donor ligands, but note Pt formally is still zero but 14-$e^-$ if one thinks of In(I) as 2$e^-$ donors filling sites like ethylene and a vacant site. The two indium ligands are trans to each other, forming a $In_2Pt$ core. X-ray diffraction confirmed a In–Pt bond distance consistent with a dative bond. No significant In–In interaction was noted, with the two In atoms well separated. **172** parallels earlier (BDI)Al and (BDI)Ga coordinated to Pt(dppe), showing that In(I) can likewise form stable adducts with Pt(0) species.

Furthermore, by using organometallic Pt(II) precursors, Jones showed that the number of indium ligands attached to Pt can be tuned. The reaction of (Giso)In with bis(aryl)Pt(II) complexes $(C_6F_4R)_2Pt$ (R = H or OMe) yielded In–Pt aggregates whose nuclearity depended on the stoichiometry. With roughly 2 equivalents of (Giso)In per Pt, a dinuclear species $((Giso)In)_2Pt(C_6F_4R)_2$ **173** (R = H, OMe) was formed, whereas with 3 equivalents, a trinuclear indium array $((Giso)In)_3Pt(C_6F_4R)_2$ **174** (R = H, OMe) could be isolated (Figure 56C).[88a] These complexes feature one Pt center bound by two aryl groups and either two or three (Giso)In ligands. In **173**, Pt is likely 4-coordinate (two aryl anions, two In(I) donors), while in **174** the Pt might be 5-coordinate with an additional indium interacting (or one indium might be bridging between two others). Such structures highlight that indium ligands can oligomerize around a metal center, especially if the metal has available coordination sites and if no strong co-ligands (like phosphines or CO) are present to cap the coordination number. The ability to accommodate 3 In(I) L-type ligands on Pt(II) to form $In_3Pt$ cores is striking and speaks to In(I)'s softer Lewis basic character. Multiple indiums can coordinate without overwhelming the metal's electron count in the way multiple CO or phosphines might.

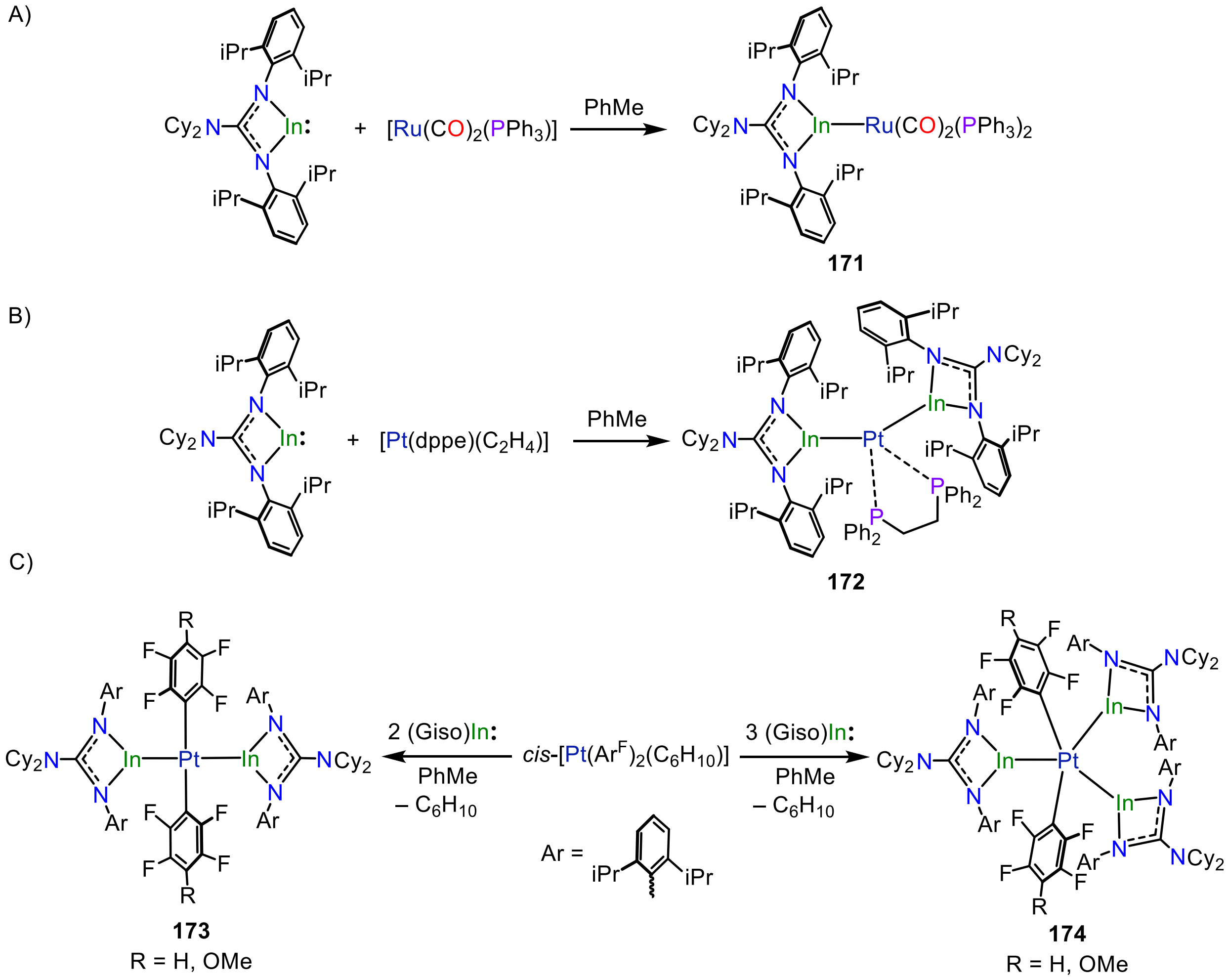

**Figure 56.** Formation of In–Ru and In–Pt complexes based on four-membered ring N-heterocyclic carbene analogue.

A recent example by Takaya and colleagues involves a terpyridine-diphosphine pincer ligand supporting an In–Rh bond. The ligand 6,6''-bis(di-phenylphosphino)-2,2':6',2''-terpyridine was used to stabilize indium in a dichloride complex. When this indium(III) dichloride (bearing the pincer) was reacted with 0.5 equiv of the Rh(I) dimer $[RhCl(COE)_2]_2$ (COE = cyclooctene), they isolated a bimetallic complex $(PPh_2)_2$-terpyridine)(Cl)InRh(Cl) **175** (Figure 57A).[95] In **175**, the indium, bound by the tridentate terpyridine diphosphine and one chloride, is directly bonded to a RhCl fragment. Essentially, one chloride bridge from Rh to In has formed, giving a structure that can be described as In(III)–Rh(I) with a single metal–metal bond and each metal carrying a chloride. The rigid pincer enforces a particular geometry, bringing In and Rh in close proximity. Structurally, the In–Rh bond length and geometry were clearly delineated in the X-ray crystal structure. This complex is a rare example where a pincer ligand links a Group 9 and Group 13 metal in one molecule, showcasing that with appropriate supporting ligands, even fairly electrophilic metal centers like Rh(I) can be coordinated by an In(I) fragment. Here the In is formally +3, but upon bonding to Rh it likely shares electron density, acting like an In(I) donor. The choice of a chelating bis-phosphine terpyridine ensured that the indium center remained inert to disproportionation and available to bind Rh.

Complementary to In–TM adducts, Krossing recently disclosed a discrete In→Ag dication formulated as $[(phen)_2In–Ag(\eta^3-C_6H_5F)]^{2+}$ **176** (Figure 57A, counter-anion $[Al(OC(CF_3)_3)_4]^-$).[126] In this species the tetragonally pyramidal $[(phen)_2In]^+$ fragment acts as a neutral L-type donor via its 5s lone pair, while $Ag^+$ is stabilized by a π-bound fluorobenzene ($\eta^3-C_6H_5F$) within the weakly coordinating environment of the anion $[Al(OC(CF_3)_3)_4]^-$. The short In→Ag contact and near-linear N–In→Ag vector confirm a genuine indium-centered donation to a late-metal cation. Conceptually, **176** demonstrates that pre-organised N,N-chelation at In(I) can out-compete disproportionation and deliver a heterometallic L-type linkage under non-reducing conditions. Practically, it showcases a cationic In→TM adduct that is held together exclusively by neutral donor ligation plus a weakly coordinating anion, thereby expanding the transferable, L-type In(I) toolbox for building coinage-metal assemblies and probing ligand-centered vs metal-centered electronic delocalization in mixed-metal cations.

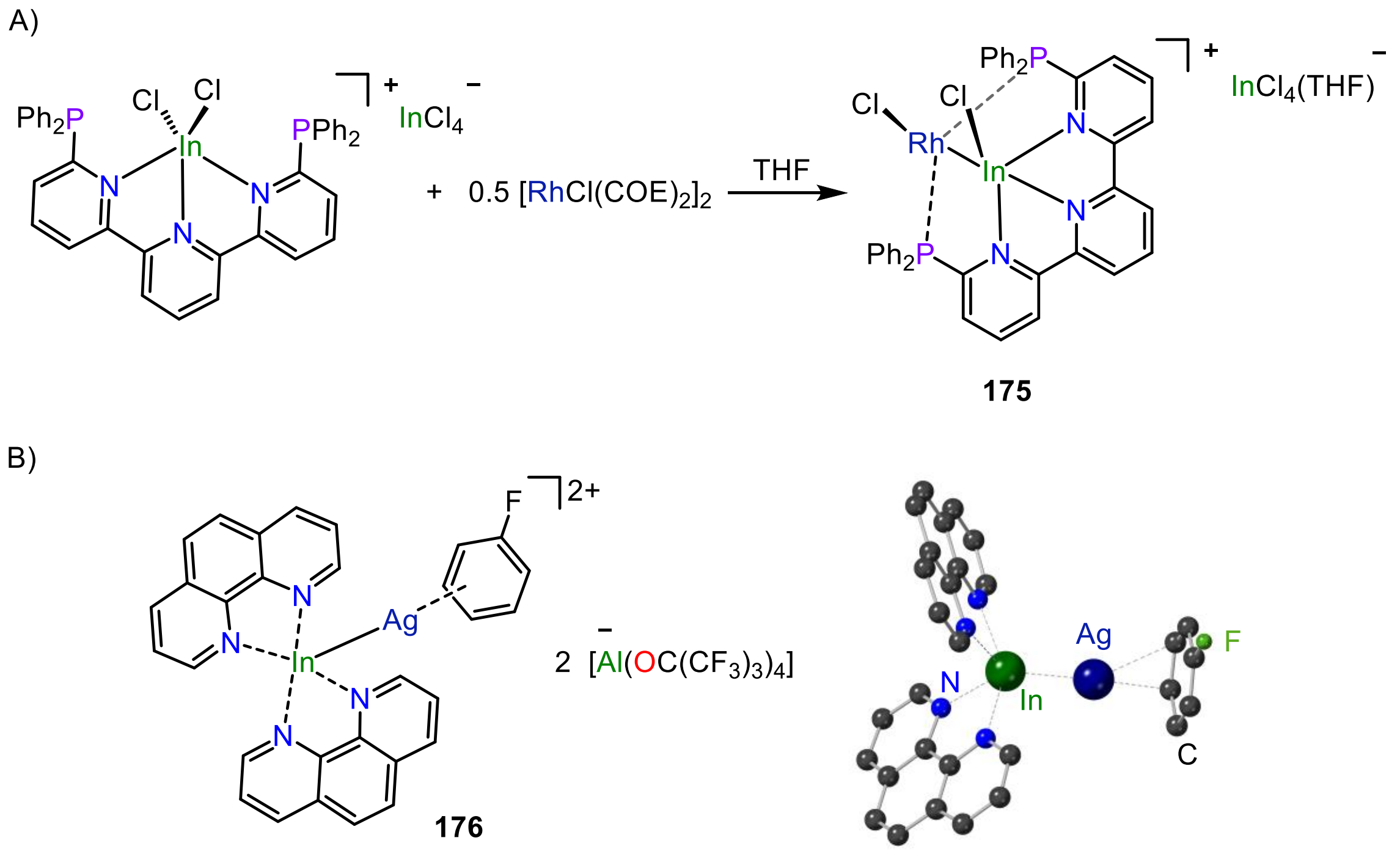

**Figure 57.** In–Rh and In–Ag complexes by pincer and chelating pyridine ligands.

Across all these examples, a unifying theme is that steric bulk and/or chelating support are crucial to tame the reactivity of In(I) and allow it to function as a ligand. Bulky aryls (e.g. terphenyls, mesityl), anionic chelates (BDI, Tp*, guanidinates), and intramolecular Lewis base stabilization all help prevent the indium from oligomerizing or disproportionating. Consequently, the indium can deploy its lone pair to coordinate transition metals, forming a spectrum of complexes from simple 2-center, 2-electron bonds (analogous to classical donor ligands) to multi-center clusters and insertion products. In addition, many of these In–TM complexes exhibit noteworthy bonding features. In several cases, especially with low-valent early or late transition metals, there is evidence of metal-to-indium π-backbonding, reinforcing the isolobal analogy between RIn and CO. This is seen in electron-rich metals (Group 4 metallocenes, Ni/Fe carbonyls, etc.) where donation into indium's acceptor orbitals can occur. On the other hand, indium's σ-donor strength is generally less than that of carbonyl or nitrosyl, but comparable to bulky phosphines; thus, it stabilizes metals in low formal oxidation states by contributing electron density.

Overall, the chemistry of bulky In(I) ligands with transition metals is a burgeoning field that has moved well beyond simple carbonyl substitution. From early studies that mirrored CO analogues, it has expanded to rich coordination chemistry and bond activation chemistry. In(I) ligands enable the construction of heterometallic complexes that are otherwise difficult to obtain, and they serve as useful models for understanding metal–metal interactions, isolobal analogies, and the periodic trends going down Group 13. The highly detailed studies on Ga(I) and Al(I) analogs set the stage, but In(I)'s unique properties (softer Lewis base, larger radius, accessible vacant orbitals) give rise to distinctly indium-centric outcomes, including higher cluster nuclearities and insertion processes, all while still functioning as an L-type ligand in the formal sense. These investigations not only achieve comprehensive coverage of Group 13 L-type metalloligand behavior but also push the boundaries of inorganic synthesis, revealing new facets of metal–metal bonding and reactivity that involve the often-underappreciated element indium.

## 5 Tl(I) as an L-type donor in transition-metal complexes? A reality check

Tl(I) is the heaviest Group-13 congener and exhibits the strongest inert-pair effect with the $6s^2$ electrons remaining contracted/low-lying and the 6p acceptors being diffuse. Both features disfavor neutral two-electron donation to transition metals compared with Al(I)/Ga(I)/In(I).[10b, 127] Accordingly, well-defined Tl(I)→TM L-type complexes are exceedingly rare. In practice, Tl(I) is encountered mostly as a closed-shell cation in metallophilic/outer-sphere assemblies (e.g., Pt···Tl, Au···Tl),[128] or as a Z-type (Lewis-acid) partner at electron-rich metals (reverse-dative M→Tl σ-donation).[129]

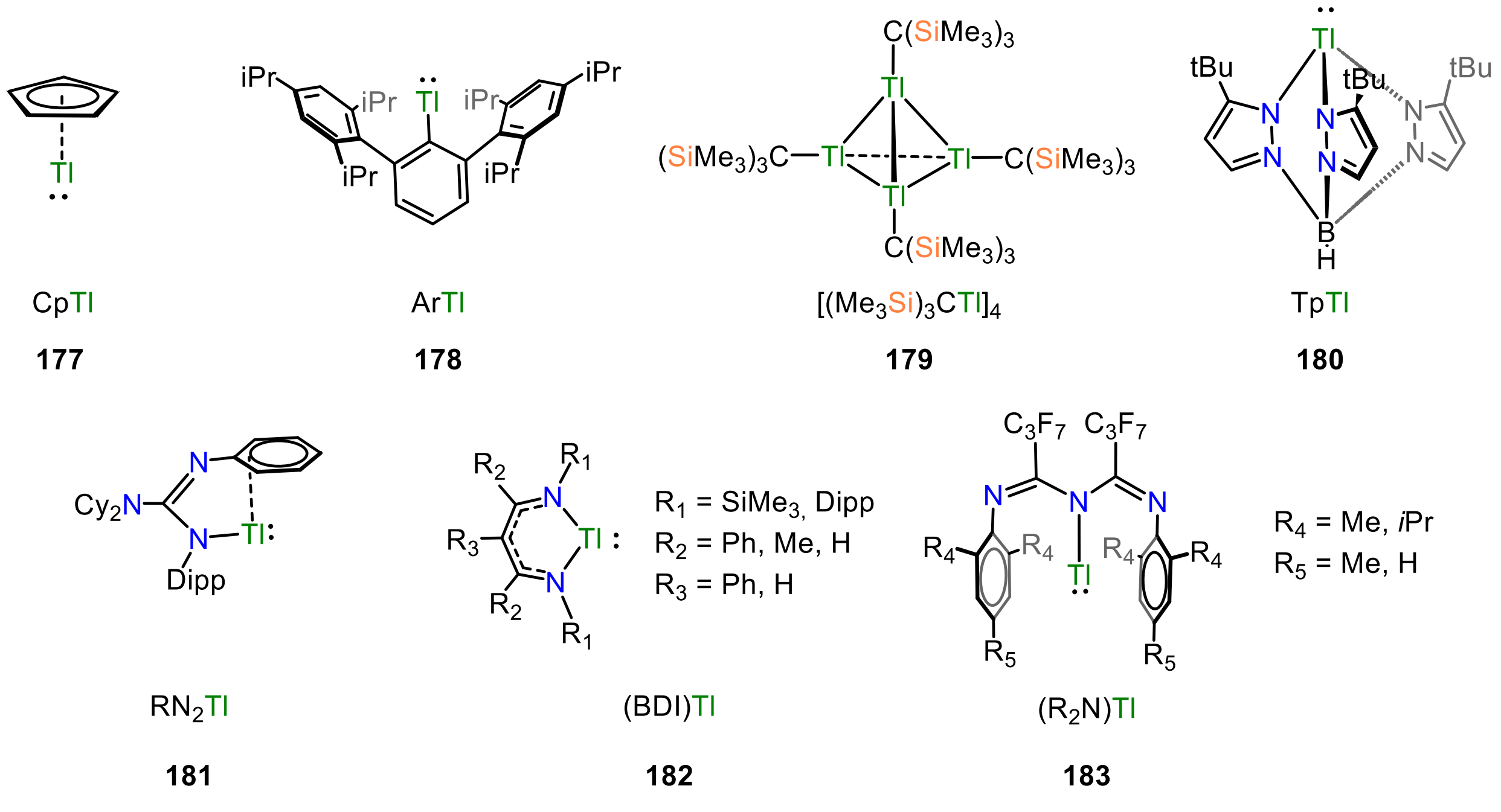

**Figure 58.** Sterically encumbered Tl(I) synthons/precursors.

To isolate monomeric Tl(I) that could, in principle, donate, extreme steric protection is required (e.g., TlCp / TlCp* **177**, super-bulky aryl-Tl(I) **178**, TlTp **180**, (BDI)Tl(I) **182**, and related fluorinated frameworks **183**, Figure 58).[100, 106, 130] Such species are useful precursors/transfer agents. In a few cases, formally donor behavior is observable, for instance, ArTl monomer acts as a Lewis base toward $B(C_6F_5)_3$, giving ArTl→$B(C_6F_5)_3$ with a near-linear $C_{ipso}$–Tl–B linkage.[127] Yet even here, Tl(I) resists further two-electron chemistry. The same ArTl refusesTl(I)→Tl(III) oxidation under conditions where lighter congeners react.[127] Overall, in sharp contrast with Ga(I) or Al(I) (which readily show oxidative addition/insertion at the TM center), Tl(I) more often forms weak/labile dative contacts or adopts bridging/contact-ion roles in multi-center or electrostatic frameworks. Thus, Tl(I) is best viewed as a transmetalation/outer-sphere partner to TMs,[131] rather than a practical neutral Tl(I)→TM L-type metalloligand.

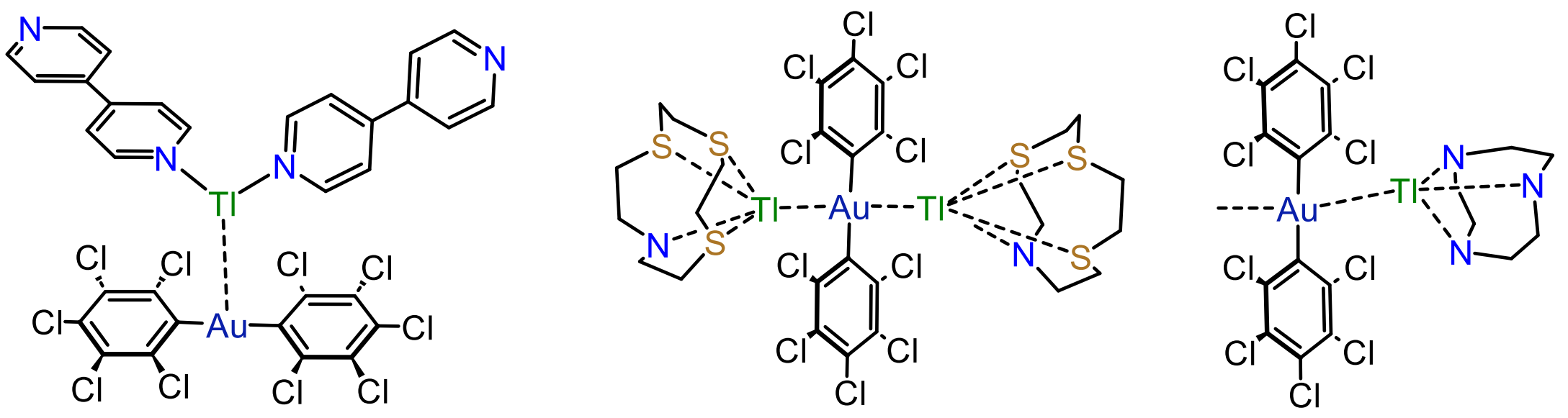

**184** **185** **186**

**Figure 59.** Selected Au→Tl assemblies illustrating soft-acid Tl(I) bound environments

Historically, Tl(I) transition-metal carbonyls are metalate salts such as $Tl[Co(CO)_4]$, where $Tl^+$ is a counter-cation; bonding to the TM is ionic, not neutral Tl→TM dative.[132] For group-10 metals (Ni, Pd, Pt), reverse-dative M→Tl σ-interactions have been unambiguously established.[129] In these examples Tl(I) behaves as a Z-type ligand at electron-rich TM centers. While numerous Tl–Pt and Tl–Au assemblies display short M⋯Tl contacts, red-shifted absorptions, and enhanced luminescence, these properties arise from metallophilic/Z-type interactions, not from Tl→TM donation.[128e, 133]

Representative Tl–Au complexes **184–186** (Figure 59), illustrate soft-acid Tl(I) environments, with applications in photophysical applications, such as in volatile organic vapor sensors (VOC), light-emitting devices (LEDs).[128, 134] with applications in photophysics, such as in volatile organic vapor sensors (VOC), light-emitting devices (LED's).

## 6 Conclusions and future outlook

Across the case studies assembled here, a coherent picture emerges, low-valent Group-13 M(I) fragments can be fashioned into neutral L-type metalloligands that genuinely recalibrate the structure and behavior of transition-metal complexes. A robust periodic logic underpins the whole field. Aluminylenes (Al(I)), are the most accessible and strongest σ-donors in the series, gallylenes (Ga(I)) and indylenes (In(I)) are competent L-donors but demand tighter steric and electronic control and display a greater tendency to adopt μ-bridging roles and to nucleate clusters, and Tl(I) systems have not yet been convincingly shown to sustain a neutral Tl→TM L-type linkage, appearing instead most reliably as a Z-type (reverse-dative) or metallophilic partners. Both terminal and bridging coordination modes are synthetically accessible, the latter often unlocks multi-center bonding and cluster manifolds that classical ligands rarely access. Throughout, rigorous CBC (L/X/Z) assignment and explicit oxidation-state accounting are indispensable, because closely related scaffolds can drift between neutral (L-type) and anionic (X-type) behavior depending on how the M(I) fragment is delivered.

Viewed element by element, the M–TM patterns are striking. Al(I) donors act as flagship L-ligands, furnishing robust Al→TM σ-bonds and well-defined μ-Al bridges that enable two-site, metal–ligand cooperation in H–H, Si–H and C–H activation and have already underpinned first catalytic cycles at earth-abundant metals. Ga(I) systems combine strong σ-donation with selectively tunable π-acceptance, stabilize polyhydrides, and participate in cooperative bond activation and photochemical ligand reorganization. In(I) donors extend carbonyl-analog In→TM coordination yet preferentially occupy bridging positions in clusters and, under designed conditions, insert into M–M or M–X links, and in coinage-metal settings they carry distinctive photophysical signatures that can be harnessed as functional handles. Tl(I) donors remain the frontier in the group for genuine L-type ligation. To date, Tl(I) most reliable contributions are Z-type or metallophilic, where its heavy-atom characteristics can still be put to use as structural or optical modulators rather than as neutral donors.

Conceptually and practically, Group-13 L-type ligation offers a principled route to endow base-metal complexes with new capabilities. By delivering a second, chemically distinct metal site adjacent to the d-block center and by modulating electron density through strong σ-donation combined with controllable π-acceptance, these metalloligand M(I) donors open complementary reaction pathways that monometallic, purely spectator-ligand platforms do not access as readily. In practice this has already been translated into selective activation of challenging σ-bonds, cluster-mediated transformations that exploit μ-element bridges for electron redistribution, and photochemical responses arising from metallophilic or ligand-centered excited states. The design canons distilled here, using peripheral bulk to steer terminal versus μ-bridging engagement, matching M(I) donors to electrophilic, coordinatively unsaturated TM fragments, and engineering ambiphilicity to stabilize but not immobilize the L-bond, form a transferable playbook for building heterometallic platforms that prioritize function over form.

Group-13 L-type donors have matured from curiosities to strategic levers for designing next-generation heterometallic platforms. Looking ahead, unlocking new ligand architectures, co-evolved jointly with mechanistic targets would undoubtedly advance the field. On the ligand side, the goal should shift from generic "very bulky" to electronically modular constructs, chelate-on-chelate and pincer/BDI hybrids, judiciously placed hard/soft donors, and designed hemilability, that allow σ-donation and π-acceptance to be tuned independently while suppressing oligomerization in heavier donors. On the reactivity side, cooperative catalysis should become an explicit design requirement rather than a serendipitous finding. Reactions that intrinsically require two-site metal–metal designs, with the Group-13 center acting as a Lewis base and latent hydride/alkyl carrier and the transition metal serving as the redox engine, offer a coherent path to small-molecule and related σ-bond activations that mononuclear analogs struggle to achieve. In parallel, paired synthetic strategies should be developed to map L↔X manifolds with precision. Engineering switchable neutral↔anionic M(I) routes (e.g., aluminylene–TM↔aluminyl–TM) will capture cross-over pathways, quantify when bonding crosses classification, and reveal new ambiphilicity signatures that can be leveraged in base-metal cooperativity and catalysis.

## 7 Acknowledgements

We thank the Spanish Ministerio de Ciencia, Innovación y Universidades MICIN/AEI/10.13039/501100011033 (PID2022-142270OB-I00) and MICIN/NextGenerationEU/PRTR (CNS2022-136087 for A.J.M.M., CNS2024-154163 for M.A.F), the European Research and Development Fund (ERDF) and the University of Huelva (EPIT1442023). M.A.F acknowledges the Ministry of Science and Innovation for a Ramón y Cajal Research Contract (RYC2023-042945-I).